\documentclass[11pt]{article}
\usepackage{amsmath,amsfonts,amssymb,graphicx,epsfig,pdflscape,multirow,theorem,gensymb}
\usepackage[matrix,arrow,curve]{xy} 
\numberwithin{equation}{section}
\graphicspath{{./eps/}}
\allowdisplaybreaks

\usepackage[nosort]{cite}
\usepackage[usenames]{color}
\usepackage{pstricks}
\usepackage[backref=false]{hyperref}
\hypersetup{
colorlinks=true,
citecolor=red,
linkcolor=darkblue
}
\definecolor{darkblue}{rgb}{0,0,.8}
\definecolor{red}{rgb}{1,0,0}
\newcommand{\nc}{\newcommand}
\nc{\fh}{\hat{f}}
\nc{\muh}{\hat{\mu}}
\nc{\nuh}{\hat{\nu}}
\nc{\bib}{\bibitem}
\nc{\al}{\alpha}
\nc{\mch}{\mathrm{ch}}
\nc{\g}{\gamma}
\nc{\G}{\Gamma}
\nc{\D}{\Delta}
\nc{\eps}{\epsilon}
\nc{\la}{\lambda}
\nc{\La}{\Lambda}
\nc{\var}{\varphi}
\nc{\pa}{\partial}
\nc{\nn}{\nonumber \\ }
\nc{\hf}{\frac{1}{2}}
\nc{\dz}{\frac{dz}{2\pi i}}
\nc{\bin}[2]{\left(\!\!\!\begin{array}{c} {#1}\\ {#2} \end{array}\!\!\!\right)}
\nc{\be}{\begin{equation}}
\nc{\ee}{\end{equation}}
\nc{\bea}{\begin{eqnarray}}
\nc{\eea}{\end{eqnarray}}
\nc{\bra}[1]{\langle {#1}|}
\nc{\ket}[1]{|{#1}\rangle}
\def\Re{\mathop{\rm Re}\nolimits}

\def\cosec{\mathop{\rm cosec}}
\nc{\chit}{\raisebox{0.25ex}{$\chi$}}
\nc{\ch}{{\rm ch}}
\nc{\Bb}{\mbox{\boldmath $B$}}
\nc{\Fb}{\mbox{\boldmath $F$}}
\nc{\Hb}{\mbox{\boldmath $H$}}
\nc{\Ib}{\mbox{\boldmath $I$}}
\nc{\Jb}{\mbox{\boldmath $J$}}
\nc{\Rb}{\mbox{\boldmath $R$}}
\nc{\Tb}{\mbox{\boldmath $T$}}

\nc{\Hc}{{\cal H}}
\nc{\Rc}{{\cal R}}
\nc{\Lc}{{\cal L}}
\nc{\Vc}{{\cal V}}
\nc{\rbar}{\bar{r}}
\nc{\qbar}{\bar{q}}
\nc{\kbar}{\bar{k}}
\def\floor#1{\lfloor #1\rfloor}
\nc{\sbin}[2]{\left\{\!\!\!\begin{array}{c} {#1}\\ {#2} \end{array}\!\!\!\right\}}
\nc{\sbinlr}[2]{\Big\langle\!\!\begin{array}{c} {#1}\\ {#2} \end{array}\!\!\Big\rangle}
\def\vvdots{\mathinner{\mkern1mu\raise1pt\vbox{\kern7pt\hbox{.}}\mkern2mu
  \raise4pt\hbox{.}\mkern2mu\raise7pt\hbox{.}\mkern1mu}}
\nc{\gauss}[2]{\left[\!\!\begin{array}{c} {#1}\\ {#2} \end{array}\!\!\right]_{\!q}}
\nc{\gaussbar}[2]{\left[\!\!\begin{array}{c} {#1}\\ {#2} \end{array}\!\!\right]_{\!\qbar}}
\nc{\perm}[2]{\left[\!\!\begin{array}{c} {#1}\\ \\ {#2} \end{array}\!\!\right]}
\nc{\bino}[2]{\left(\!\!\begin{array}{c} {#1}\\ {#2} \end{array}\!\!\right)}
\def\half {\mbox{$\textstyle \frac{1}{2}$}}
\def\vec#1{\mbox {\boldmath $#1$}}
\def\svec#1{\mbox {\scriptsize\boldmath $#1$}}
\definecolor{lightblue}{rgb}{.7,.7,1}
\definecolor{purple}{rgb}{1,0,1}
\definecolor{lightlightblue}{rgb}{.92,.92,1}
\definecolor{lightyellow}{rgb}{.99,.99,.85}

\def\leftarc#1{\psarc[linecolor=black,linewidth=1.5pt]#1{.5}{90}{270}}
\def\rightarc#1{\psarc[linecolor=black,linewidth=1.5pt]#1{.5}{-90}{90}}

\def\facegrid#1#2{
\psframe[fillstyle=solid,fillcolor=lightlightblue,linewidth=0pt]#1#2
\psgrid[gridlabels=0pt,subgriddiv=1]#1#2}

\definecolor{sky}{rgb}{.4,.8,1}
\definecolor{orange}{rgb}{1,.4,0}
\definecolor{green}{rgb}{0,1,0}
\definecolor{brightblue}{rgb}{0,1,1}
\definecolor{mediumblue}{rgb}{.34,.446,1}
\definecolor{greyblue}{rgb}{.345,.345,.7424}
\definecolor{bglightblue}{rgb}{.625,.845,.917}
\definecolor{graphblue}{rgb}{.719,.918,.996}
\definecolor{lightlightblue}{rgb}{.89,.89,1}
\definecolor{lightpurple}{rgb}{1,.88,1}
\definecolor{WeekColor}{rgb}{1,.7,.9}
\definecolor{lightblue}{rgb}{.55,.55,1}
\definecolor{midblue}{rgb}{.7,.7,1}
\definecolor{lightlightblue}{rgb}{.89,.89,1}
\definecolor{lightestblue}{rgb}{.96,.96,1}

\def\qbin#1#2#3{{\genfrac{[}{]}{0pt}{}{#1}{#2}}_{#3}}
\def\sqbin#1#2#3{\mbox{$\genfrac{[}{]}{0pt}{}{#1}{#2}$}_{#3}}
\def\sc#1{\mbox{\scriptsize $#1$}}
\def\sm#1{\mbox{\small $#1$}}
\def\disp{\displaystyle}
\def\Tr{\mathop{\mbox{Tr}}}

\definecolor{apricot}{rgb}{.99,.87,.77}
\def\var{
\pspolygon[linewidth=.25pt,fillstyle=solid,fillcolor=lightlightblue](0,0)(1,0)(1,1)(0,1)(0,0)
\psline[linewidth=1.pt,arrowsize=9pt,linecolor=blue]{>->}(-.15,.5)(1.15,.5)
\psline[linewidth=1.pt,arrowsize=9pt,linecolor=blue]{>->}(.5,-.15)(.5,1.15)}

\def\vcs{
\pspolygon[linewidth=.25pt,fillstyle=solid,fillcolor=lightpurple](0,0)(1,0)(1,1)(0,1)(0,0)
\psline[linewidth=1.pt,arrowsize=9pt,linecolor=blue]{<-<}(-.15,.5)(1.15,.5)
\psline[linewidth=1.pt,arrowsize=9pt,linecolor=blue]{>->}(.5,-.15)(.5,1.15)}

\def\va{
\pspolygon[linewidth=.25pt,fillstyle=solid,fillcolor=lightlightblue](0,0)(1,0)(1,1)(0,1)(0,0)
\psline[linewidth=1.pt,arrowsize=9pt]{>->}(-.15,.5)(1.15,.5)
\psline[linewidth=1.pt,arrowsize=9pt]{>->}(.5,-.15)(.5,1.15)}
\def\vb{
\pspolygon[linewidth=.25pt,fillstyle=solid,fillcolor=lightlightblue](0,0)(1,0)(1,1)(0,1)(0,0)
\psline[linewidth=1.pt,arrowsize=9pt]{<-<}(-.15,.5)(1.15,.5)
\psline[linewidth=1.pt,arrowsize=9pt]{<-<}(.5,-.15)(.5,1.15)}
\def\vc{
\pspolygon[linewidth=.25pt,fillstyle=solid,fillcolor=lightlightblue](0,0)(1,0)(1,1)(0,1)(0,0)
\psline[linewidth=1.pt,arrowsize=9pt]{<-<}(-.15,.5)(1.15,.5)
\psline[linewidth=1.pt,arrowsize=9pt]{>->}(.5,-.15)(.5,1.15)}
\def\vd{
\pspolygon[linewidth=.25pt,fillstyle=solid,fillcolor=lightlightblue](0,0)(1,0)(1,1)(0,1)(0,0)
\psline[linewidth=1.pt,arrowsize=9pt]{>->}(-.15,.5)(1.15,.5)
\psline[linewidth=1.pt,arrowsize=9pt]{<-<}(.5,-.15)(.5,1.15)}
\def\ve{
\pspolygon[linewidth=.25pt,fillstyle=solid,fillcolor=lightlightblue](0,0)(1,0)(1,1)(0,1)(0,0)
\psline[linewidth=1.pt,arrowsize=9pt]{<->}(-.15,.5)(1.15,.5)
\psline[linewidth=1.pt,arrowsize=9pt]{>-<}(.5,-.15)(.5,1.15)}
\def\vf{
\pspolygon[linewidth=.25pt,fillstyle=solid,fillcolor=lightlightblue](0,0)(1,0)(1,1)(0,1)(0,0)
\psline[linewidth=1.pt,arrowsize=9pt]{>-<}(-.15,.5)(1.15,.5)
\psline[linewidth=1.pt,arrowsize=9pt]{<->}(.5,-.15)(.5,1.15)}
\def\vva{
\pspolygon[linewidth=.25pt,fillstyle=solid,fillcolor=lightpurple](0,0)(1,0)(1,1)(0,1)(0,0)
\psline[linewidth=1.pt,arrowsize=9pt]{>->}(-.15,.5)(1.15,.5)
\psline[linewidth=1.pt,arrowsize=9pt]{>->}(.5,-.15)(.5,1.15)}
\def\vvb{
\pspolygon[linewidth=.25pt,fillstyle=solid,fillcolor=lightpurple](0,0)(1,0)(1,1)(0,1)(0,0)
\psline[linewidth=1.pt,arrowsize=9pt]{<-<}(-.15,.5)(1.15,.5)
\psline[linewidth=1.pt,arrowsize=9pt]{<-<}(.5,-.15)(.5,1.15)}
\def\vvc{
\pspolygon[linewidth=.25pt,fillstyle=solid,fillcolor=lightpurple](0,0)(1,0)(1,1)(0,1)(0,0)
\psline[linewidth=1.pt,arrowsize=9pt]{<-<}(-.15,.5)(1.15,.5)
\psline[linewidth=1.pt,arrowsize=9pt]{>->}(.5,-.15)(.5,1.15)}
\def\vvd{
\pspolygon[linewidth=.25pt,fillstyle=solid,fillcolor=lightpurple](0,0)(1,0)(1,1)(0,1)(0,0)
\psline[linewidth=1.pt,arrowsize=9pt]{>->}(-.15,.5)(1.15,.5)
\psline[linewidth=1.pt,arrowsize=9pt]{<-<}(.5,-.15)(.5,1.15)}
\def\vve{
\pspolygon[linewidth=.25pt,fillstyle=solid,fillcolor=lightpurple](0,0)(1,0)(1,1)(0,1)(0,0)
\psline[linewidth=1.pt,arrowsize=9pt]{<->}(-.15,.5)(1.15,.5)
\psline[linewidth=1.pt,arrowsize=9pt]{>-<}(.5,-.15)(.5,1.15)}
\def\vvf{
\pspolygon[linewidth=.25pt,fillstyle=solid,fillcolor=lightpurple](0,0)(1,0)(1,1)(0,1)(0,0)
\psline[linewidth=1.pt,arrowsize=9pt]{>-<}(-.15,.5)(1.15,.5)
\psline[linewidth=1.pt,arrowsize=9pt]{<->}(.5,-.15)(.5,1.15)}
\def\pa{
\pspolygon[linewidth=.25pt,fillstyle=solid,fillcolor=lightlightblue](0,0)(1,0)(1,1)(0,1)(0,0)
\psarc[linewidth=.5pt,linecolor=red](0,0){.1}{0}{90}}
\def\pb{
\pspolygon[linewidth=.25pt,fillstyle=solid,fillcolor=lightlightblue](0,0)(1,0)(1,1)(0,1)(0,0)
\psline[linewidth=2pt](0,.5)(.4,.5)
\psline[linewidth=2pt](.5,0)(.5,.4)
\psline[linewidth=2pt](.6,.5)(1,.5)
\psline[linewidth=2pt](.5,.6)(.5,1)
\psarc[linewidth=2pt](.6,.4){.1}{90}{180}
\psarc[linewidth=2pt](.4,.6){.1}{270}{0}
\psarc[linewidth=.5pt,linecolor=red](0,0){.1}{0}{90}}
\def\pc{
\pspolygon[linewidth=.25pt,fillstyle=solid,fillcolor=lightlightblue](0,0)(1,0)(1,1)(0,1)(0,0)
\psline[linewidth=2pt](0,.5)(1,.5)
\psarc[linewidth=.5pt,linecolor=red](0,0){.1}{0}{90}}
\def\pd{
\pspolygon[linewidth=.25pt,fillstyle=solid,fillcolor=lightlightblue](0,0)(1,0)(1,1)(0,1)(0,0)
\psline[linewidth=2pt](.5,0)(.5,1)
\psarc[linewidth=.5pt,linecolor=red](0,0){.1}{0}{90}}
\def\pe{
\pspolygon[linewidth=.25pt,fillstyle=solid,fillcolor=lightlightblue](0,0)(1,0)(1,1)(0,1)(0,0)
\psline[linewidth=2pt](0,.5,)(.4,.5)
\psline[linewidth=2pt](.5,.6)(.5,1)
\psarc[linewidth=2pt](.4,.6){.1}{270}{0}
\psarc[linewidth=.5pt,linecolor=red](0,0){.1}{0}{90}}
\def\pf{
\pspolygon[linewidth=.25pt,fillstyle=solid,fillcolor=lightlightblue](0,0)(1,0)(1,1)(0,1)(0,0)
\psline[linewidth=2pt](.6,.5)(1,.5)
\psline[linewidth=2pt](.5,0)(.5,.4)
\psarc[linewidth=2pt](.6,.4){.1}{90}{180}
\psarc[linewidth=.5pt,linecolor=red](0,0){.1}{0}{90}}

\def\qa{
\pspolygon[linewidth=.25pt,fillstyle=solid,fillcolor=lightpurple](0,0)(1,0)(1,1)(0,1)(0,0)
\psarc[linewidth=.5pt,linecolor=red](1,0){.1}{90}{180}}
\def\qb{
\pspolygon[linewidth=.25pt,fillstyle=solid,fillcolor=lightpurple](0,0)(1,0)(1,1)(0,1)(0,0)
\psline[linewidth=2pt](0,.5)(.4,.5)
\psline[linewidth=2pt](.5,0)(.5,.4)
\psline[linewidth=2pt](.6,.5)(1,.5)
\psline[linewidth=2pt](.5,.6)(.5,1)
\psarc[linewidth=2pt](.4,.4){.1}{0}{90}
\psarc[linewidth=2pt](.6,.6){.1}{180}{270}
\psarc[linewidth=.5pt,linecolor=red](1,0){.1}{90}{180}}
\def\qc{
\pspolygon[linewidth=.25pt,fillstyle=solid,fillcolor=lightpurple](0,0)(1,0)(1,1)(0,1)(0,0)
\psline[linewidth=2pt](0,.5)(1,.5)
\psarc[linewidth=.5pt,linecolor=red](1,0){.1}{90}{180}}
\def\qd{
\pspolygon[linewidth=.25pt,fillstyle=solid,fillcolor=lightpurple](0,0)(1,0)(1,1)(0,1)(0,0)
\psline[linewidth=2pt](.5,0)(.5,1)
\psarc[linewidth=.5pt,linecolor=red](1,0){.1}{90}{180}}

\def\qg{
\pspolygon[linewidth=.25pt,fillstyle=solid,fillcolor=lightpurple](0,0)(1,0)(1,1)(0,1)(0,0)
\psline[linewidth=2pt](.6,.5,)(1,.5)
\psline[linewidth=2pt](.5,.6)(.5,1)
\psarc[linewidth=2pt](.6,.6){.1}{180}{270}
\psarc[linewidth=.5pt,linecolor=red](1,0){.1}{90}{180}}
\def\qh{
\pspolygon[linewidth=.25pt,fillstyle=solid,fillcolor=lightpurple](0,0)(1,0)(1,1)(0,1)(0,0)
\psline[linewidth=2pt](.5,0)(.5,.4)
\psline[linewidth=2pt](0,.5)(.4,.5)
\psarc[linewidth=2pt](.4,.4){.1}{0}{90}
\psarc[linewidth=.5pt,linecolor=red](1,0){.1}{90}{180}}

\def\daDimer{
\pspolygon[linewidth=1.5pt,fillstyle=solid,fillcolor=lightyellow](0,0)(.5,-.5)(1.5,.5)(1,1)(0,0)
\pspolygon[linewidth=.25pt,fillstyle=solid,fillcolor=lightlightblue](0,0)(1,0)(1,1)(0,1)(0,0)
\psline[linewidth=1.5pt](0,0)(1,1)
\psline[linewidth=1.5pt](0,1)(.5,.5)}
\def\dbDimer{
\pspolygon[linewidth=1.5pt,fillstyle=solid,fillcolor=lightyellow](0,0)(-.5,.5)(.5,1.5)(1,1)(0,0)
\pspolygon[linewidth=.25pt,fillstyle=solid,fillcolor=lightlightblue](0,0)(1,0)(1,1)(0,1)(0,0)
\psline[linewidth=1.5pt](0,0)(1,1)
\psline[linewidth=1.5pt](.5,.5)(1,0)}
\def\dcDimer{
\pspolygon[linewidth=1.5pt,fillstyle=solid,fillcolor=lightyellow](0,1)(-.5,.5)(.5,-.5)(1,0)(0,0)
\pspolygon[linewidth=.25pt,fillstyle=solid,fillcolor=lightlightblue](0,0)(1,0)(1,1)(0,1)(0,0)
\psline[linewidth=1.5pt](0,1)(1,0)
\psline[linewidth=1.5pt](1,1)(.5,.5)}
\def\ddDimer{
\pspolygon[linewidth=1.5pt,fillstyle=solid,fillcolor=lightyellow](0,1)(.5,1.5)(1.5,.5)(1,0)(0,0)
\pspolygon[linewidth=.25pt,fillstyle=solid,fillcolor=lightlightblue](0,0)(1,0)(1,1)(0,1)(0,0)
\psline[linewidth=1.5pt](0,1)(1,0)
\psline[linewidth=1.5pt](.5,.5)(0,0)}
\def\deDimer{
\pspolygon[linewidth=1.5pt,fillstyle=solid,fillcolor=lightyellow](0,0)(.5,-.5)(1.5,.5)(1,1)(0,0)
\pspolygon[linewidth=1.5pt,fillstyle=solid,fillcolor=lightyellow](0,0)(-.5,.5)(.5,1.5)(1,1)(0,0)
\pspolygon[linewidth=.25pt,fillstyle=solid,fillcolor=apricot](0,0)(1,0)(1,1)(0,1)(0,0)
\psline[linewidth=1.5pt](0,0)(1,1)}
\def\deeDimer{
\pspolygon[linewidth=1.5pt,fillstyle=solid,fillcolor=lightyellow](0,1)(-.5,.5)(.5,-.5)(1,0)(0,0)
\pspolygon[linewidth=1.5pt,fillstyle=solid,fillcolor=lightyellow](0,1)(.5,1.5)(1.5,.5)(1,0)(0,0)
\pspolygon[linewidth=.25pt,fillstyle=solid,fillcolor=apricot](0,0)(1,0)(1,1)(0,1)(0,0)
\psline[linewidth=1.5pt](1,0)(0,1)}

\def\da{
\pspolygon[linewidth=.25pt,fillstyle=solid,fillcolor=lightlightblue](0,0)(1,0)(1,1)(0,1)(0,0)
\psline[linewidth=1.5pt](0,0)(1,1)
\psline[linewidth=1.5pt](0,1)(.5,.5)}
\def\db{
\pspolygon[linewidth=.25pt,fillstyle=solid,fillcolor=lightlightblue](0,0)(1,0)(1,1)(0,1)(0,0)
\psline[linewidth=1.5pt](0,0)(1,1)
\psline[linewidth=1.5pt](.5,.5)(1,0)}
\def\dc{
\pspolygon[linewidth=.25pt,fillstyle=solid,fillcolor=lightlightblue](0,0)(1,0)(1,1)(0,1)(0,0)
\psline[linewidth=1.5pt](0,1)(1,0)
\psline[linewidth=1.5pt](1,1)(.5,.5)}
\def\dd{
\pspolygon[linewidth=.25pt,fillstyle=solid,fillcolor=lightlightblue](0,0)(1,0)(1,1)(0,1)(0,0)
\psline[linewidth=1.5pt](0,1)(1,0)
\psline[linewidth=1.5pt](.5,.5)(0,0)}
\def\de{
\pspolygon[linewidth=.25pt,fillstyle=solid,fillcolor=apricot](0,0)(1,0)(1,1)(0,1)(0,0)
\psline[linewidth=1.5pt](0,0)(1,1)}
\def\dee{
\pspolygon[linewidth=.25pt,fillstyle=solid,fillcolor=apricot](0,0)(1,0)(1,1)(0,1)(0,0)
\psline[linewidth=1.5pt](1,0)(0,1)}
\def\df{
\pspolygon[linewidth=.25pt,fillstyle=solid,fillcolor=lightlightblue](0,0)(1,0)(1,1)(0,1)(0,0)
\psline[linewidth=1.5pt](0,0)(1,1)
\psline[linewidth=1.5pt](0,1)(1,0)}

\begin{document}

\topmargin -5mm
\oddsidemargin 5mm

\begin{titlepage}
\setcounter{page}{0}

\vspace{8mm}
\begin{center}
{\huge {\bf Yang-Baxter Solution of Dimers\\[8pt] as a Free-Fermion Six-Vertex Model}}

\vspace{10mm}
{\Large Paul A. Pearce, Alessandra Vittorini-Orgeas}\\[.3cm]
{\em School of Mathematics and Statistics, University of Melbourne}\\
{\em Parkville, Victoria 3010, Australia}\\[.4cm]
papearce@unimelb.edu.au, alessandra.vittorini@unimelb.edu.au

\end{center}

\vspace{10mm}
\centerline{{\bf{Abstract}}}
\vskip.4cm
\noindent
It is shown that dimers is Yang-Baxter \mbox{integrable} as a six-vertex model at the  free-fermion point with crossing parameter $\lambda=\tfrac{\pi}{2}$. A one-to-many mapping of vertex onto dimer configurations allows the free-fermion solutions to be applied to the anisotropic dimer model on a square lattice where the dimers are rotated by $45\degree$ compared to their usual orientation. This dimer model is exactly solvable in geometries of arbitrary finite size. In this paper, we establish and solve inversion identities for dimers with periodic boundary conditions on the cylinder. In the particle representation, the local face tile operators give a representation of the fermion algebra and the fermion particle trajectories play the role of nonlocal (logarithmic) degrees of freedom. In a suitable gauge, the dimer model is described by the Temperley-Lieb algebra with loop fugacity $\beta=2\cos\lambda=0$. At the isotropic point, the exact solution allows for the explicit counting of $45\degree$ rotated dimer configurations on a periodic $M\times N$ rectangular lattice.  We show that the modular invariant partition function on the torus is the same as symplectic fermions and critical dense polymers. We also show that nontrivial Jordan cells appear for the dimer Hamiltonian on the strip with vacuum boundary conditions. We therefore argue that, in the continuum scaling limit, the dimer model gives rise to a {\em logarithmic} conformal field theory with central charge $c=-2$, minimal conformal weight $\Delta_{\text{min}}=-1/8$ and effective central charge $c_{\text{eff}}=1$. 

\end{titlepage}
\newpage
\renewcommand{\thefootnote}{\arabic{footnote}}
\setcounter{footnote}{0}

\tableofcontents

\newpage
\section{Introduction}

The six-vertex model~\cite{Pauling,Lieb,Sutherland,LiebWu} is a ferroelectric model on the square lattice that is Yang-Baxter integrable~\cite{BaxBook}. 
If the six vertex weights satisfy an additional free-fermion condition (\ref{FreeFermion}), the model reduces to a non-interacting 
free-fermion model~\cite{FanWu,Felderhof}. In contradistinction, the dimer (domino tiling) model~\cite{Roberts,FowlerRush} captures the molecular freedom in densely arranging (adsorbing) non-overlapping diatomic molecules (dimers) on a lattice substrate. In 1961, the dimer model on the square lattice was solved exactly~\cite{Kasteleyn1961,TempFisher,Fisher1961} by Pfaffians. 
While the original solutions were by Pfaffians, the idea of using a transfer matrix and free fermions was initiated in \cite{LiebTransfer}. After more than 50 years, the dimer model continues to be the subject of extensive study~\cite{IzOganHu2003,IzPRHu2005,IzPR2007,RasRuelle2012,Nigro2012,Allegra2015,MDRR2015,MDRR2016} primarily to understand the finite-size effects of boundary conditions and steric effects under the influence of infinitely repulsive hard-core local interactions. These effects are manifest in the related problems of Aztec diamonds~\cite{Aztec} and the six-vertex model with domain wall boundary conditions~\cite{DWBC,KorepinZJ}. 
For these systems, even the thermodynamic limit can fail to exist. 
Although such boundary conditions clearly break conformal invariance in the continuum scaling limit there are other boundary conditions, such as periodic and free boundary conditions, which are conformally invariant. For dimers with the latter boundary conditions it has been argued~\cite{IzPRHu2005,IzPR2007} that the system is best described as a logarithmic Conformal Field Theory (CFT) with central charge $c=-2$ rather than the $c=1$ Gaussian free field usually associated~\cite{Kenyon} with dimers and the six-vertex model at the free fermion point.

Although it is not immediately apparent, it is known that the six-vertex free-fermion and dimer models are related, at least at the level of configurations.  
In fact, a one-to-many mapping exists~\cite{Baxter1972,KorepinZJ,FerrariSpohn} from six-vertex configurations to dimer configurations. 
Notably, this maps  six-vertex configurations onto dimer configurations where the dimers are rotated by $45\degree$ compared to their usual orientation parallel to the bonds of the square lattice. As shown in Figure~\ref{Tconfigs}, in this orientation, each dimer covers two sites of the medial lattice whose sites consist of midpoints of the bonds of the original square lattice. For an $M\times N$ rectangular lattice with periodic boundary conditions, there are thus $2MN$ sites on the medial lattice covered by $MN$ dimers. 

In this paper, we solve exactly the anisotropic dimer model on the square lattice for dimers with the 45 degree rotated orientation. 
This is achieved, using Yang-Baxter integrability, by viewing the dimer model as a free-fermion six-vertex model and solving the associated inversion identity~\cite{Felderhof,BaxBook,OPW1996} satisfied by the transfer matrices. 
By considering periodic boundary conditions on the cylinder and closing to a torus by taking a matrix trace, we obtain the modular invariant partition function (MIPF). 
Notably, the MIPF of dimers precisely agrees with symplectic fermions~\cite{Symplectic} and critical dense polymers~\cite{SaleurSuper,PRpolymers,PRVKac,PRV,MDPR2013}. The latter is nontrivial because, as a special six-vertex model, the usual matrix trace is used to close the cylinder to a torus for the dimer model whereas a modified trace~\cite{MDPR2013} is needed for critical dense polymers. 
Remarkably, we find that the spectra of dimers agrees sector-by-sector with the spectra of critical dense polymers. 
However, an important difference with periodic boundary conditions is that the transfer matrix of critical dense polymers exhibits Jordan cells~\cite{PRV,MDSA2013} on the cylinder whereas the transfer matrix of dimers (six-vertex free-fermion model) is normal and diagonalizable so it does not exhibit Jordan cells. 
With vacuum boundary conditions on the strip, we show that the situation is the other way around --- the double row transfer matrices of critical dense polymers do not exhibit Jordan cells whereas the Hamiltonian and double row transfer matrices of dimers (six-vertex free-fermion model) do exhibit Jordan cells (see for example \cite{GST2014,GHNS2015,MDRRSA2015}). 
We therefore argue that dimers is a logarithmic CFT with central charge $c=-2$ and effective central charge $c_\text{eff}=1$.

It is to be stressed that our dimer model is exactly solvable in arbitrary {\em finite} geometries on the strip, cylinder and torus with a  variety of different integrable boundary conditions. On the one hand, for periodic boundary conditions, we obtain a {\em finitized\/} modular invariant partition function for dimers. On the other hand, at the isotropic point, we obtain explicit formulas for the counting of dimer configurations on a {\em finite} $M\times N$ periodic rectangular lattice. 

The layout of the paper is as follows. In Section~\ref{SecLattice} we describe the dimer model, with dimers rotated by $45\degree$\!, as a free-fermion six-vertex model and give the equivalence of the face tiles in the vertex, particle and dimer representations. We also describe the relations between the free-fermion algebra, the Temperley-Lieb algebra and the Yang-Baxter equation. We end this section by showing that the residual entropy is not changed by rotating the orientation of the dimers and agrees with the known result~\cite{Fisher1961}. In Section~\ref{SecCylTorus}, we introduce the periodic row transfer matrices and show that the associated dimer Hamiltonian reduces to the usual free-fermion hopping Hamiltonian. In this section, we also obtain the exact eigenvalues of the transfer matrices on finite cylinders. Following closely \cite{PRV}, we calculate the finite-size spectra in the $\mathbb{Z}_4$, Ramond and Neveu-Schwarz sectors. The $N$ even sectors are combined  to show that the MIPF is that of symplectic fermions~\cite{Symplectic}. This completes the CFT description of dimers on the torus. Next, in Section~\ref{FiniteRect}, we consider the isotropic point and obtain explicit formulas for the counting of rotated dimer configurations on a periodic $M\times N$ rectangular lattice. Although it displays the same asymptotic growth, the precise counting of these configurations differs from the counting in the usual orientation~\cite{Kasteleyn1961,IzOganHu2003}. 
Finally, in Section~5, we give a brief summary of the analysis of the double row transfer matrices of dimers on the strip with vacuum boundary conditions. Most importantly, for small system sizes, we show that the Hamiltonian coincides with the Hamiltonian of the $U_q(sl(2))$-invariant XX Hamiltonian and exhibits nontrivial rank-2 Jordan cells.

\section{Dimers as a Free-Fermion Six-Vertex Model}
\label{SecLattice}

\subsection{Face tiles and equivalence of vertex, particle and dimer representations}

The allowed six-vertex (arrow conserving) face configurations and the equivalent tiles in the particle (even and odd rows) and dimer~\cite{KorepinZJ} representations are shown in Figure~\ref{vpd}. For $N$ columns, the vertex (arrow) degrees of freedom $\sigma_j=\pm 1$ and the particle occupation numbers $a_j=\half(1-\sigma_j)=0,1$ live on the medial lattice with $j=1,2,\ldots,N$.  
The Boltzmann weights of the six-vertex tiles are
\bea
a(u)=\rho\,\frac{\sin(\lambda-u)}{\sin\lambda},\quad b(u)=\rho\,\frac{\sin u}{\sin\lambda},\quad c_1(u)=\rho g,\quad c_2(u)=\frac{\rho}{g},\qquad \lambda\in (0,\pi),\quad \rho\in\mathbb{R}\label{6Vwts}
\eea
The spectral parameter $u$ plays the role of spatial anisotropy with $u=\frac{\lambda}{2}$ being the isotropic point. 
Geometrically~\cite{KimP}, varying $u$ effectively distorts a square tile into a rhombus with an opening anisotropy angle $\vartheta=\frac{\pi u}{\lambda}$. 
The arbitrary parameter $\rho$ is an overall normalization. Assuming boundary conditions such that there are an equal number of sources and sinks of horizontal arrows (vertices $c_1$ and $c_2$) along any row, the transfer matrix entries (\ref{Tmatrix}) are all independent of the gauge factor~$g$. 

At the free-fermion point ($\lambda=\frac{\pi}{2}$), the six-vertex face weights reduce to
\bea
a(u)=\rho\cos u,\quad b(u)=\rho\sin u,\quad c_1(u)=\rho g,\quad c_2(u)=\frac{\rho}{g},\qquad  \rho\in\mathbb{R}\label{FreeFermion}
\eea
These weights satisfy the free-fermion condition 
\bea
a(u)^2+b(u)^2=c_1(u)c_2(u)\label{FreeFermionCond}
\eea
As shown in Section~\ref{SectTL}, with the special choice of gauge $g=z=e^{iu}$, the tiles 
give a representation of the free-fermion algebra with generators $\{f_j,f_j^\dagger\}$ and, consequently, also a representation of the Temperley-Lieb algebra with generators $\{e_j\}$ and loop fugacity $\beta=2\cos\lambda=0$. Explicitly, the face transfer operators are
\bea
X_j(u)=\rho(\cos u\,I + \sin u\,e_j),\qquad j=1,2,\ldots,N\label{faceOps}
\eea
This Temperley-Lieb model is directly equivalent to an anisotropic dimer model as shown in Figures~\ref{vpd}, \ref{Tconfigs} and \ref{TheDimers}. 
A dimer weight is assigned to the unique square face which is half-covered by the dimer as shown in Figure~\ref{TheDimers}. The statistical weights assigned to ``horizontal" and ``vertical" dimers are
\bea
\zeta_h(u)=a(u)=\rho\cos u,\qquad \zeta_v(u)=b(u)=\rho\sin u
\eea
Setting $g=\rho$, and allowing for the facts that (i) the $c_1$ face has two allowed configurations and (ii) no dimer covers the $c_2$ face, it follows that
\bea
c_1(u)=\zeta_h(u)^2+\zeta_v(u)^2=\rho^2(\cos^2 u+\sin^2 u)=\rho^2,\qquad c_2(u)=1
\eea

Alternatively, fixing $\rho=g=\sqrt{2}$  at the isotropic point ($u=\frac{\lambda}{2}=\frac{\pi}{4}$) gives
\bea
a(\tfrac{\pi}{4})=1,\qquad b(\tfrac{\pi}{4})=1,\qquad c_1(\tfrac{\pi}{4})=2,\qquad c_2(\tfrac{\pi}{4})=1
\eea
It follows that, with this normalization and any gauge $g$, the partition function at the isotropic point gives the correct counting of distinct dimer configurations. 

\psset{unit=1.5cm}
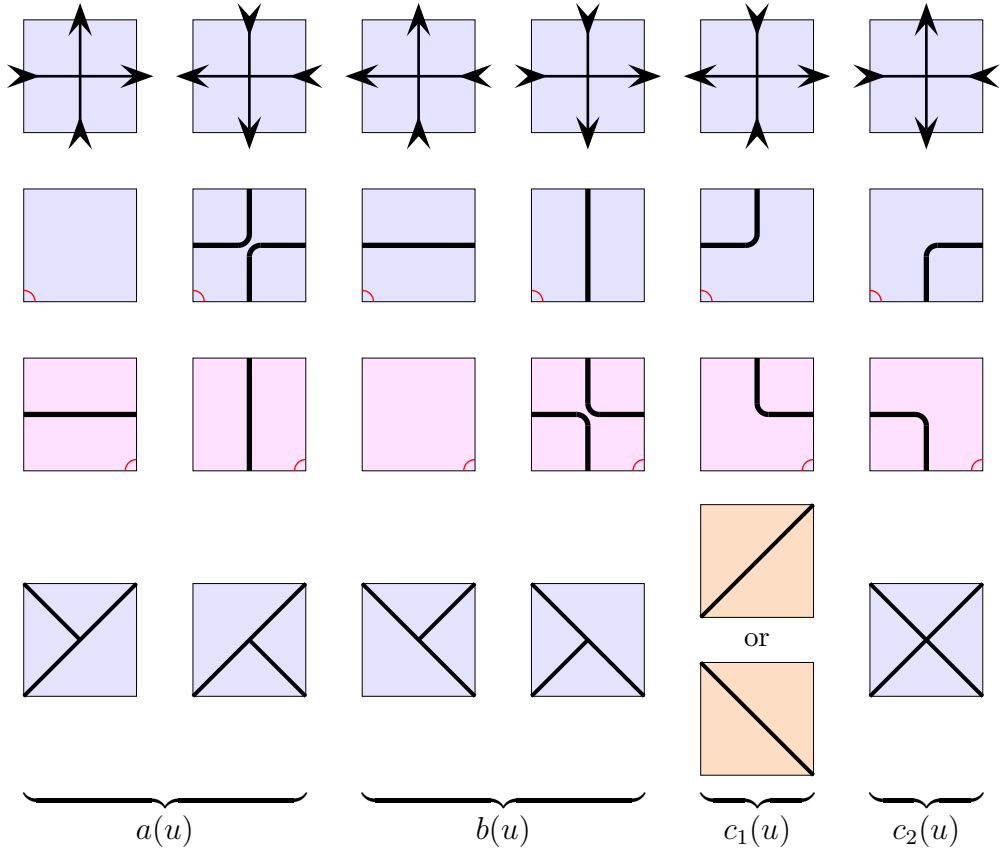
\begin{figure}[p]
\begin{center}
\begin{pspicture}(0,0)(8.5,7.2)
\rput(0,6){\va}
\rput(1.5,6){\vb}
\rput(3,6){\vc}
\rput(4.5,6){\vd}
\rput(6,6){\ve}
\rput(7.5,6){\vf}
\rput(0,4.5){\pa}
\rput(1.5,4.5){\pb}
\rput(3,4.5){\pc}
\rput(4.5,4.5){\pd}
\rput(6,4.5){\pe}
\rput(7.5,4.5){\pf}
\rput(0,3){\qc}
\rput(1.5,3){\qd}
\rput(3,3){\qa}
\rput(4.5,3){\qb}
\rput(6,3){\qg}
\rput(7.5,3){\qh}
\rput(0,1){\da}
\rput(1.5,1){\db}
\rput(3,1){\dc}
\rput(4.5,1){\dd}
\rput(6.5,1.5){\mbox{or}}
\rput(6,1.7){\de}
\rput(6,.3){\dee}
\rput(7.5,1){\df}
\rput(1.25,-.1){$\underbrace{\hspace{3.75cm}}_{\mbox{\large $a(u)$}}$}
\rput(4.25,-.1){$\underbrace{\hspace{3.75cm}}_{\mbox{\large $b(u)$}}$}
\rput(6.5,-.1){$\underbrace{\hspace{1.5cm}}_{\mbox{\large $c_1(u)$}}$}
\rput(8,-.1){$\underbrace{\hspace{1.5cm}}_{\mbox{\large $c_2(u)$}}$}
\end{pspicture}
\end{center}
\caption{\label{vpd}Equivalent face tiles of the six-vertex model in the vertex, particle (even and odd rows) and dimer representations. On the strip, the odd and even rows alternate. For periodic boundary conditions, all rows are odd. The heavy particle lines are drawn whenever the arrows disagree with the reference state as shown in Figure~\ref{RefStates}. The particles move up and to the right on odd rows and up and to the left on even rows. 
}
\end{figure}

\begin{figure}[p]
\begin{center}
\begin{pspicture}(0,0)(6,1)
\rput(0,0){\var}
\rput(1,0){\var}
\rput(2,0){\var}
\rput(3,0){\var}
\rput(4,0){\var}
\rput(5,0){\var}
\end{pspicture}

\bigskip

\begin{pspicture}(0,0)(6,2)
\rput(0,0){\var}
\rput(1,0){\var}
\rput(2,0){\var}
\rput(3,0){\var}
\rput(4,0){\var}
\rput(5,0){\var}
\rput(0,1){\vcs}
\rput(1,1){\vcs}
\rput(2,1){\vcs}
\rput(3,1){\vcs}
\rput(4,1){\vcs}
\rput(5,1){\vcs}
\end{pspicture}
\end{center}
\caption{\label{RefStates}Reference states for the single and double row transfer matrices for mapping onto the particle representation. The reference arrows point up and to the right for the single row transfer matrices. For the double row transfer matrices, the reference arrows point up and right on odd rows and up and left on even rows.}
\end{figure}
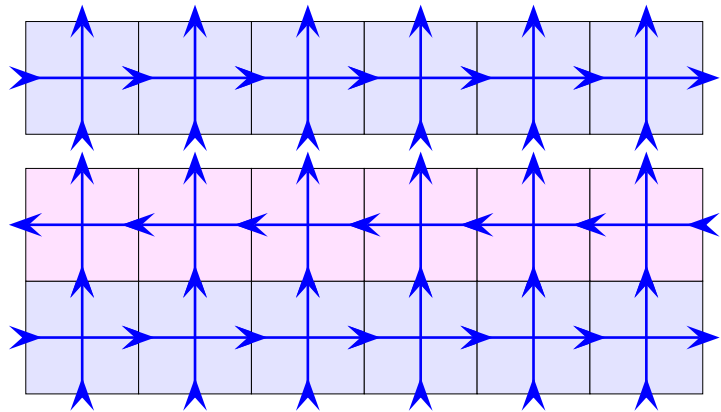

\psset{unit=1.5cm}
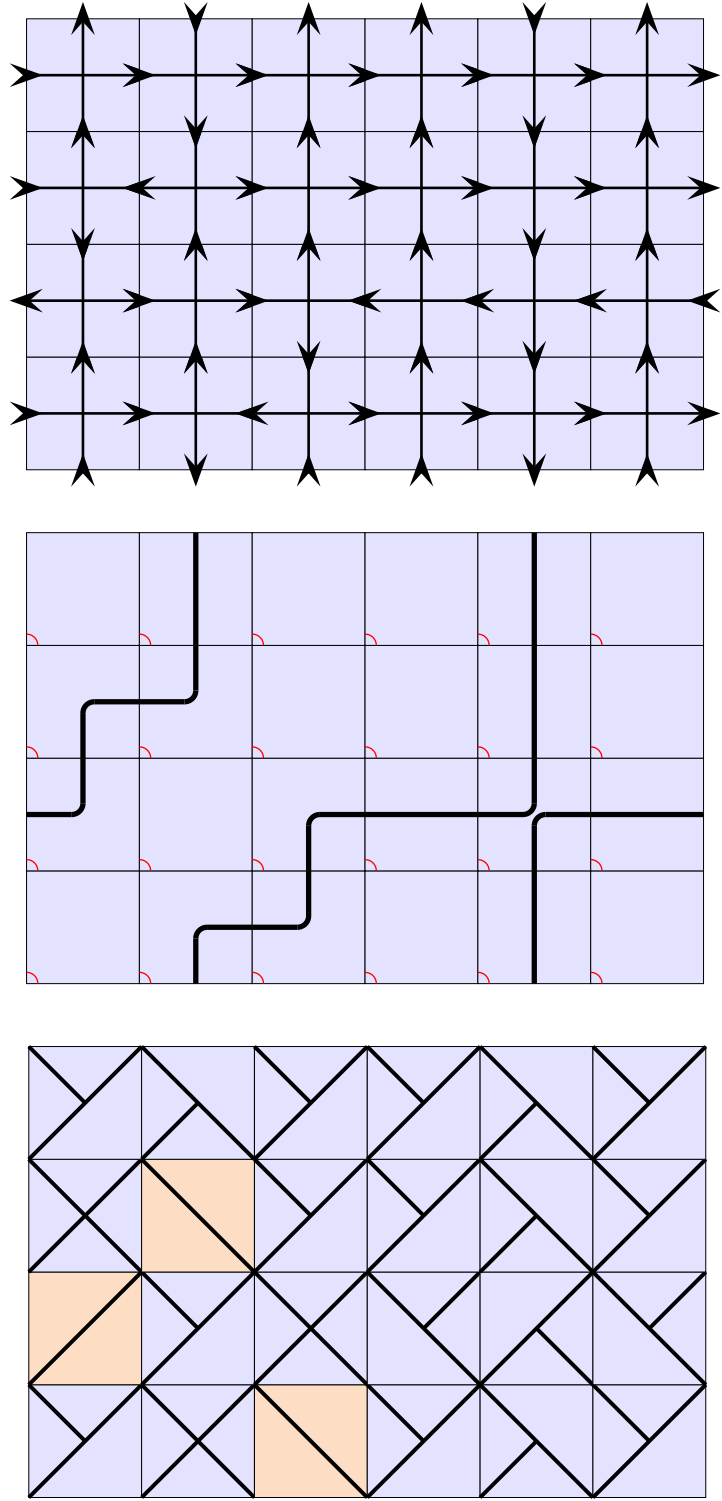
\begin{figure}[p]
\begin{center}
\begin{pspicture}(0,0)(6,4)
\rput(0,0){\va}
\rput(1,0){\vf}
\rput(2,0){\ve}
\rput(3,0){\va}
\rput(4,0){\vd}
\rput(5,0){\va}
\rput(0,1){\ve}
\rput(1,1){\va}
\rput(2,1){\vf}
\rput(3,1){\vc}
\rput(4,1){\vb}
\rput(5,1){\vc}
\rput(0,2){\vf}
\rput(1,2){\ve}
\rput(2,2){\va}
\rput(3,2){\va}
\rput(4,2){\vd}
\rput(5,2){\va}
\rput(0,3){\va}
\rput(1,3){\vd}
\rput(2,3){\va}
\rput(3,3){\va}
\rput(4,3){\vd}
\rput(5,3){\va}
\end{pspicture}
\mbox{}\vspace{.8cm}\mbox{}\!\!
\begin{pspicture}(0,0)(6,4)
\rput(0,0){\pa}
\rput(1,0){\pf}
\rput(2,0){\pe}
\rput(3,0){\pa}
\rput(4,0){\pd}
\rput(5,0){\pa}
\rput(0,1){\pe}
\rput(1,1){\pa}
\rput(2,1){\pf}
\rput(3,1){\pc}
\rput(4,1){\pb}
\rput(5,1){\pc}
\rput(0,2){\pf}
\rput(1,2){\pe}
\rput(2,2){\pa}
\rput(3,2){\pa}
\rput(4,2){\pd}
\rput(5,2){\pa}
\rput(0,3){\pa}
\rput(1,3){\pd}
\rput(2,3){\pa}
\rput(3,3){\pa}
\rput(4,3){\pd}
\rput(5,3){\pa}
\end{pspicture}
\mbox{}\vspace{.8cm}\mbox{}\!\!
\begin{pspicture}(0,0)(6,4)
\rput(0,0){\da}
\rput(1,0){\df}
\rput(2,0){\dee}
\rput(3,0){\da}
\rput(4,0){\dd}
\rput(5,0){\da}
\rput(0,1){\de}
\rput(1,1){\da}
\rput(2,1){\df}
\rput(3,1){\dc}
\rput(4,1){\db}
\rput(5,1){\dc}
\rput(0,2){\df}
\rput(1,2){\dee}
\rput(2,2){\da}
\rput(3,2){\da}
\rput(4,2){\dd}
\rput(5,2){\da}
\rput(0,3){\da}
\rput(1,3){\dd}
\rput(2,3){\da}
\rput(3,3){\da}
\rput(4,3){\dd}
\rput(5,3){\da}
\end{pspicture}
\end{center}
\caption{Typical periodic arrow configuration on a $6\times 4$ rectangle corresponding to four applications of the single row transfer matrix. The associated particle and (one of the $2^3=8$) possible periodic dimer configurations are also shown. The boundary conditions are periodic such that the left/right edges and top/bottom edges are identified. The excess of up arrows over down arrows (2 in this case) is conserved. Particles travel up and to the right. They can wind around the torus but do not cross. An $M\times N$ rectangular lattice is covered by $MN$ dimers. Each dimer covers two adjacent sites of the medial lattice.\label{Tconfigs}}
\end{figure}

\psset{unit=1.2cm}
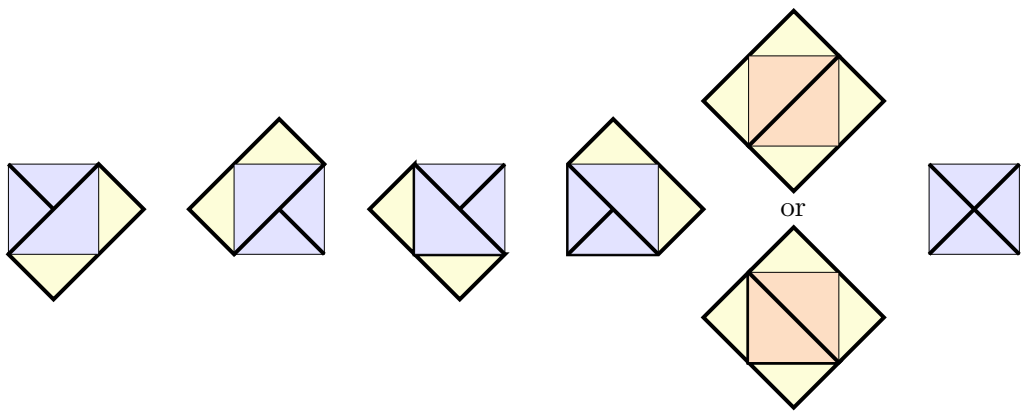
\begin{figure}[p]
\begin{center}
\begin{pspicture}(0,-.4)(10.5,3.6)
\rput(-.2,1){\daDimer}
\rput(2.3,1){\dbDimer}
\rput(4.3,1){\dcDimer}
\rput(6,1){\ddDimer}
\rput(8.5,1.5){\mbox{or}}
\rput(8,2.2){\deDimer}
\rput(8,-.2){\deeDimer}
\rput(10,1){\df}
\end{pspicture}\qquad\mbox{}
\end{center}
\caption{\label{TheDimers} Face configurations showing (in light yellow) the one or two dimers associated with each face. No dimers are associated with the last face.}
\end{figure}
\definecolor{apricot}{rgb}{1,0.9,0.7}
\definecolor{darkapricot}{rgb}{1.2,0.95,0.9}
%
\def\da{
\pspolygon[linewidth=.25pt,fillstyle=solid,fillcolor=lightlightblue](0,0)(1,0)(1,1)(0,1)(0,0)
\psline[linewidth=1.5pt](0,0)(1,1)
\psline[linewidth=1.5pt](0,1)(.5,.5)}
\def\db{
\pspolygon[linewidth=.25pt,fillstyle=solid,fillcolor=lightlightblue](0,0)(1,0)(1,1)(0,1)(0,0)
\psline[linewidth=1.5pt](0,0)(1,1)
\psline[linewidth=1.5pt](.5,.5)(1,0)}
\def\dc{
\pspolygon[linewidth=.25pt,fillstyle=solid,fillcolor=lightlightblue](0,0)(1,0)(1,1)(0,1)(0,0)
\psline[linewidth=1.5pt](0,1)(1,0)
\psline[linewidth=1.5pt](1,1)(.5,.5)}
\def\dd{
\pspolygon[linewidth=.25pt,fillstyle=solid,fillcolor=lightlightblue](0,0)(1,0)(1,1)(0,1)(0,0)
\psline[linewidth=1.5pt](0,1)(1,0)
\psline[linewidth=1.5pt](.5,.5)(0,0)}
\def\de{
\pspolygon[linewidth=.25pt,fillstyle=solid,fillcolor=apricot](0,0)(1,0)(1,1)(0,1)(0,0)
\psline[linewidth=1.5pt](0,0)(1,1)}
\def\dee{
\pspolygon[linewidth=.25pt,fillstyle=solid,fillcolor=apricot](0,0)(1,0)(1,1)(0,1)(0,0)
\psline[linewidth=1.5pt](1,0)(0,1)}
\def\df{
\pspolygon[linewidth=.25pt,fillstyle=solid,fillcolor=lightlightblue](0,0)(1,0)(1,1)(0,1)(0,0)
\psline[linewidth=1.5pt](0,0)(1,1)
\psline[linewidth=1.5pt](0,1)(1,0)}

\def\da{
\pspolygon[linewidth=.25pt,fillstyle=solid,fillcolor=lightlightblue](0,0)(1,0)(1,1)(0,1)(0,0)
\psline[linewidth=1.5pt](0,0)(1,1)
\psline[linewidth=1.5pt](0,1)(.5,.5)}
\def\db{
\pspolygon[linewidth=.25pt,fillstyle=solid,fillcolor=lightlightblue](0,0)(1,0)(1,1)(0,1)(0,0)
\psline[linewidth=1.5pt](0,0)(1,1)
\psline[linewidth=1.5pt](.5,.5)(1,0)}
\def\dc{
\pspolygon[linewidth=.25pt,fillstyle=solid,fillcolor=lightlightblue](0,0)(1,0)(1,1)(0,1)(0,0)
\psline[linewidth=1.5pt](0,1)(1,0)
\psline[linewidth=1.5pt](1,1)(.5,.5)}
\def\dd{
\pspolygon[linewidth=.25pt,fillstyle=solid,fillcolor=lightlightblue](0,0)(1,0)(1,1)(0,1)(0,0)
\psline[linewidth=1.5pt](0,1)(1,0)
\psline[linewidth=1.5pt](.5,.5)(0,0)}
\def\de{
\pspolygon[linewidth=.25pt,fillstyle=solid,fillcolor=apricot](0,0)(1,0)(1,1)(0,1)(0,0)
\psline[linewidth=1.5pt](0,0)(1,1)}
\def\dee{
\pspolygon[linewidth=.25pt,fillstyle=solid,fillcolor=apricot](0,0)(1,0)(1,1)(0,1)(0,0)
\psline[linewidth=1.5pt](1,0)(0,1)}
\def\df{
\pspolygon[linewidth=.25pt,fillstyle=solid,fillcolor=lightlightblue](0,0)(1,0)(1,1)(0,1)(0,0)
\psline[linewidth=1.5pt](0,0)(1,1)
\psline[linewidth=1.5pt](0,1)(1,0)}
\def\fracdb{
\pspolygon[linewidth=.25pt,fillstyle=solid,fillcolor=lightlightblue](0,0)(0.5,-0.5)(1,0)(0,0)}
\def\fracdl{
\pspolygon[linewidth=.25pt,fillstyle=solid,fillcolor=lightlightblue](0,0)(-0.5,0.5)(0,1)(0,0)}
\def\fracdr{
\pspolygon[linewidth=.25pt,fillstyle=solid,fillcolor=lightlightblue](0,0)(0.5,0.5)(0,1)(0,0)}
\def\fracdt{
\pspolygon[linewidth=.25pt,fillstyle=solid,fillcolor=lightlightblue](0,0)(0.5,0.5)(1,0)(0,0)}
\def\fracdby{
\pspolygon[linewidth=.25pt,fillstyle=solid,fillcolor=apricot](0,0)(0.5,-0.5)(1,0)(0,0)}
\def\fracdly{
\pspolygon[linewidth=.25pt,fillstyle=solid,fillcolor=apricot](0,0)(-0.5,0.5)(0,1)(0,0)}
\def\fracdry{
\pspolygon[linewidth=.25pt,fillstyle=solid,fillcolor=apricot](0,0)(0.5,0.5)(0,1)(0,0)}
\def\fracdty{
\pspolygon[linewidth=.25pt,fillstyle=solid,fillcolor=apricot](0,0)(0.5,0.5)(1,0)(0,0)}
\def\drightup{
\pspolygon[linewidth=2.0pt,fillstyle=solid,fillcolor=lightyellow](0,0)(0.5,-0.5)(1.5,0.5)(1,1)(0,0)}
\def\dleftup{
\pspolygon[linewidth=2.0pt,fillstyle=solid,fillcolor=lightyellow](1,0)(0.5,-0.5)(-0.5,0.5)(0,1)(1,0)}
\def\drightupy{
\pspolygon[linewidth=2.0pt,fillstyle=solid,fillcolor=apricot](0,0)(0.5,-0.5)(1.5,0.5)(1,1)(0,0)}
\def\dleftupy{
\pspolygon[linewidth=2.0pt,fillstyle=solid,fillcolor=apricot](1,0)(0.5,-0.5)(-0.5,0.5)(0,1)(1,0)}
\def\drightupyy{
\pspolygon[linewidth=2.0pt,fillstyle=solid,fillcolor=darkapricot](0,0)(0.5,-0.5)(1.5,0.5)(1,1)(0,0)}
\def\dleftupyy{
\pspolygon[linewidth=2.0pt,fillstyle=solid,fillcolor=darkapricot](1,0)(0.5,-0.5)(-0.5,0.5)(0,1)(1,0)}
\psset{unit=.85cm}
\begin{figure}[p]
\begin{center}
\begin{pspicture}(0,0)(18.5,12.)
\rput(0,1){\df}
\rput(1,1){\de}
\rput(0,2){\de}
\rput(1,2){\df}
\rput(0,3){\fracdty}
\rput(1,1){\fracdby}
\rput(2,1){\fracdry}
\rput(0,2){\fracdly}
\rput(1,1){\drightupyy}
\rput(0.5,1.5){\drightupyy}
\rput(0,2){\drightupy}
\rput(-.5,2.5){\drightupy}
\rput(3.5,1){\de}
\rput(4.5,1){\df}
\rput(3.5,2){\df}
\rput(4.5,2){\de}
\rput(3.5,1){\fracdly}
\rput(5.5,2){\fracdry}
\rput(3.5,1){\fracdby}
\rput(4.5,3){\fracdty}
\rput(3.5,1){\drightupyy}
\rput(3,1.5){\drightupyy}
\rput(4.5,2){\drightupy}
\rput(4.,2.5){\drightupy}
\rput(6.5,1){\dd}
\rput(7.5,1){\dd}
\rput(6.5,2){\db}
\rput(7.5,2){\db}
\rput(6.5,3){\fracdt}
\rput(7.5,3){\fracdt}
\rput(8.5,1){\fracdr}
\rput(6.5,2){\fracdl}
\rput(7,1.5){\dleftup}
\rput(8,1.5){\dleftup}
\rput(6.,2.5){\drightup}
\rput(7.,2.5){\drightup}
\rput(9.5,1){\db}
\rput(10.5,1){\db}
\rput(9.5,2){\dd}
\rput(10.5,2){\dd}
\rput(9.5,3){\fracdt}
\rput(10.5,3){\fracdt}
\rput(11.5,2){\fracdr}
\rput(9.5,1){\fracdl}
\rput(9.,1.5){\drightup}
\rput(10.,1.5){\drightup}
\rput(10,2.5){\dleftup}
\rput(11,2.5){\dleftup}
\rput(13,1){\db}
\rput(14,1){\dc}
\rput(13,2){\db}
\rput(14,2){\dc}
\rput(14,1){\fracdb}
\rput(13,3){\fracdt}
\rput(13,1){\fracdl}
\rput(13,2){\fracdl}
\rput(12.5,1.5){\drightup}
\rput(14,1){\dleftup}
\rput(12.5,2.5){\drightup}
\rput(14,2){\dleftup}
\rput(16.5,1){\dc}
\rput(17.5,1){\db}
\rput(16.5,2){\dc}
\rput(17.5,2){\db}
\rput(16.5,1){\fracdb}
\rput(17.5,3){\fracdt}
\rput(16.5,2){\fracdl}
\rput(16.5,1){\fracdl}
\rput(16.5,1){\dleftup}
\rput(17.,1.5){\drightup}
\rput(16.5,2){\dleftup}
\rput(17.,2.5){\drightup}
\rput(0,4.5){\dd}
\rput(1,4.5){\da}
\rput(0,5.5){\dd}
\rput(1,5.5){\da}
\rput(0,6.5){\fracdt}
\rput(1,4.5){\fracdb}
\rput(2,4.5){\fracdr}
\rput(2,5.5){\fracdr}
\rput(0.5,5){\dleftup}
\rput(1,4.5){\drightup}
\rput(0.5,6){\dleftup}
\rput(1,5.5){\drightup}
\rput(3.5,4.5){\da}
\rput(4.5,4.5){\dd}
\rput(3.5,5.5){\da}
\rput(4.5,5.5){\dd}
\rput(5.5,4.5){\fracdr}
\rput(5.5,5.5){\fracdr}
\rput(3.5,4.5){\fracdb}
\rput(4.5,6.5){\fracdt}
\rput(3.5,4.5){\drightup}
\rput(5,5){\dleftup}
\rput(3.5,5.5){\drightup}
\rput(5,6){\dleftup}
\rput(6.5,4.5){\dd}
\rput(7.5,4.5){\da}
\rput(6.5,5.5){\db}
\rput(7.5,5.5){\dc}
\rput(6.5,6.5){\fracdt}
\rput(7.5,4.5){\fracdb}
\rput(8.5,4.5){\fracdr}
\rput(6.5,5.5){\fracdl}
\rput(7,5){\dleftup}
\rput(7.5,4.5){\drightup}
\rput(6.,6){\drightup}
\rput(7.5,5.5){\dleftup}
\rput(9.5,4.5){\dc}
\rput(10.5,4.5){\db}
\rput(9.5,5.5){\da}
\rput(10.5,5.5){\dd}
\rput(9.5,4.5){\fracdb}
\rput(10.5,6.5){\fracdt}
\rput(11.5,5.5){\fracdr}
\rput(9.5,4.5){\fracdl}
\rput(9.5,4.5){\dleftup}
\rput(10.,5){\drightup}
\rput(9.5,5.5){\drightup}
\rput(11,6){\dleftup}
\rput(13,4.5){\db}
\rput(14,4.5){\dc}
\rput(13,5.5){\dd}
\rput(14,5.5){\da}
\rput(14,4.5){\fracdb}
\rput(13,6.5){\fracdt}
\rput(13,4.5){\fracdl}
\rput(15,5.5){\fracdr}
\rput(12.5,5){\drightup}
\rput(14,4.5){\dleftup}
\rput(13.5,6){\dleftup}
\rput(14,5.5){\drightup}
\rput(16.5,4.5){\da}
\rput(17.5,4.5){\dd}
\rput(16.5,5.5){\dc}
\rput(17.5,5.5){\db}
\rput(16.5,4.5){\fracdb}
\rput(17.5,6.5){\fracdt}
\rput(18.5,4.5){\fracdr}
\rput(16.5,5.5){\fracdl}
\rput(16.5,4.5){\drightup}
\rput(18,5){\dleftup}
\rput(16.5,5.5){\dleftup}
\rput(17.,6){\drightup}
\rput(0,8){\da}
\rput(1,8){\da}
\rput(0,9){\da}
\rput(1,9){\da}
\rput(0,8){\fracdb}
\rput(1,8){\fracdb}
\rput(2,8){\fracdr}
\rput(2,9){\fracdr}
\rput(0,8){\drightup}
\rput(1,8){\drightup}
\rput(0,9){\drightup}
\rput(1,9){\drightup}
\rput(3.5,8){\db}
\rput(4.5,8){\db}
\rput(3.5,9){\db}
\rput(4.5,9){\db}
\rput(3.5,8){\fracdl}
\rput(3.5,9){\fracdl}
\rput(3.5,10){\fracdt}
\rput(4.5,10){\fracdt}
\rput(3.,8.5){\drightup}
\rput(4.,8.5){\drightup}
\rput(3.,9.5){\drightup}
\rput(4.,9.5){\drightup}
\rput(6.5,8){\dc}
\rput(7.5,8){\dc}
\rput(6.5,9){\dc}
\rput(7.5,9){\dc}
\rput(6.5,8){\fracdb}
\rput(7.5,8){\fracdb}
\rput(6.5,9){\fracdl}
\rput(6.5,8){\fracdl}
\rput(6.5,8){\dleftup}
\rput(7.5,8){\dleftup}
\rput(6.5,9){\dleftup}
\rput(7.5,9){\dleftup}
\rput(9.5,8){\dd}
\rput(10.5,8){\dd}
\rput(9.5,9){\dd}
\rput(10.5,9){\dd}
\rput(9.5,10){\fracdt}
\rput(10.5,10){\fracdt}
\rput(11.5,8){\fracdr}
\rput(11.5,9){\fracdr}
\rput(10,8.5){\dleftup}
\rput(11,8.5){\dleftup}
\rput(10,9.5){\dleftup}
\rput(11,9.5){\dleftup}
\rput(13,8){\da}
\rput(14,8){\da}
\rput(13,9){\dc}
\rput(14,9){\dc}
\rput(13,8){\fracdb}
\rput(14,8){\fracdb}
\rput(15,8){\fracdr}
\rput(13,9){\fracdl}
\rput(13,8){\drightup}
\rput(14,8){\drightup}
\rput(13,9){\dleftup}
\rput(14,9){\dleftup}
\rput(16.5,8){\dc}
\rput(17.5,8){\dc}
\rput(16.5,9){\da}
\rput(17.5,9){\da}
\rput(16.5,8){\fracdb}
\rput(17.5,8){\fracdb}
\rput(18.5,9){\fracdr}
\rput(16.5,8){\fracdl}
\rput(16.5,8){\dleftup}
\rput(17.5,8){\dleftup}
\rput(16.5,9){\drightup}
\rput(17.5,9){\drightup}
\end{pspicture}
\end{center}
\caption{\label{2x2Dimers} The 24 periodic configurations of rotated dimers on $2\times 2$ square lattice. Each of the apricot shaded blocks of two dimers can occur in 2 local configurations related by a rotation through $90\degree$.}
\end{figure}
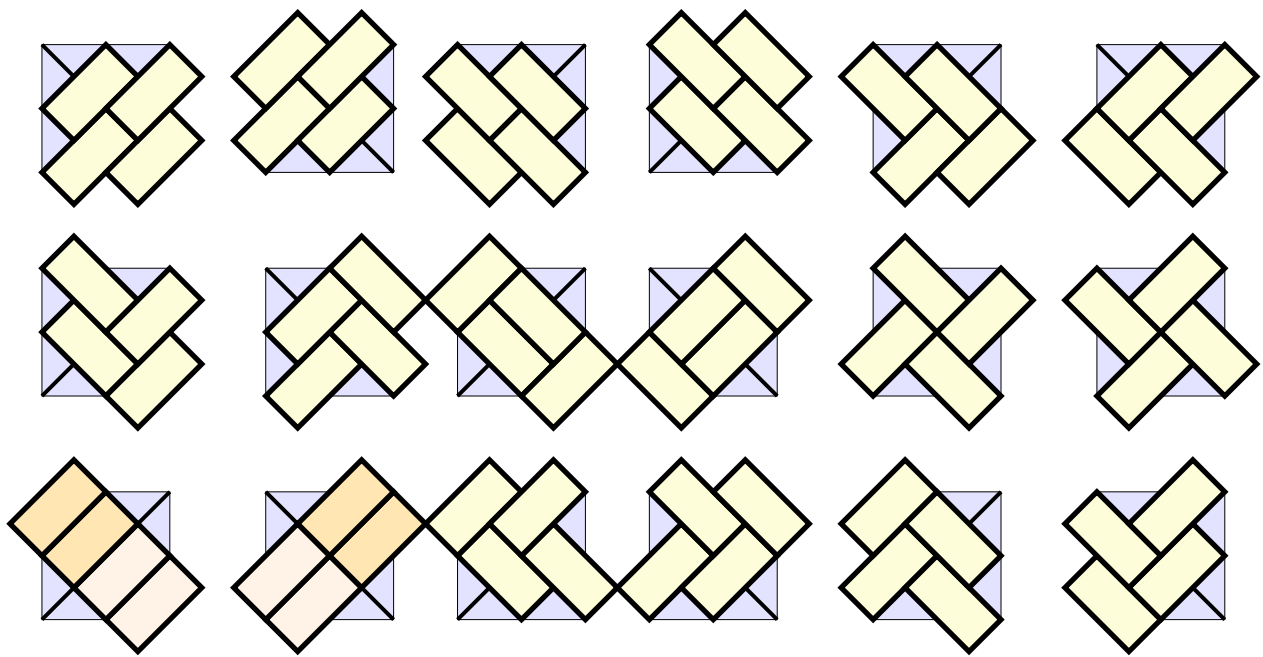

\definecolor{tile}{rgb}{1,.9,.7}

\def\tilea{\;
\begin{pspicture}[shift=-.65](0,0)(1.5,1.5)
\pspolygon[linewidth=.5pt,fillstyle=solid,fillcolor=tile](.75,0)(1.5,.75)(.75,1.5)(0,.75)(.75,0)
\end{pspicture}\;}

\def\tileb{\;
\begin{pspicture}[shift=-.65](0,0)(1.5,1.5)
\pspolygon[linewidth=.5pt,fillstyle=solid,fillcolor=tile](.75,0)(1.5,.75)(.75,1.5)(0,.75)(.75,0)
\psarc[linewidth=1.25pt](0,.75){.575}{-45}{45}
\psarc[linewidth=1.25pt](1.5,.75){.575}{135}{225}
\end{pspicture}\;}

\def\tilec{\;
\begin{pspicture}[shift=-.65](0,0)(1.5,1.5)
\pspolygon[linewidth=.5pt,fillstyle=solid,fillcolor=tile](.75,0)(1.5,.75)(.75,1.5)(0,.75)(.75,0)
\psline[linewidth=1.25pt](.375,.375)(1.125,1.125)
\end{pspicture}\;}

\def\tiled{\;
\begin{pspicture}[shift=-.65](0,0)(1.5,1.5)
\pspolygon[linewidth=.5pt,fillstyle=solid,fillcolor=tile](.75,0)(1.5,.75)(.75,1.5)(0,.75)(.75,0)
\psline[linewidth=1.25pt](1.125,.375)(.375,1.125)
\end{pspicture}\;}

\def\tilee{\;
\begin{pspicture}[shift=-.65](0,0)(1.5,1.5)
\pspolygon[linewidth=.5pt,fillstyle=solid,fillcolor=tile](.75,0)(1.5,.75)(.75,1.5)(0,.75)(.75,0)
\psarc[linewidth=1.25pt](0,.75){.575}{-45}{45}
\end{pspicture}\;}

\def\tilef{\;
\begin{pspicture}[shift=-.65](0,0)(1.5,1.5)
\pspolygon[linewidth=.5pt,fillstyle=solid,fillcolor=tile](.75,0)(1.5,.75)(.75,1.5)(0,.75)(.75,0)
\psarc[linewidth=1.25pt](1.5,.75){.575}{135}{225}
\end{pspicture}\;}

\def\diam#1#2#3#4#5{\ \ 
\begin{pspicture}[shift=-.65](0,0)(1.53,1.5)
\pspolygon[linewidth=.5pt,fillstyle=solid,fillcolor=lightlightblue](.75,0)(1.5,.75)(.75,1.5)(0,.75)(.75,0)
\rput[t](.75,0){\scriptsize $#1$}
\rput[l](1.5,.75){\scriptsize $#2$}
\rput[b](.75,1.5){\scriptsize $#3$}
\rput[r](0,.75){\scriptsize $#4$}
\rput(.75,.75){\small $#5$}
\rput[tr](.32,.375){\small $j$}
\rput[tl](1.125,.375){\small $j\!\!+\!\!1$}
\psarc[linewidth=1pt,linecolor=red](.75,0){.15}{45}{135}
\end{pspicture}}

\def\diamt#1#2#3#4#5{\ \ 
\begin{pspicture}[shift=-.65](0,0)(1.53,1.5)
\pspolygon[linewidth=.5pt,fillstyle=solid,fillcolor=lightpurple](.75,0)(1.5,.75)(.75,1.5)(0,.75)(.75,0)
\rput[t](.75,0){\scriptsize $#1$}
\rput[l](1.5,.75){\scriptsize $#2$}
\rput[b](.75,1.5){\scriptsize $#3$}
\rput[r](0,.75){\scriptsize $#4$}
\rput(.75,.75){\small $#5$}
\rput[tr](.32,.375){\small $j$}
\rput[tl](1.125,.375){\small $j\!\!+\!\!1$}
\psarc[linewidth=1pt,linecolor=red](1.5,.75){.15}{135}{225}
\end{pspicture}}

In addition to the vertex and dimer representations, the six-vertex free-fermion model admits a particle representation as shown in Figure~\ref{Tconfigs}. A reference state on the cylinder and strip is fixed as in Figure~\ref{RefStates}. An edge of a given vertex is a segment of a particle trajectory (and has particle occupation number $a_j=1$) if its arrow points in the opposite direction to that of the reference state. Otherwise, if the edge arrow points in the same direction as the reference state, the edge is not a segment of a particle trajectory (and the particle occupation is $a_j=0$). The segments of particle trajectories live on the medial lattice and are indicated with heavy lines in Figure~\ref{RefStates}. The number of particles is conserved and their trajectories are non-intersecting. On the cylinder (which is glued at the top and bottom to form the torus), the particle trajectories are constrained to move up and to the right through the lattice. 
The particle representation is the simplest of the three representations and is convenient for coding in Mathematica~\cite{Wolfram} and for manipulations in the diagrammatic planar algebra so we usually work in the particle representation.

With suitable face weights, the mapping between the six-vertex model, particle representation and dimers also holds for $\lambda\ne \frac{\pi}{2}$ and the model is still Yang-Baxter integrable. 
The difference is that, in the free-fermion case $\lambda=\frac{\pi}{2}$, the particles are non-interacting whereas, for $\lambda\ne \frac{\pi}{2}$, the particles interact. In terms of dimers, for $\lambda\ne \frac{\pi}{2}$, there are anisotropic 3-dimer interactions for the faces of vertices 1 through 4. 
In general, for $\lambda=\frac{(p'-p)\pi}{p'}$ with $p,p'$ coprime, the (non-intersecting) fermion particle trajectories play the role of nonlocal degrees of freedom in these logarithmic Conformal Field Theories (CFTs). 
The $\mathbb{Z}_2$ arrow reversal symmetry of the vertex model implies a particle-hole duality in the particle representation.

\subsection{Free-fermion, Temperley-Lieb algebras and Yang-Baxter equation}
\label{SectTL}
In this section, we consider the free-fermion model (\ref{FreeFermion}) with $\lambda=\frac{\pi}{2}$ and set $g=z=e^{iu}$ and $\rho=1$. 
The overall normalization $\rho=\sqrt{2}$ is easily reinstated, as needed, to count dimer configurations at the isotropic point ($u=\frac{\pi}{4}$). 

\subsubsection{Free-fermion algebra}
As elements of a planar algebra~\cite{Jones}, the face operators of the free-fermion six-vertex model decompose~\cite{BEPR2015} in the particle representation into a sum of contributions from six elementary tiles
\psset{unit=.75cm}
\setlength{\unitlength}{.75cm}
\begin{align}
X_j(u)&=\!\!\!\!\diam{}{}{}{}u=a(u)\Bigg(\!\!\tilea \!+\!\tileb\!\!\Bigg)+b(u)\Bigg(\!\!\tilec\!+\!\tiled\!\!\Bigg)+c_1(u)\tilee\!+c_2(u)\tilef\label{faceDecomp}
\end{align}
Multiplication of the tiles in the planar algebra is given by local tensor contraction of indices ($a,b,c,\ldots=0,1$) specifying the particle occupation numbers on the centers of the tile edges. Regarding the elementary tiles as operators acting on an upper (zigzag) row particle configuration to produce a lower (zigzag) row particle configuration, we write them respectively as
\bea
E_j=n_j^{00},\ n_j^{11},\ f_j^\dagger f_{j+1},\ f_{j+1}^\dagger f_j,\ n_j^{10},\ n_j^{01},\qquad\quad 
n_j^{00}+n_j^{11}+n_j^{10}+n_j^{01}=I
\label{elemOps}
\eea
The four operators $n_j^{ab}$ are (diagonal) orthogonal projection operators which factorize into single-site orthogonal projectors corresponding to left and right  half (triangular) tiles
\bea
n_j^{ab}=n_j^an_{j+1}^b,\qquad n_j^a n_j^b=\delta_{ab}\,n_j^a,\qquad n_j^0+n_j^1=I,\qquad a,b=0,1
\eea
Here $n_j=n_j^1$ is the number operator counting single-site occupancy at position $j$ and $n_j^0$ is the dual number operator counting the single-site vacancies at position $j$. The operators $f_j$ and $f_j^\dagger$ are single-site particle annihilation and creation operators respectively which satisfy the Canonical Anticommutation Relations (CAR) for fermions
\bea
\{f_j,f_k\}=\{f_j^\dagger,f_k^\dagger\}=0,\qquad \{f_j,f_k^\dagger\}=\delta_{jk},\qquad n_j^1=f_j^\dagger f_j,\qquad n_j^0=f_j f_j^\dagger=1-f_j^\dagger f_j
\eea

It follows that all of the elementary tile operators can be written as combinations of bilinears in the fermion operators $f_j$ and $f_j^\dagger$. 
Diagrammatically, the particle hopping terms $f_j^\dagger f_{j+1}$ and $f_{j+1}^\dagger f_j$ factorize into left and right half (triangular) tiles 
\bea
f_j^\dagger f_{j+1}=\begin{pspicture}[shift=-.65](0,0)(1.5,1.5)
\pspolygon[linewidth=.5pt,fillstyle=solid,fillcolor=tile](.75,0)(1.5,.75)(.75,1.5)(0,.75)
\psline[linewidth=1.25pt](.375,.375)(1.125,1.125)
\psline[linewidth=.5pt,linestyle=dashed,dash=3pt 2pt](.75,0)(.75,1.5)
\rput[tr](.32,.375){\small $j$}
\rput[tl](1.125,.375){\small $j\!\!+\!\!1$}
\end{pspicture}\qquad\quad
f_{j+1}^\dagger f_j=\begin{pspicture}[shift=-.65](0,0)(1.5,1.5)
\pspolygon[linewidth=.5pt,fillstyle=solid,fillcolor=tile](.75,0)(1.5,.75)(.75,1.5)(0,.75)(.75,0)
\psline[linewidth=1.25pt](1.125,.375)(.375,1.125)
\psline[linewidth=.5pt,linestyle=dashed,dash=2pt 2pt](.75,0)(.75,1.5)
\rput[tr](.32,.375){\small $j$}
\rput[tl](1.125,.375){\small $j\!\!+\!\!1$}
\end{pspicture}
\eea
so that the fermion generators are represented by half (triangular) tiles
\bea
f_j^\dagger=\begin{pspicture}[shift=-.65](0,0)(.75,.75)
\pspolygon[linewidth=.0pt,fillstyle=solid,fillcolor=tile](.75,0)(.75,1.5)(0,.75)
\psline[linewidth=1.25pt](.375,.375)(.75,.75)
\rput[tr](.32,.375){\small $j$}
\end{pspicture}
\;=\;\begin{pspicture}[shift=-.65](.75,0)(1.5,1.5)
\pspolygon[linewidth=.5pt,fillstyle=solid,fillcolor=tile](.75,0)(1.5,.75)(.75,1.5)(.75,0)
\psline[linewidth=1.25pt](1.125,.375)(.75,.75)
\rput[tl](1.125,.375){\small $j$}
\end{pspicture},\qquad f_j=\begin{pspicture}[shift=-.65](0,0)(.75,.75)
\pspolygon[linewidth=.0pt,fillstyle=solid,fillcolor=tile](.75,0)(.75,1.5)(0,.75)
\psline[linewidth=1.25pt](.375,1.125)(.75,.75)
\rput[tr](.32,.375){\small $j$}
\end{pspicture}
\;=\;\begin{pspicture}[shift=-.65](.75,0)(1.5,1.5)
\pspolygon[linewidth=.5pt,fillstyle=solid,fillcolor=tile](.75,0)(1.5,.75)(.75,1.5)(.75,0)
\psline[linewidth=1.25pt](1.125,1.125)(.75,.75)
\rput[tl](1.125,.375){\small $j$}
\end{pspicture}\label{fermiTiles}
\eea
The action of the fermion operators (\ref{fermiTiles}) on states is given by
\begin{align}
f_j^\dagger\big|\cdots,n_{j-1},n_j,n_{j+1},\cdots\big\rangle&=(-1)^{\sum_{1\le k<j}n_k}(1-n_j)\big|\cdots,n_{j-1},(1-n_j),n_{j+1},\cdots\big\rangle\\
f_j\big|\cdots,n_{j-1},n_j,n_{j+1},\cdots\big\rangle&=(-1)^{\sum_{1\le k<j}n_k}\,n_j\big|\cdots,n_{j-1},(1-n_j),n_{j+1},\cdots\big\rangle
\end{align}
The presence of the ``Jordan-Wigner string" $(-1)^{\sum_{1\le k<j}n_k}$ ensures the complete antisymmetry of the states and reflects the non-locality of the fermion operators.
Notice that a particle can only hop into a site that is vacant and that the action of the hopping terms on states is
\begin{align}
f_j^\dagger f_{j+1}\big|\cdots,n_{j-1},0,1,n_{j+2},\cdots\big\rangle&=\big|\cdots,n_{j-1},1,0,n_{j+2},\cdots\big\rangle\\
f_{j+1}^\dagger f_j\big|\cdots,n_{j-1},1,0,n_{j+2},\cdots\big\rangle&=\big|\cdots,n_{j-1},0,1,n_{j+2},\cdots\big\rangle
\end{align}
since the contributions from the Jordan-Wigner strings cancel. This action applies for both open and periodic boundary conditions with the cyclic boundary condition $f_{j+N}=f_j$ on the fermions. Although the Jordan-Wigner strings seem to break  translation invariance for periodic boundary conditions, this invariance is restored~\cite{Felderhof} for operators composed of an even number of fermion operators.

\subsubsection{Temperley-Lieb algebra}
To realise a Temperley-Lieb algebra, let us introduce $x=e^{i\lambda}=i$ and the generators
\begin{subequations}
\begin{align}
e_j&=x\tilee+x^{-1} \tilef+ \tilec+\tiled\label{TLgen}\\
&=x f_j^\dagger f_j (1-f_{j+1}^\dagger f_{j+1}) +x^{-1}(1-f_j^\dagger f_j) f_{j+1}^\dagger f_{j+1} + f_j^\dagger f_{j+1}+f_{j+1}^\dagger f_j\\
&=x f_j^\dagger f_j +x^{-1} f_{j+1}^\dagger f_{j+1} + f_j^\dagger f_{j+1}+f_{j+1}^\dagger f_j\label{TLgenf}
\end{align}
\end{subequations}
The quartic (interacting) terms vanish, since $\beta=x+x^{-1}=0$, leaving bilinears in fermion operators. 
Using the planar algebra of tiles, it is easily shown that these operators yield a representation of the Temperley-Lieb algebra~\cite{TempLieb}
\bea
e_j^2=\beta e_j=0,\qquad e_je_{j\pm 1}e_j=e_j,\qquad \beta=2\cos\lambda=x+x^{-1}=0
\eea
Equivalently this follows, purely from fermionic algebra, by writing the generators in terms of the fermionic operators $f_j$ and $f_j^\dagger$ as in (\ref{TLgenf}).

\def\abar{\bar{a}}
\def\Xface#1#2#3#4{
\psset{unit=.5cm}\;
\begin{pspicture}[shift=-.85](0,0)(2,2)
\pspolygon[linewidth=.25pt,fillstyle=solid,fillcolor=lightlightblue](1,0)(2,1)(1,2)(0,1)
\rput[tr](.5,.5){\small $#1$}
\rput[tl](1.5,.5){\small $#2$}
\rput[bl](1.5,1.5){\small $#3$}
\rput[br](.5,1.5){\small $#4$}
\psarc[linewidth=.75pt,linecolor=red](1,0){.25}{45}{135}
\end{pspicture}\,}
\def\oface#1{
\pspolygon[linewidth=.25pt,fillstyle=solid,fillcolor=lightlightblue](0,0)(1,0)(1,1)(0,1)(0,0)
\rput(.5,.5){\small $#1$}
\psarc[linewidth=.5pt,linecolor=red](0,0){.125}{0}{90}}
\def\eface#1{
\pspolygon[linewidth=.25pt,fillstyle=solid,fillcolor=lightpurple](0,0)(1,0)(1,1)(0,1)(0,0)
\rput(.5,.5){\small $#1$}
\psarc[linewidth=.5pt,linecolor=red](1,0){.125}{90}{180}}
\def\ltri{
\begin{pspicture}(0,0)
\pspolygon[linewidth=.25pt,fillstyle=solid,fillcolor=lightlightblue](-.5,0)(0,1)(-.5,2)
\psline[linewidth=1.5pt](-.5,1.5)(0,1.5)
\end{pspicture}}
\def\rtri{
\begin{pspicture}(0,0)
\pspolygon[linewidth=.25pt,fillstyle=solid,fillcolor=lightlightblue](.5,0)(0,1)(.5,2)
\psline[linewidth=1.5pt](0,1.5)(.5,1.5)
\end{pspicture}}
\nc{\spos}[2]{\makebox(0,0)[#1]{$\sm{#2}$}}
\nc{\botrightrefnodots}[4]{
\begin{pspicture}(1.5,2)
\rput(-.5,0){\psline[linewidth=.5pt,fillstyle=solid,fillcolor=lightlightblue](2,0)(2,2)(1,1)}
\rput(0,0){\psline[linewidth=.5pt,fillstyle=solid,fillcolor=lightlightblue](0,.5)(.5,0)(1,.5)(.5,1)}
\put(1.5,0){\line(0,1){2}}
\put(1.5,1){\line(-1,1){0.5}}
\put(1.5,1){\line(-1,-1){1}}\put(1.5,0){\line(-1,1){1}}
\put(1.5,2){\line(-1,-1){1.5}}\put(0,0.5){\line(1,-1){0.5}}
\put(1.4,0.5){\spos{r}{#1}}\put(1.4,1.5){\spos{r}{#2}}
\put(0.5,0.5){\spos{}{#3}}
\put(1,1){\spos{}{#4}}
\psarc[linecolor=red](0,.5){.1}{-45}{45}
\psarc[linecolor=red](1,.5){.1}{45}{135}
\end{pspicture}}
\nc{\toprightrefnodots}[4]{
\begin{pspicture}(1.5,2)
\rput(-.5,0){\psline[linewidth=.5pt,fillstyle=solid,fillcolor=lightlightblue](2,0)(2,2)(1,1)}
\rput(0,1){\psline[linewidth=.5pt,fillstyle=solid,fillcolor=lightlightblue](0,.5)(.5,0)(1,.5)(.5,1)}
\put(1.5,0){\line(0,1){2}}
\put(1.5,1){\line(-1,1){1}}\put(1.5,1){\line(-1,-1){0.5}}
\put(1.5,0){\line(-1,1){1.5}}\put(1.5,2){\line(-1,-1){1}}
\put(0,1.5){\line(1,1){0.5}}
\put(1.4,0.5){\spos{r}{#1}}\put(1.4,1.5){\spos{r}{#2}}
\put(1,1){\spos{}{#3}}\put(0.5,1.5){\spos{}{#4}}
\psarc[linecolor=red](0,1.5){.1}{-45}{45}
\psarc[linecolor=red](1,.5){.1}{45}{135}
\end{pspicture}}

\def\righttri#1#2#3{
\begin{pspicture}(0,0)
\pspolygon[linewidth=.25pt,fillstyle=solid,fillcolor=lightlightblue](0,1)(1,0)(1,2)
\rput[br](.5,1.5){\small $#1$}
\rput[tr](.5,.5){\small $#2$}
\rput(.6,1){\small $#3$}
\end{pspicture}}

\def\diam#1#2#3#4#5{
\begin{pspicture}(0,0)(2,2)
\pspolygon[linewidth=.25pt,fillstyle=solid,fillcolor=lightlightblue](1,0)(2,1)(1,2)(0,1)
\rput[tr](.5,.5){\small $#1$}
\rput[tl](1.5,.5){\small $#2$}
\rput[bl](1.5,1.5){\small $#3$}
\rput[br](.5,1.5){\small $#4$}
\rput(1,1){\small $#5$}
\psarc[linewidth=1pt,linecolor=red](1,0){.15}{45}{135}
\end{pspicture}}
\def\vdiam#1#2#3#4#5{\ \ 
\begin{pspicture}[shift=-.85](0,-.25)(1.53,1.75)
\pspolygon[linewidth=.5pt,fillstyle=solid,fillcolor=lightlightblue](.75,0)(1.5,.75)(.75,1.5)(0,.75)(.75,0)
\rput[t](.75,0){\scriptsize $#1$}
\rput[l](1.5,.75){\scriptsize $#2$}
\rput[b](.75,1.5){\scriptsize $#3$}
\rput[r](0,.75){\scriptsize $#4$}
\rput(.75,.75){\small $#5$}
\end{pspicture}}

\subsubsection{Yang-Baxter equation}
Using the above algebra, it is straightforward to confirm that, in the gauge with $g=z=e^{iu}$, the face transfer operators (\ref{faceDecomp}) of the free-fermion six vertex model now take the form
\psset{unit=.6cm}
\bea
X_j(u)=\!\!\raisebox{-.5cm}{\diam{}{}{}{}u}\,=\cos u\,I + \sin u\,e_j
\label{faceOps2}
\eea
It immediately follows, from standard arguments~\cite{BaxterInv82}, that they  
satisfy the Yang-Baxter Equation (YBE)
\begin{subequations}
\label{YBE}
\bea
X_j(u)X_{j+1}(u+v)X_j(v)=X_{j+1}(v)X_j(u+v)X_{j+1}(u)
\eea
\psset{unit=.8cm}
\bea
\raisebox{-1.5cm}{
\begin{pspicture}(0,0)(3,3.8)
\rput(1,1){\diam ab{}{}u}
\rput(1,3){\diam {}{}efv}
\rput(2,2){\diam {}cd{}{u\!+\!v}}
\rput(.5,1.5){\pscircle[fillstyle=solid,fillcolor=black](.05,-.015){.07}}
\rput(.5,2.5){\pscircle[fillstyle=solid,fillcolor=black](.05,-.015){.07}}
\rput(1.5,1.5){\pscircle[fillstyle=solid,fillcolor=black](.05,-.015){.07}}
\rput(1.5,2.5){\pscircle[fillstyle=solid,fillcolor=black](.05,-.015){.07}}
\psline[linestyle=dashed](.55,1.5)(.55,2.5)
\end{pspicture}}\quad
=\quad
\raisebox{-1.5cm}{
\begin{pspicture}(0,0)(3,3.8)
\rput(2,3){\diam {}{}deu}
\rput(2,1){\diam bc{}{}v}
\rput(1,2){\diam a{}{}f{u\!+\!v}}
\rput(1.5,1.5){\pscircle[fillstyle=solid,fillcolor=black](.05,-.015){.07}}
\rput(1.5,2.5){\pscircle[fillstyle=solid,fillcolor=black](.05,-.015){.07}}
\rput(2.5,1.5){\pscircle[fillstyle=solid,fillcolor=black](.05,-.015){.07}}
\rput(2.5,2.5){\pscircle[fillstyle=solid,fillcolor=black](.05,-.015){.07}}
\psline[linestyle=dashed](2.55,1.5)(2.55,2.5)
\end{pspicture}}
\eea
\end{subequations}

The initial condition and inversion relation are
\bea
X_j(0)=I,\qquad X_j(u)X_j(-u)=\cos^2 u\,I
\eea
In the usual vertex model terminology, $X_j(u)$ is the $\check R$-matrix and not the $R$-matrix.

\subsection{Free energy and residual entropy}

Since the free-fermion six-vertex model is Yang-Baxter integrable, its partition function per site
\bea
\rho\,\kappa(u)=\rho\exp(-f_\text{bulk}(u))
\eea
can be obtained by solving~\cite{BaxterInv82} the inversion relation $\kappa(u)\kappa(-u)=\cos^2 u$ or by using Euler-Maclaurin approach as in \cite{PRpolymers}. 
The two equivalent integrals for the bulk free energy are
\bea
f_\text{bulk}(u)=-\int_{-\infty}^\infty \frac{\sinh ut \sinh(\frac{\pi}{2}-u)t}{t\sinh \pi t\cosh \frac{\pi t}{2}}\,dt
=\half\log 2-\frac{1}{\pi}\int_0^{\pi/2} \log(\cosec t+\sin 2u)dt
\eea
Setting $\rho=\sqrt{2}$ and $u=\frac{\pi}{4}$ gives the known~\cite{Fisher1961} molecular freedom $W$ and residual entropy $S$ of dimers on the square lattice as
\bea
W=e^S=\sqrt{2}\exp(-f_\text{bulk}(\tfrac{\pi}{4}))=\exp(\tfrac{2G}{\pi})=1.791\,622\,812\ldots,\qquad S=\tfrac{2G}{\pi}=.583\,121\,808\ldots
\eea
where the molecular freedom $W$ and Catalan's constant $G$ are given by
\bea
W=\sqrt{2}\,\kappa(\tfrac{\pi}{4})=\lim_{M,N\to\infty}(Z_{M\times N})^{\frac{1}{MN}},\qquad G=\half\int_0^{\pi/2} \log(1+\cosec t)dt=.915\,965\,594\ldots\label{CatalanG}
\eea

\section{Solution on a Cylinder and Torus and Finite-Size Spectra}
\label{SecCylTorus}

\subsection{Commuting single row transfer matrices}

The single row transfer matrix of the free-fermion model in the particle representation is defined by
\bea
\psset{unit=1.2cm}
\vec T(u)=
\raisebox{-.75cm}{
\begin{pspicture}(0,-.2)(6,1.2)
\rput(0,0){\oface u}
\rput(1,0){\oface u}
\rput(2,0){\oface u}
\rput(3,0){\oface u}
\rput(4,0){\oface u}
\rput(5,0){\oface u}
\rput[t](.5,-.1){\small $a_1$}
\rput[t](1.5,-.1){\small $a_2$}
\rput[t](3.5,-.1){\small $\cdots$}
\rput[t](5.5,-.1){\small $a_N$}
\rput[b](.5,1.1){\small $b_1$}
\rput[b](1.5,1.1){\small $b_2$}
\rput[b](3.5,1.1){\small $\cdots$}
\rput[b](5.5,1.1){\small $b_N$}
\multirput(0,0)(1,0){7}{\pscircle[fillstyle=solid,fillcolor=black](0,.5){.05}}
\end{pspicture}}\label{Tmatrix}
\eea
where there are $N$ columns and the left and right edges are identified. The occupation numbers on the vertical edges (auxiliary space) are summed out. 
The quantum space consists of $2^N$ row configurations $\vec a=\{a_1,a_2,\ldots,a_N\}$ of particle occupation numbers. 
The Yang-Baxter equation (\ref{YBE}) implies~\cite{BaxBook} that the single row transfer matrices commute
\bea
[\vec T(u),\vec T(v)]=0
\eea
From the crossing relation and commutation, it follows that the transfer matrices are normal
\bea
\vec T(u)^T=\vec T(\lambda-u)\quad\Rightarrow\quad [\vec T(u),\vec T(u)^T]=0
\eea
The transfer matrices are therefore simultaneously diagonalizable by a similarity transformation with a matrix $\vec S$ whose columns are the common $u$-independent right eigenvectors of $\vec T(u)$.

In the six-vertex arrow (or spin) representation, the total magnetization 
\bea
S_z=\sum_{j=1}^N \sigma_j=-N,-N+2,\ldots, N-2, N
\eea
 is conserved under the action of the transfer matrix. The magnetization $S_z$ is thus a good quantum number separating the spectrum into sectors. For dimers, it therefore plays the role of the variation index in \cite{RasRuelle2012,MDRR2015,MDRR2016}. 
By the $\mathbb{Z}_2$ up-down symmetry, 
the spectrum for the sectors $S_z=m$ and $S_z=-m$ coincide for $m>0$. So all these eigenvalues 
are exactly doubly degenerate. We will see that, for $N$ even, the lowest energy (ground) state is unique and occurs in the sector 
$S_z=0$ so that there are $\half N$ up arrows (spins) and $\half N$ down arrows (spins).
More generally, the number of down spins is $d=\half(N-S_z)$, the number of up spins is thus $N-d=\half(N+S_z)$ and 
the counting of states in the $S_z$ sector is given by the binomial $\genfrac{(}{)}{0pt}{}{N}{d}$ 
with $S_z=N$ mod 2. 
The number of particles $d=\sum_{j=1}^N a_j$ coincides with the number of down arrows and is also conserved. 
The transfer matrix and vector space of states thus decompose as
\bea
\vec T(u)=\mathop{\bigoplus}_{d=0}^N \vec T_d(u),\qquad \mathop{\mbox{dim}} {\cal V}^{(N)}=\sum_{d=0}^N \mathop{\mbox{dim}}{\cal V}^{(N)}_d=\sum_{d=0}^N
\genfrac{(}{)}{0pt}{}{N}{d}=2^N=\mbox{dim}\,(\mathbb{C}^2)^{\otimes N}\label{Tdecomp}
\eea 
Comparing the spectra sector-by-sector with critical dense polymers~\cite{PRV} gives a precise matching if the number of defects $\ell$ is identified as
\bea
\ell=|N-2d|=|S_z|=\begin{cases}
0,2,4,\ldots,N,&\mbox{$N$ even}\\
1,3,5,\ldots,N,&\mbox{$N$ odd}\end{cases}\label{matching}
\eea

Taking the logarithmic derivative of the single row transfer matrix (\ref{Tmatrix}) gives the Hermitian free-fermion Hamilitonian
\begin{subequations}
\label{FFHam}
\begin{align}
{\cal H}&=-\sum_{j=1}^N e_j=-\sum_{j=1}^N (x f_j^\dagger f_j +x^{-1} f_{j+1}^\dagger f_{j+1} + f_j^\dagger f_{j+1}+f_{j+1}^\dagger f_j)\\
&=-\sum_{j=1}^N(f_j^\dagger f_{j+1}+f_{j+1}^\dagger f_j)
\end{align}
\end{subequations}
Using $\beta=x+x^{-1}=0$ and cyclic symmetry, the first two fermionic terms cancel leaving the expected free-fermion hopping Hamiltonian. Since the Hamiltonian is a quadratic form in fermi operators it can be diagonalized by standard free-fermion techniques. Here we diagonalize the transfer matrices using inversion identities to make clear the relation to critical dense polymers. 
The methods we use, based on Yang-Baxter integrability, functional equations and physical combinatorics, are more general and can also be applied to percolation~\cite{MDKP2016} and to the six-vertex model at other roots of unity.

\subsection{Inversion identities on the cylinder}

The inversion identities for the single row transfer matrix of the free-fermion six vertex model with periodic boundary conditions are 
\begin{subequations}
\label{InvIdCyl}
\begin{align}
\vec T(u)\vec T(u+\lambda)&=
\big(\cos^{2N} u- \sin^{2N} u\big) I,\qquad &\mbox{$N$ odd}\label{InvIdCylOdd}\\[4pt]
\vec T_d(u)\vec T_d(u+\lambda)&=\big(\cos^{2N} u+ \sin^{2N} u+2(-1)^d\sin^N u\cos^N u\big) I\nonumber\\
&=(\cos^N u+(-1)^{(N-\ell)/2} \sin^N u)^2 I,\quad &\mbox{$N$ even}\label{InvIdCylEven}
\end{align}
\end{subequations}
These are specializations of the elliptic inversion identities~\cite{Felderhof} of the eight-vertex free-fermion model. 
The derivation following \cite{Felderhof} is given in Appendix~\ref{CylInvProof}. 
For $N$ even, the inversion identity is different in the sectors with even and odd parity for $d$. \mbox{Alternatively},
these inversion identities can be written as 
\bea  
\vec T(u)\vec T(u+\lambda)=\big(\cos^{2N} u+(-1)^N \sin^{2N} u\big) I+(\sin u\cos u)^N J
\eea
where $J=0$ for $N$ odd and, for $N$ even, $J$ is a diagonal matrix with entries $\pm 2$ alternating in the different $d$ sectors. 
The commuting transfer matrices $\vec T(u)$ admit a common set of right eigenvectors independent of $u$. They can therefore be simultaneously diagonalized by a similarity transformation. It therefore follows that the inversion identities are satisfied by the individual eigenvalues. 

As in the case of critical dense polymers, $J$ is related to the braid transfer matrices $\vec B^+$ and its inverse $\vec B^-$ by
\bea
\vec B^{\pm}=\lim_{u\to \pm i\infty} \frac{\vec T(u)}{\sin^N(u+\lambda/2)},\qquad (\vec B^{\pm})^2=
\begin{cases}2I+(-1)^{N/2}J,&\mbox{$N$ even}\\ 2I,&\mbox{$N$ odd}
\end{cases}
\eea

\subsection{Exact eigenvalues}

The inversion identities (\ref{InvIdCylOdd}) and (\ref{InvIdCylEven}), for $N$ odd and $N$ even, precisely coincide with the $N$ odd and $N$ even inversion identities of critical dense polymers~\cite{PRV}. 
In the rest of Section~3, we summarize the solution of these inversion identities and the relevant results of \cite{PRV}. From the Temperley-Lieb equivalence, it is expected that the spectra of the free-fermion six-vertex model and critical dense polymers agree up to the possibility of different degeneracies. The new content of the following subsections is that we find empirically that the spectra of these two models precisely coincide sector-by-sector as matched by the identification (\ref{matching}) of the number of defects $\ell$ of critical dense polymers with $|S_z|$. In particular, the central charge $c=-2$ and conformal weights $\Delta=-\frac{1}{8}, 0, \frac{3}{8}$ of dimers are given by the same calculations using the Euler-Maclaurin formula carried out previously for critical dense polymers so we do not repeat these calculations.

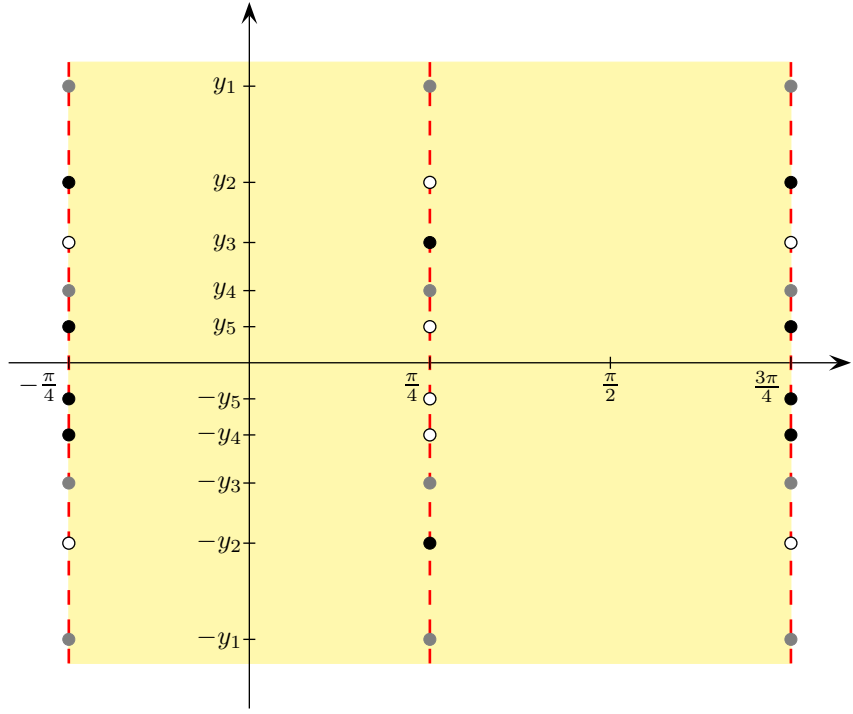
\begin{figure}[htb]
\psset{unit=.8cm}
\setlength{\unitlength}{.8cm}
\begin{center}
\begin{pspicture}[shift=-.45](-.25,.5)(14,12)
\psframe[linecolor=yellow!40!white,linewidth=0pt,fillstyle=solid,
  fillcolor=yellow!40!white](1,1)(13,11)
\psline[linecolor=black,linewidth=.5pt,arrowsize=6pt]{->}(4,.25)(4,12)
\psline[linecolor=black,linewidth=.5pt,arrowsize=6pt]{->}(0,6)(14,6)
\psline[linecolor=red,linewidth=1pt,linestyle=dashed,dash=.25 .25](1,1)(1,11)
\psline[linecolor=red,linewidth=1pt,linestyle=dashed,dash=.25 .25](7,1)(7,11)
\psline[linecolor=red,linewidth=1pt,linestyle=dashed,dash=.25 .25](13,1)(13,11)
\psline[linecolor=black,linewidth=.5pt](1,5.9)(1,6.1)
\psline[linecolor=black,linewidth=.5pt](7,5.9)(7,6.1)
\psline[linecolor=black,linewidth=.5pt](10,5.9)(10,6.1)
\psline[linecolor=black,linewidth=.5pt](13,5.9)(13,6.1)
\rput(.5,5.6){\small $-\frac{\pi}{4}$}
\rput(6.7,5.6){\small $\frac{\pi}{4}$}
\rput(10,5.6){\small $\frac{\pi}{2}$}
\rput(12.6,5.6){\small $\frac{3\pi}{4}$}
\psline[linecolor=black,linewidth=.5pt](3.9,6.6)(4.1,6.6)
\psline[linecolor=black,linewidth=.5pt](3.9,7.2)(4.1,7.2)
\psline[linecolor=black,linewidth=.5pt](3.9,8.0)(4.1,8.0)
\psline[linecolor=black,linewidth=.5pt](3.9,9.0)(4.1,9.0)
\psline[linecolor=black,linewidth=.5pt](3.9,10.6)(4.1,10.6)
\psline[linecolor=black,linewidth=.5pt](3.9,5.4)(4.1,5.4)
\psline[linecolor=black,linewidth=.5pt](3.9,4.8)(4.1,4.8)
\psline[linecolor=black,linewidth=.5pt](3.9,4.0)(4.1,4.0)
\psline[linecolor=black,linewidth=.5pt](3.9,3.0)(4.1,3.0)
\psline[linecolor=black,linewidth=.5pt](3.9,1.4)(4.1,1.4)
\rput(3.6,6.6){\small $y_5$}
\rput(3.6,7.2){\small $y_4$}
\rput(3.6,8.0){\small $y_3$}
\rput(3.6,9.0){\small $y_2$}
\rput(3.6,10.6){\small $y_1$}
\rput(3.5,5.4){\small $-y_5$}
\rput(3.5,4.8){\small $-y_4$}
\rput(3.5,4.0){\small $-y_3$}
\rput(3.5,3.0){\small $-y_2$}
\rput(3.5,1.4){\small $-y_1$}
\psarc[linecolor=black,linewidth=0pt,fillstyle=solid,fillcolor=black](1,6.6){.1}{0}{360}
\psarc[linecolor=gray,linewidth=0pt,fillstyle=solid,fillcolor=gray](1,7.2){.1}{0}{360}
\psarc[linecolor=black,linewidth=.5pt,fillstyle=solid,fillcolor=white](1,8.0){.1}{0}{360}
\psarc[linecolor=black,linewidth=0pt,fillstyle=solid,fillcolor=black](1,9.0){.1}{0}{360}
\psarc[linecolor=gray,linewidth=0pt,fillstyle=solid,fillcolor=gray](1,10.6){.1}{0}{360}
\psarc[linecolor=black,linewidth=.5pt,fillstyle=solid,fillcolor=white](7,6.6){.1}{0}{360}
\psarc[linecolor=gray,linewidth=0pt,fillstyle=solid,fillcolor=gray](7,7.2){.1}{0}{360}
\psarc[linecolor=black,linewidth=0pt,fillstyle=solid,fillcolor=black](7,8.0){.1}{0}{360}
\psarc[linecolor=black,linewidth=.5pt,fillstyle=solid,fillcolor=white](7,9.0){.1}{0}{360}
\psarc[linecolor=gray,linewidth=0pt,fillstyle=solid,fillcolor=gray](7,10.6){.1}{0}{360}
\psarc[linecolor=black,linewidth=0pt,fillstyle=solid,fillcolor=black](13,6.6){.1}{0}{360}
\psarc[linecolor=gray,linewidth=0pt,fillstyle=solid,fillcolor=gray](13,7.2){.1}{0}{360}
\psarc[linecolor=black,linewidth=.5pt,fillstyle=solid,fillcolor=white](13,8.0){.1}{0}{360}
\psarc[linecolor=black,linewidth=0pt,fillstyle=solid,fillcolor=black](13,9.0){.1}{0}{360}
\psarc[linecolor=gray,linewidth=0pt,fillstyle=solid,fillcolor=gray](13,10.6){.1}{0}{360}
\psarc[linecolor=black,linewidth=0pt,fillstyle=solid,fillcolor=black](1,5.4){.1}{0}{360}
\psarc[linecolor=black,linewidth=0pt,fillstyle=solid,fillcolor=black](1,4.8){.1}{0}{360}
\psarc[linecolor=gray,linewidth=0pt,fillstyle=solid,fillcolor=gray](1,4.0){.1}{0}{360}
\psarc[linecolor=black,linewidth=.5pt,fillstyle=solid,fillcolor=white](1,3.0){.1}{0}{360}
\psarc[linecolor=gray,linewidth=0pt,fillstyle=solid,fillcolor=gray](1,1.4){.1}{0}{360}
\psarc[linecolor=black,linewidth=.5pt,fillstyle=solid,fillcolor=white](7,5.4){.1}{0}{360}
\psarc[linecolor=black,linewidth=.5pt,fillstyle=solid,fillcolor=white](7,4.8){.1}{0}{360}
\psarc[linecolor=gray,linewidth=0pt,fillstyle=solid,fillcolor=gray](7,4.0){.1}{0}{360}
\psarc[linecolor=black,linewidth=0pt,fillstyle=solid,fillcolor=black](7,3.0){.1}{0}{360}
\psarc[linecolor=gray,linewidth=0pt,fillstyle=solid,fillcolor=gray](7,1.4){.1}{0}{360}
\psarc[linecolor=black,linewidth=0pt,fillstyle=solid,fillcolor=black](13,5.4){.1}{0}{360}
\psarc[linecolor=black,linewidth=0pt,fillstyle=solid,fillcolor=black](13,4.8){.1}{0}{360}
\psarc[linecolor=gray,linewidth=0pt,fillstyle=solid,fillcolor=gray](13,4.0){.1}{0}{360}
\psarc[linecolor=black,linewidth=.5pt,fillstyle=solid,fillcolor=white](13,3.0){.1}{0}{360}
\psarc[linecolor=gray,linewidth=0pt,fillstyle=solid,fillcolor=gray](13,1.4){.1}{0}{360}
\end{pspicture}
\end{center}
\caption{A typical pattern of zeros in the complex $u$-plane for the $\ell$ even sectors. Here,
$N=12$ and $\ell=|S_z|=2$. The ordinates of the locations of the zeros $u_j$ are
$y_j=-\half \log \tan{\frac{j\pi}{N}}, j=1,2,\ldots,N/2-1$. At each position $j$, there is either 
two 1-strings with $\Re u_j=\pi/4$, two 2-strings with real parts $\Re u_j=-\pi/4, 3\pi/4$ or 
one 1-string and one 2-string. A double zero is indicated by a black circle, a single zero 
by a grey circle and an unoccupied position by an open circle.\label{uplaneR}}
\end{figure}

For $N$ and $\ell=|S_z|$ even, 
the transfer matrix eigenvalue spectra breaks up into Ramond and Neveu-Schwarz sectors according to the even or odd parity of $\ell/2=|S_z|/2$. As in (5.24) and (5.25) of \cite{PRV}, the eigenvalues $T(u)$ factor into elementary contributions arising from zeros in the complex $u$-plane in the form of single or double 1-strings on the line $\Re u=\frac{\pi}{4}$ at ordinates
\bea
y_j=\begin{cases}
  -\half \log \tan{\frac{\half(j-\half)\pi}{N}},& \mbox{$\mathbb{Z}_4$:\ \ $N, \ell$ odd},\quad j=1,2,\ldots,N\\[6pt]
  \disp-\half \log \tan{\frac{(j\!-\!\half)\pi}{N}},\quad & \mbox{R:\ \ $N, \ell/2$ even},\quad  j=1,2,\ldots,N/2\\[6pt]
  \disp-\half \log \tan{\frac{j\pi}{N}}, & \mbox{NS:\ \ $N$ even, $\ell/2$ odd},\quad  j=1,2,\ldots,N/2-1
  \end{cases}\label{ordinates}
\eea
A typical pattern of zeros is shown in Figure~\ref{uplaneR}. 
From \cite{PRV}, the elementary excitation energy of a single 1-string at position $j$ in the upper or lower half plane is
\bea
E_j\;=\;\begin{cases}
  \half(j-\half),&\mbox{$\mathbb{Z}_4$:\ \ $N$, $\ell$ odd}\\
  j-\half,&\mbox{R:\ \ $N, \ell/2$ even}\\
  j,&\mbox{NS:\ \ $N$ even, $\ell/2$ odd}
\end{cases}
\eea
For $N$ odd, the contributions from 1-strings in each half-plane are encoded in one-column diagrams as shown in Figures~\ref{EsigmaZ4} and \ref{binomZ4}.
For $N$ even, the contributions from the single or double 1-strings in each half-plane are encoded in double-column diagrams as shown in Figures~\ref{EsigmaR1}--\ref{binomR2}. 
Each column of a two-column diagram has a 0- or 1-string at a given position or height, labelled by $j$. These combine to encode no 1-string, a single 1-string or a double 1-string at each height $j$ or position (\ref{ordinates}) in the pattern of zeros. By convention, zeros on the real $u$-axis are regarded as being in the upper half plane.

Explicitly, as shown in \cite{PRV,MDPR2013}, the solution of the inversion identity (\ref{InvIdCylEven}) by factorization of the eigenvalues yields
\begin{subequations}
\label{expSoln}
\begin{align}
\mbox{}\hspace{-8pt}T(u)&=\epsilon\,\frac{(-i)^{N/2}e^{-Niu}}{2^{N-1/2}}
  \prod_{j=1}^{N} {\Big(e^{2i u}+i\epsilon_j  \tan{\frac{(2j-1)\pi}{4N}}\Big)},\quad \mbox{$\mathbb{Z}_4$:\ \ $N,\ell$ odd}\\
\mbox{}\hspace{-8pt}T(u)&=\disp \frac{\epsilon(-i)^{\frac{N}{2}}e^{-Niu}}{2^{N-1}} \prod_{j=1}^{N} 
\Big( e^{2iu}  + i \epsilon_j \tan \frac{(2j\!-\!1)\pi}{2N}\Big) 
 , \ \ \mbox{R:\ \ $N,\ell/2$ even}\\
\mbox{}\hspace{-8pt}T(u)&= \disp\frac{\epsilon(-i)^{\frac{N}{2}}\!Ne^{-Niu}}{2^{N-1}} \mathop{\prod_{j=1}^{N}}_{j\ne N/2}\!\!
\Big( e^{2iu}  + i \epsilon_j \tan \frac{j\pi}{N}\Big) 
,\ \ \mbox{NS:\ $N$ even, $\ell/2$ odd}
\end{align}
\end{subequations}
where $\epsilon_j=\pm 1$. The overall sign $\epsilon=\pm 1$ of each eigenvalue is not fixed by the inversion relation. 
These sign factors $\epsilon$ are fixed by \cite{MDPR2013}
\bea
\epsilon=(-1)^{\frac{N-s}{4}},\qquad\epsilon^\text{R}=\epsilon^\text{NS}=(-1)^{\floor{\frac{|s|+2}{4}}}, 
\qquad s=S_z\label{epsfix}
\eea 
Separating the zeros in the upper and lower half planes leads to
\begin{subequations}
\begin{align}
T(u)&=\frac{\mu\, (-i)^{N/2}}{2^{N-1/2}e^{Niu}}
  \prod_{j=1}^{\frac{N+1}{2}}{\Big(e^{2i u}+i\epsilon_j  \tan{\frac{(2j-1)\pi}{4N}}\Big)}
  \prod_{j=1}^{\frac{N-1}{2}}{\Big(\bar\epsilon_j e^{2i u}+i \cot{\frac{(2j-1)\pi}{4N}}\Big)},\quad \mbox{$\mathbb{Z}_4$:\ \ $N,\ell$ odd}
\label{T(u)Nodd}\\
T(u)&\;=\;\frac{\mu(-i)^{\frac{N}{2}}e^{-Niu}}{2^{N-1}}
\prod_{j=1}^{\floor{(N+2)/4}}{\Big(e^{2i u}+i\epsilon_j  \tan{\frac{(2j-1)\pi}{2N}}\Big)}
{\Big(\bar{\epsilon}_j e^{2i u}+i  \cot{\frac{(2j-1)\pi}{2N}}\Big)} \nonumber\\
&\times\prod_{j=1}^{\floor{N/4}}{\Big(e^{2i u}+i\mu_j  \tan{\frac{(2j-1)\pi}{2N}}\Big)}
\Big(\bar{\mu}_j e^{2i u}+i \cot{\frac{(2j-1)\pi}{2N}}\Big), \qquad\mbox{R:\ \ $N,\ell/2$ even}
\label{T(u)even}\\
T(u)&\;=\;\frac{\mu(-i)^{\frac{N-2}{2}}Ne^{(2-N)iu}}{2^{N-1}}
\prod_{j=1}^{\floor{N/4}}{\Big(e^{2i u}+i\epsilon_j  \tan{\frac{j\pi}{N}}\Big)}
{\Big(\bar{\epsilon}_j e^{2i u}+i  \cot{\frac{j\pi}{N}}\Big)} \nonumber\\
&\times\prod_{j=1}^{\floor{(N-2)/4}}{\Big(e^{2i u}+i\mu_j  \tan{\frac{j\pi}{N}}\Big)}
\Big(\bar{\mu}_j e^{2i u}+i \cot{\frac{j\pi}{N}}\Big), \qquad\qquad\qquad\mbox{NS:\ \ $N$ even, $\ell/2$ odd}
\label{T(u)odd}
\end{align}
\end{subequations}
where $\mu, \epsilon_j,\bar\epsilon_j,\mu_j,\bar\mu_j=\pm 1$. Up to the overall choice of sign $\mu$, there are either $2^N$ 
or $2^{N-2}$ possible eigenvalues allowing for all excitations but they are not all physical and are subject to selection rules. 
The number of 1-strings $m_j$ plus the number of 2-strings $n_j$ at any given position is
\bea
m_j+n_j=\begin{cases}1,&\mathbb{Z}_4\\ 2,&\mbox{R, NS}\end{cases}
\eea

\subsection{Patterns of zeros and selection rules}

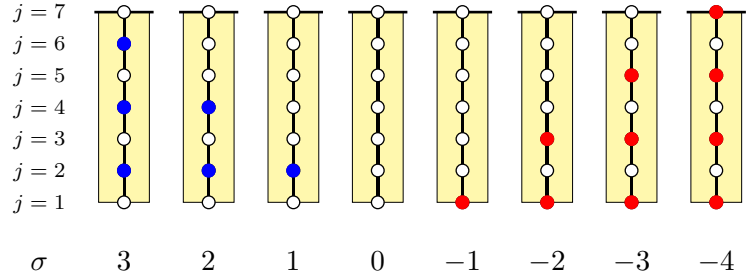
\begin{figure}[htbp]
\setlength{\unitlength}{.8pt}
\psset{unit=.8pt}
\begin{center}
\begin{pspicture}(-50,-20)(280,100)
\thicklines
\multirput(0,0)(40,0){8}{\psframe[linewidth=0pt,fillstyle=solid,
  fillcolor=yellow!40!white](-12,0)(12,90)}
\multiput(0,0)(40,0){8}{\line(0,10){90}}
\multirput(0,0)(40,0){8}{\psline[linewidth=1pt](-14,90)(14,90)}
\multiput(0,0)(40,0){8}{\multiput(0,0)(0,15){7}{\psarc[linecolor=black,linewidth=.5pt,fillstyle=solid,
  fillcolor=white](0,0){3}{0}{360}}}
\multiput(0,0)(0,30){3}{\psarc[linecolor=blue,linewidth=.5pt,fillstyle=solid,
  fillcolor=blue](0,15){3}{0}{360}}
\multiput(40,0)(0,30){2}{\psarc[linecolor=blue,linewidth=.5pt,fillstyle=solid,
  fillcolor=blue](0,15){3}{0}{360}}
\psarc[linecolor=blue,linewidth=.5pt,fillstyle=solid,fillcolor=blue](80,15){3}{0}{360}
\psarc[linecolor=red,linewidth=.5pt,fillstyle=solid,fillcolor=red](160,0){3}{0}{360}
\multiput(200,0)(0,30){2}{\psarc[linecolor=red,linewidth=.5pt,fillstyle=solid,
  fillcolor=red](0,0){3}{0}{360}}
\multiput(240,0)(0,30){3}{\psarc[linecolor=red,linewidth=.5pt,fillstyle=solid,
  fillcolor=red](0,0){3}{0}{360}}
\multiput(280,0)(0,30){4}{\psarc[linecolor=red,linewidth=.5pt,fillstyle=solid,
  fillcolor=red](0,0){3}{0}{360}}
\rput[B](-40,-32){$\sigma$}
\rput[B](0,-32){$3$}
\rput[B](40,-32){$2$}
\rput[B](80,-32){$1$}
\rput[B](120,-32){$0$}
\rput[B](160,-32){$-1$}
\rput[B](200,-32){$-2$}
\rput[B](240,-32){$-3$}
\rput[B](280,-32){$-4$}
\rput[B](-40,-3){\scriptsize $j=1$}
\rput[B](-40,12){\scriptsize $j=2$}
\rput[B](-40,27){\scriptsize $j=3$}
\rput[B](-40,42){\scriptsize $j=4$}
\rput[B](-40,57){\scriptsize $j=5$}
\rput[B](-40,72){\scriptsize $j=6$}
\rput[B](-40,87){\scriptsize $j=7$}
\end{pspicture}
\end{center}
\smallskip
\caption{$\mathbb{Z}_4$ sectors ($N$, $\ell$ odd): Minimal configurations of 
single-columns with energy $E(\sigma)=\mbox{$\half\sigma(\sigma+\half)$}$.
The quantum number $\sigma=\floor{n/2}-m$ is given by the excess of blue (even $j$) over 
red (odd $j$) 1-strings.  At each empty position $j$, there is a 2-string. This analyticity strip 
is in the upper-half complex $u$-plane rotated by 180 degrees so that position $j=1$ 
(furthest from the real axis) is at the bottom.\label{EsigmaZ4}}
\end{figure}
\begin{figure}[p]
\setlength{\unitlength}{.8pt}
\psset{unit=.8pt}
\begin{center}
\begin{pspicture}(-50,-30)(400,340)
\thicklines
\multirput(0,0)(40,0){11}{\psframe[linewidth=0pt,fillstyle=solid,
  fillcolor=yellow!40!white](-12,0)(12,90)}
\multirput(0,0)(40,0){11}{\psline[linewidth=1pt](-14,90)(14,90)}
\multirput(80,120)(40,0){7}{\psframe[linewidth=0pt,fillstyle=solid,
  fillcolor=yellow!40!white](-12,0)(12,90)}
\multirput(80,120)(40,0){7}{\psline[linewidth=1pt](-14,90)(14,90)}
\multirput(160,240)(40,0){3}{\psframe[linewidth=0pt,fillstyle=solid,
  fillcolor=yellow!40!white](-12,0)(12,90)}
\multirput(160,240)(40,0){3}{\psline[linewidth=1pt](-14,90)(14,90)}
%
\multiput(0,0)(40,0){11}{\multiput(0,0)(0,15){7}{\psarc[linecolor=black,linewidth=.5pt,fillstyle=solid,
  fillcolor=white](0,0){3}{0}{360}}}
\multiput(80,120)(40,0){7}{\multiput(0,0)(0,15){7}{\psarc[linecolor=black,
  linewidth=.5pt,fillstyle=solid,fillcolor=white](0,0){3}{0}{360}}}
\multiput(160,240)(40,0){3}{\multiput(0,0)(0,15){7}{\psarc[linecolor=black,
  linewidth=.5pt,fillstyle=solid,fillcolor=white](0,0){3}{0}{360}}}
\psarc[linecolor=red,linewidth=.5pt,fillstyle=solid,fillcolor=red](0,0){3}{0}{360}
\psarc[linecolor=red,linewidth=.5pt,fillstyle=solid,fillcolor=red](0,30){3}{0}{360}
\psarc[linecolor=red,linewidth=.5pt,fillstyle=solid,fillcolor=red](40,0){3}{0}{360}
\psarc[linecolor=red,linewidth=.5pt,fillstyle=solid,fillcolor=red](40,60){3}{0}{360}
\psarc[linecolor=red,linewidth=.5pt,fillstyle=solid,fillcolor=red](80,30){3}{0}{360}
\psarc[linecolor=red,linewidth=.5pt,fillstyle=solid,fillcolor=red](80,60){3}{0}{360}
\psarc[linecolor=red,linewidth=.5pt,fillstyle=solid,fillcolor=red](80,120){3}{0}{360}
\psarc[linecolor=red,linewidth=.5pt,fillstyle=solid,fillcolor=red](80,210){3}{0}{360}
\psarc[linecolor=red,linewidth=.5pt,fillstyle=solid,fillcolor=red](120,0){3}{0}{360}
\psarc[linecolor=blue,linewidth=.5pt,fillstyle=solid,fillcolor=blue](120,15){3}{0}{360}
\psarc[linecolor=red,linewidth=.5pt,fillstyle=solid,fillcolor=red](120,30){3}{0}{360}
\psarc[linecolor=red,linewidth=.5pt,fillstyle=solid,fillcolor=red](120,60){3}{0}{360}
\psarc[linecolor=red,linewidth=.5pt,fillstyle=solid,fillcolor=red](120,150){3}{0}{360}
\psarc[linecolor=red,linewidth=.5pt,fillstyle=solid,fillcolor=red](120,210){3}{0}{360}
\psarc[linecolor=red,linewidth=.5pt,fillstyle=solid,fillcolor=red](160,0){3}{0}{360}
\psarc[linecolor=blue,linewidth=.5pt,fillstyle=solid,fillcolor=blue](160,45){3}{0}{360}
\psarc[linecolor=red,linewidth=.5pt,fillstyle=solid,fillcolor=red](160,30){3}{0}{360}
\psarc[linecolor=red,linewidth=.5pt,fillstyle=solid,fillcolor=red](160,60){3}{0}{360}
\psarc[linecolor=red,linewidth=.5pt,fillstyle=solid,fillcolor=red](160,120){3}{0}{360}
\psarc[linecolor=blue,linewidth=.5pt,fillstyle=solid,fillcolor=blue](160,135){3}{0}{360}
\psarc[linecolor=red,linewidth=.5pt,fillstyle=solid,fillcolor=red](160,150){3}{0}{360}
\psarc[linecolor=red,linewidth=.5pt,fillstyle=solid,fillcolor=red](160,210){3}{0}{360}
\psarc[linecolor=red,linewidth=.5pt,fillstyle=solid,fillcolor=red](160,300){3}{0}{360}
\psarc[linecolor=red,linewidth=.5pt,fillstyle=solid,fillcolor=red](160,330){3}{0}{360}
\psarc[linecolor=red,linewidth=.5pt,fillstyle=solid,fillcolor=red](200,0){3}{0}{360}
\psarc[linecolor=red,linewidth=.5pt,fillstyle=solid,fillcolor=red](200,30){3}{0}{360}
\psarc[linecolor=blue,linewidth=.5pt,fillstyle=solid,fillcolor=blue](200,75){3}{0}{360}
\psarc[linecolor=red,linewidth=.5pt,fillstyle=solid,fillcolor=red](200,60){3}{0}{360}
\psarc[linecolor=red,linewidth=.5pt,fillstyle=solid,fillcolor=red](200,120){3}{0}{360}
\psarc[linecolor=blue,linewidth=.5pt,fillstyle=solid,fillcolor=blue](200,165){3}{0}{360}
\psarc[linecolor=red,linewidth=.5pt,fillstyle=solid,fillcolor=red](200,150){3}{0}{360}
\psarc[linecolor=red,linewidth=.5pt,fillstyle=solid,fillcolor=red](200,210){3}{0}{360}
\psarc[linecolor=red,linewidth=.5pt,fillstyle=solid,fillcolor=red](200,240){3}{0}{360}
\psarc[linecolor=blue,linewidth=.5pt,fillstyle=solid,fillcolor=blue](200,255){3}{0}{360}
\psarc[linecolor=red,linewidth=.5pt,fillstyle=solid,fillcolor=red](200,300){3}{0}{360}
\psarc[linecolor=red,linewidth=.5pt,fillstyle=solid,fillcolor=red](200,330){3}{0}{360}
\psarc[linecolor=red,linewidth=.5pt,fillstyle=solid,fillcolor=red](240,0){3}{0}{360}
\psarc[linecolor=red,linewidth=.5pt,fillstyle=solid,fillcolor=red](240,30){3}{0}{360}
\psarc[linecolor=blue,linewidth=.5pt,fillstyle=solid,fillcolor=blue](240,75){3}{0}{360}
\psarc[linecolor=red,linewidth=.5pt,fillstyle=solid,fillcolor=red](240,90){3}{0}{360}
\psarc[linecolor=red,linewidth=.5pt,fillstyle=solid,fillcolor=red](240,150){3}{0}{360}
\psarc[linecolor=blue,linewidth=.5pt,fillstyle=solid,fillcolor=blue](240,135){3}{0}{360}
\psarc[linecolor=red,linewidth=.5pt,fillstyle=solid,fillcolor=red](240,180){3}{0}{360}
\psarc[linecolor=red,linewidth=.5pt,fillstyle=solid,fillcolor=red](240,210){3}{0}{360}
\psarc[linecolor=red,linewidth=.5pt,fillstyle=solid,fillcolor=red](240,240){3}{0}{360}
\psarc[linecolor=blue,linewidth=.5pt,fillstyle=solid,fillcolor=blue](240,285){3}{0}{360}
\psarc[linecolor=red,linewidth=.5pt,fillstyle=solid,fillcolor=red](240,300){3}{0}{360}
\psarc[linecolor=red,linewidth=.5pt,fillstyle=solid,fillcolor=red](240,330){3}{0}{360}
\psarc[linecolor=red,linewidth=.5pt,fillstyle=solid,fillcolor=red](280,30){3}{0}{360}
\psarc[linecolor=blue,linewidth=.5pt,fillstyle=solid,fillcolor=blue](280,45){3}{0}{360}
\psarc[linecolor=red,linewidth=.5pt,fillstyle=solid,fillcolor=red](280,60){3}{0}{360}
\psarc[linecolor=red,linewidth=.5pt,fillstyle=solid,fillcolor=red](280,90){3}{0}{360}
\psarc[linecolor=red,linewidth=.5pt,fillstyle=solid,fillcolor=red](280,120){3}{0}{360}
\psarc[linecolor=red,linewidth=.5pt,fillstyle=solid,fillcolor=red](280,180){3}{0}{360}
\psarc[linecolor=blue,linewidth=.5pt,fillstyle=solid,fillcolor=blue](280,195){3}{0}{360}
\psarc[linecolor=red,linewidth=.5pt,fillstyle=solid,fillcolor=red](280,210){3}{0}{360}
\psarc[linecolor=red,linewidth=.5pt,fillstyle=solid,fillcolor=red](320,0){3}{0}{360}
\psarc[linecolor=blue,linewidth=.5pt,fillstyle=solid,fillcolor=blue](320,15){3}{0}{360}
\psarc[linecolor=red,linewidth=.5pt,fillstyle=solid,fillcolor=red](320,30){3}{0}{360}
\psarc[linecolor=blue,linewidth=.5pt,fillstyle=solid,fillcolor=blue](320,45){3}{0}{360}
\psarc[linecolor=red,linewidth=.5pt,fillstyle=solid,fillcolor=red](320,60){3}{0}{360}
\psarc[linecolor=red,linewidth=.5pt,fillstyle=solid,fillcolor=red](320,90){3}{0}{360}
\psarc[linecolor=red,linewidth=.5pt,fillstyle=solid,fillcolor=red](320,150){3}{0}{360}
\psarc[linecolor=red,linewidth=.5pt,fillstyle=solid,fillcolor=red](320,180){3}{0}{360}
\psarc[linecolor=blue,linewidth=.5pt,fillstyle=solid,fillcolor=blue](320,195){3}{0}{360}
\psarc[linecolor=red,linewidth=.5pt,fillstyle=solid,fillcolor=red](320,210){3}{0}{360}
\psarc[linecolor=red,linewidth=.5pt,fillstyle=solid,fillcolor=red](360,0){3}{0}{360}
\psarc[linecolor=blue,linewidth=.5pt,fillstyle=solid,fillcolor=blue](360,15){3}{0}{360}
\psarc[linecolor=red,linewidth=.5pt,fillstyle=solid,fillcolor=red](360,30){3}{0}{360}
\psarc[linecolor=blue,linewidth=.5pt,fillstyle=solid,fillcolor=blue](360,75){3}{0}{360}
\psarc[linecolor=red,linewidth=.5pt,fillstyle=solid,fillcolor=red](360,60){3}{0}{360}
\psarc[linecolor=red,linewidth=.5pt,fillstyle=solid,fillcolor=red](360,90){3}{0}{360}
\psarc[linecolor=red,linewidth=.5pt,fillstyle=solid,fillcolor=red](400,0){3}{0}{360}
\psarc[linecolor=blue,linewidth=.5pt,fillstyle=solid,fillcolor=blue](400,45){3}{0}{360}
\psarc[linecolor=red,linewidth=.5pt,fillstyle=solid,fillcolor=red](400,30){3}{0}{360}
\psarc[linecolor=blue,linewidth=.5pt,fillstyle=solid,fillcolor=blue](400,75){3}{0}{360}
\psarc[linecolor=red,linewidth=.5pt,fillstyle=solid,fillcolor=red](400,60){3}{0}{360}
\psarc[linecolor=red,linewidth=.5pt,fillstyle=solid,fillcolor=red](400,90){3}{0}{360}
\rput[B](-43,-30){$\qbin75q\;=$}
\multiput(15,-30)(40,0){10}{$+$}
\rput[B](0,-30){$1$}
\rput[B](40,-30){$q$}
\rput[B](80,-30){$2q^2$}
\rput[B](120,-30){$2q^3$}
\rput[B](160,-30){$3q^4$}
\rput[B](200,-30){$3q^5$}
\rput[B](240,-30){$3q^6$}
\rput[B](280,-30){$2q^7$}
\rput[B](320,-30){$2q^8$}
\rput[B](360,-30){$q^9$}
\rput[B](400,-30){$q^{10}$}
\rput[B](-45,-3){\scriptsize $j=1$}
\rput[B](-45,12){\scriptsize $j=2$}
\rput[B](-45,27){\scriptsize $j=3$}
\rput[B](-45,42){\scriptsize $j=4$}
\rput[B](-45,57){\scriptsize $j=5$}
\rput[B](-45,72){\scriptsize $j=6$}
\rput[B](-45,87){\scriptsize $j=7$}
\end{pspicture}
\end{center}
\smallskip
\caption{$\mathbb{Z}_4$ sectors ($N$, $\ell$ odd): Combinatorial enumeration by 
single-columns of the $q$-binomial $\sqbin nmq=\sqbin 75q
=q^{-3/2} \sum q^{\sum_j m_j E_j}$. The excess of blue (even $j$) over red (odd $j$) 1-strings 
is given by the quantum number $\sigma=\floor{n/2}-m=-2$.   The elementary excitation energy 
of a 1-string at position $j$ is $E_j=\half(j-\half)$. The lowest energy configuration has energy 
$E(\sigma)=1/4+5/4=3/2=\half\sigma(\sigma+\half)$. At each empty position $j$, there is a 2-string.
This analyticity strip is in the upper-half complex $u$-plane rotated by 180 degrees so that 
position $j=1$ (furthest from the real axis) is at the bottom. The elementary excitations 
(of energy 1) are generated by either inserting two 1-strings at positions $j=1$ and $j=2$ 
or promoting a 1-string at position $j$ to position $j+2$.  Notice that 
$\sqbin nmq=\sqbin n{n-m}q$ as $q$-polynomials but they have different combinatorial 
interpretations because they have different quantum numbers $\sigma$. In the lower half-plane, 
$q$ is replaced with $\qbar$ and no rotation is required. In this example, $\ell=7$ and the value $\bar\sigma=-2$ of the 
quantum number in the lower half-plane is related to $\sigma=-2$ in the upper half-plane by 
the selection rules $\sigma+\bar\sigma=-(\ell+1)/2$ and 
$\half(\sigma-\bar\sigma)\in\mathbb{Z}$.\label{binomZ4}}
\end{figure}

\begin{figure}[p]
\setlength{\unitlength}{.8pt}
\psset{unit=.8pt}
\begin{center}
\begin{pspicture}(-50,-20)(335,50)
\thicklines
\multirput(0,0)(52,0){7}{\psframe[linewidth=0pt,fillstyle=solid,
  fillcolor=yellow!40!white](-18,-7.5)(18,37.5)}
\multirput(0,0)(52,0){7}{\psline[linewidth=1pt](-20,37.5)(20,37.5)}
\multiput(-7,0)(52,0){7}{\multiput(0,0)(0,15){3}{\psarc[linecolor=black,
  linewidth=.5pt,fillstyle=solid,fillcolor=white](0,0){3}{0}{360}}}
\multiput(7,0)(52,0){7}{\multiput(0,0)(0,15){3}{\psarc[linecolor=black,
  linewidth=.5pt,fillstyle=solid,fillcolor=white](0,0){3}{0}{360}}}
\multiput(-7,0)(0,15){3}{\psarc[linecolor=red,linewidth=.5pt,fillstyle=solid,
  fillcolor=red](0,0){3}{0}{360}}
\multiput(45,0)(0,15){2}{\psarc[linecolor=red,linewidth=.5pt,fillstyle=solid,
  fillcolor=red](0,0){3}{0}{360}}
\multiput(97,0)(0,15){1}{\psarc[linecolor=red,linewidth=.5pt,fillstyle=solid,
  fillcolor=red](0,0){3}{0}{360}}
\multiput(215,0)(0,15){1}{\psarc[linecolor=blue,linewidth=.5pt,fillstyle=solid,
  fillcolor=blue](0,0){3}{0}{360}}
\multiput(267,0)(0,15){2}{\psarc[linecolor=blue,linewidth=.5pt,fillstyle=solid,
  fillcolor=blue](0,0){3}{0}{360}}
\multiput(319,0)(0,15){3}{\psarc[linecolor=blue,linewidth=.5pt,fillstyle=solid,
  fillcolor=blue](0,0){3}{0}{360}}
\rput[B](-40,-32){$\sigma$}
\rput[B](0,-32){$-3$}
\rput[B](52,-32){$-2$}
\rput[B](104,-32){$-1$}
\rput[B](156,-32){$0$}
\rput[B](208,-32){$1$}
\rput[B](260,-32){$2$}
\rput[B](312,-32){$3$}
\rput[B](-40,-3){\scriptsize $j=1$}
\rput[B](-40,12){\scriptsize $j=2$}
\rput[B](-40,27){\scriptsize $j=3$}
\end{pspicture}
\end{center}
\smallskip
\caption{Ramond sectors ($N$ even, $\ell/2$ even): Minimal configurations of double-columns 
with energy $E(\sigma)=\half\sigma^2$. The quantum number $\sigma=\floor{n/2}-m$ 
is given by the excess of blue (right) over red (left) 1-strings.  At each position $j$, the number 
of 1-strings $m_j$ plus the number of 2-strings $n_j$ is 2. This analyticity strip is in the 
upper-half complex $u$-plane rotated by 180 degrees so that position $j=1$ (furthest from the 
real axis) is at the bottom.\label{EsigmaR1}}
\end{figure}
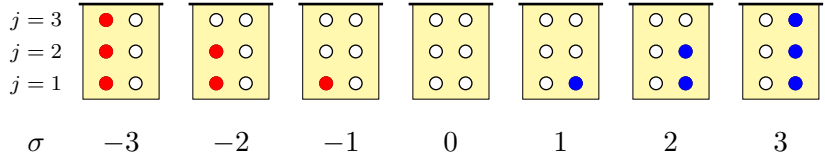
\begin{figure}[p]
\setlength{\unitlength}{.7pt}
\psset{unit=.7pt}
\begin{center}
\begin{pspicture}(-60,-30)(450,170)
\thicklines
\multirput(0,0)(52,0){9}{\psframe[linewidth=0pt,fillstyle=solid,
  fillcolor=yellow!40!white](-18,-7.5)(18,37.5)}
\multirput(0,0)(52,0){9}{\psline[linewidth=1pt](-20,37.5)(20,37.5)}
\multirput(104,60)(52,0){5}{\psframe[linewidth=0pt,fillstyle=solid,
  fillcolor=yellow!40!white](-18,-7.5)(18,37.5)}
\multirput(104,60)(52,0){5}{\psline[linewidth=1pt](-20,37.5)(20,37.5)}
\multirput(208,120)(52,0){1}{\psframe[linewidth=0pt,fillstyle=solid,
  fillcolor=yellow!40!white](-18,-7.5)(18,37.5)}
\multirput(208,120)(52,0){1}{\psline[linewidth=1pt](-20,37.5)(20,37.5)}
%
\multiput(-7,0)(52,0){9}{\multiput(0,0)(0,15){3}{\psarc[linecolor=black,
  linewidth=.5pt,fillstyle=solid,fillcolor=white](0,0){3}{0}{360}}}
\multiput(7,0)(52,0){9}{\multiput(0,0)(0,15){3}{\psarc[linecolor=black,
  linewidth=.5pt,fillstyle=solid,fillcolor=white](0,0){3}{0}{360}}}
\multiput(97,60)(52,0){5}{\multiput(0,0)(0,15){3}{\psarc[linecolor=black,
  linewidth=.5pt,fillstyle=solid,fillcolor=white](0,0){3}{0}{360}}}
\multiput(111,60)(52,0){5}{\multiput(0,0)(0,15){3}{\psarc[linecolor=black,
  linewidth=.5pt,fillstyle=solid,fillcolor=white](0,0){3}{0}{360}}}
\multiput(201,120)(52,0){1}{\multiput(0,0)(0,15){3}{\psarc[linecolor=black,
  linewidth=.5pt,fillstyle=solid,fillcolor=white](0,0){3}{0}{360}}}
\multiput(215,120)(52,0){1}{\multiput(0,0)(0,15){3}{\psarc[linecolor=black,
  linewidth=.5pt,fillstyle=solid,fillcolor=white](0,0){3}{0}{360}}}
\psarc[linecolor=blue,linewidth=.5pt,fillstyle=solid,fillcolor=blue](7,0){3}{0}{360}
\psarc[linecolor=blue,linewidth=.5pt,fillstyle=solid,fillcolor=blue](59,15){3}{0}{360}
\psarc[linecolor=red,linewidth=.5pt,fillstyle=solid,fillcolor=red](97,0){3}{0}{360}
\psarc[linecolor=blue,linewidth=.5pt,fillstyle=solid,fillcolor=blue](111,0){3}{0}{360}
\psarc[linecolor=blue,linewidth=.5pt,fillstyle=solid,fillcolor=blue](111,15){3}{0}{360}
\psarc[linecolor=blue,linewidth=.5pt,fillstyle=solid,fillcolor=blue](111,90){3}{0}{360}
\psarc[linecolor=red,linewidth=.5pt,fillstyle=solid,fillcolor=red](149,0){3}{0}{360}
\psarc[linecolor=red,linewidth=.5pt,fillstyle=solid,fillcolor=red](149,75){3}{0}{360}
\psarc[linecolor=blue,linewidth=.5pt,fillstyle=solid,fillcolor=blue](163,0){3}{0}{360}
\psarc[linecolor=blue,linewidth=.5pt,fillstyle=solid,fillcolor=blue](163,30){3}{0}{360}
\psarc[linecolor=blue,linewidth=.5pt,fillstyle=solid,fillcolor=blue](163,60){3}{0}{360}
\psarc[linecolor=blue,linewidth=.5pt,fillstyle=solid,fillcolor=blue](163,75){3}{0}{360}
\psarc[linecolor=red,linewidth=.5pt,fillstyle=solid,fillcolor=red](201,15){3}{0}{360}
\psarc[linecolor=red,linewidth=.5pt,fillstyle=solid,fillcolor=red](201,60){3}{0}{360}
\psarc[linecolor=red,linewidth=.5pt,fillstyle=solid,fillcolor=red](201,150){3}{0}{360}
\psarc[linecolor=blue,linewidth=.5pt,fillstyle=solid,fillcolor=blue](215,0){3}{0}{360}
\psarc[linecolor=blue,linewidth=.5pt,fillstyle=solid,fillcolor=blue](215,30){3}{0}{360}
\psarc[linecolor=blue,linewidth=.5pt,fillstyle=solid,fillcolor=blue](215,75){3}{0}{360}
\psarc[linecolor=blue,linewidth=.5pt,fillstyle=solid,fillcolor=blue](215,90){3}{0}{360}
\psarc[linecolor=blue,linewidth=.5pt,fillstyle=solid,fillcolor=blue](215,120){3}{0}{360}
\psarc[linecolor=blue,linewidth=.5pt,fillstyle=solid,fillcolor=blue](215,135){3}{0}{360}
\psarc[linecolor=red,linewidth=.5pt,fillstyle=solid,fillcolor=red](253,15){3}{0}{360}
\psarc[linecolor=red,linewidth=.5pt,fillstyle=solid,fillcolor=red](253,90){3}{0}{360}
\psarc[linecolor=blue,linewidth=.5pt,fillstyle=solid,fillcolor=blue](267,15){3}{0}{360}
\psarc[linecolor=blue,linewidth=.5pt,fillstyle=solid,fillcolor=blue](267,30){3}{0}{360}
\psarc[linecolor=blue,linewidth=.5pt,fillstyle=solid,fillcolor=blue](267,60){3}{0}{360}
\psarc[linecolor=blue,linewidth=.5pt,fillstyle=solid,fillcolor=blue](267,90){3}{0}{360}
\psarc[linecolor=red,linewidth=.5pt,fillstyle=solid,fillcolor=red](305,0){3}{0}{360}
\psarc[linecolor=red,linewidth=.5pt,fillstyle=solid,fillcolor=red](305,15){3}{0}{360}
\psarc[linecolor=red,linewidth=.5pt,fillstyle=solid,fillcolor=red](305,90){3}{0}{360}
\psarc[linecolor=blue,linewidth=.5pt,fillstyle=solid,fillcolor=blue](319,0){3}{0}{360}
\psarc[linecolor=blue,linewidth=.5pt,fillstyle=solid,fillcolor=blue](319,15){3}{0}{360}
\psarc[linecolor=blue,linewidth=.5pt,fillstyle=solid,fillcolor=blue](319,30){3}{0}{360}
\psarc[linecolor=blue,linewidth=.5pt,fillstyle=solid,fillcolor=blue](319,75){3}{0}{360}
\psarc[linecolor=blue,linewidth=.5pt,fillstyle=solid,fillcolor=blue](319,90){3}{0}{360}
\psarc[linecolor=red,linewidth=.5pt,fillstyle=solid,fillcolor=red](357,0){3}{0}{360}
\psarc[linecolor=red,linewidth=.5pt,fillstyle=solid,fillcolor=red](357,30){3}{0}{360}
\psarc[linecolor=blue,linewidth=.5pt,fillstyle=solid,fillcolor=blue](371,0){3}{0}{360}
\psarc[linecolor=blue,linewidth=.5pt,fillstyle=solid,fillcolor=blue](371,15){3}{0}{360}
\psarc[linecolor=blue,linewidth=.5pt,fillstyle=solid,fillcolor=blue](371,30){3}{0}{360}
\psarc[linecolor=red,linewidth=.5pt,fillstyle=solid,fillcolor=red](409,15){3}{0}{360}
\psarc[linecolor=red,linewidth=.5pt,fillstyle=solid,fillcolor=red](409,30){3}{0}{360}
\psarc[linecolor=blue,linewidth=.5pt,fillstyle=solid,fillcolor=blue](423,0){3}{0}{360}
\psarc[linecolor=blue,linewidth=.5pt,fillstyle=solid,fillcolor=blue](423,15){3}{0}{360}
\psarc[linecolor=blue,linewidth=.5pt,fillstyle=solid,fillcolor=blue](423,30){3}{0}{360}
\rput[B](-43,-32){$\qbin62q\;=$}
\multiput(21,-32)(52,0){8}{$+$}
\rput[B](0,-32){$1$}
\rput[B](52,-32){$q$}
\rput[B](104,-32){$2q^2$}
\rput[B](156,-32){$2q^3$}
\rput[B](208,-32){$3q^4$}
\rput[B](260,-32){$2q^5$}
\rput[B](312,-32){$2q^6$}
\rput[B](364,-32){$q^7$}
\rput[B](416,-32){$q^8$}
\rput[B](-45,-3){\scriptsize $j=1$}
\rput[B](-45,12){\scriptsize $j=2$}
\rput[B](-45,27){\scriptsize $j=3$}
\end{pspicture}
\end{center}
\smallskip
\caption{Ramond sectors ($N$ even, $\ell/2$ even): Combinatorial enumeration by 
double-columns of the $q$-binomial $\sqbin nmq=\sqbin 62q
=q^{-1/2} \sum q^{\sum_j m_j E_j}$.  The excess of blue (right) over red (left) 1-strings is given 
by the quantum number $\sigma=\floor{n/2}-m=1$.  The elementary excitation energy of a 
1-string at position $j$ is $E_j=(j-\half)$. The lowest energy configuration has energy 
$E(\sigma)=\half\sigma^2=\half$. At each position $j$, there are $m_j$ 1-strings and 
$n_j=2-m_j$ 2-strings. This analyticity strip is in the upper-half complex $u$-plane rotated by 
180 degrees so that position $j=1$ (furthest from the real axis) is at the bottom. The elementary 
excitations (of energy 1) are generated by either inserting a left-right pair of 1-strings at position 
$j=1$ or promoting a 1-string at position $j$ to position $j+1$.  Notice that 
$\sqbin nmq=\sqbin n{n-m}q$ as $q$-polynomials but they admit different combinatorial 
interpretations because they have different quantum numbers $\sigma$. In the lower half-plane, 
$q$ is replaced with $\qbar$ and no rotation is required. The value $\bar\sigma$ of the quantum 
number in the lower half-plane is related to $\sigma$ in the upper half-plane by the selection 
rules $\sigma+\bar\sigma=\ell/2$ and $\half(\sigma-\bar\sigma)\in\mathbb{Z}$.
\label{binomR1}}
\end{figure}

\begin{figure}[p]
\setlength{\unitlength}{.8pt}
\psset{unit=.8pt}
\begin{center}
\begin{pspicture}(-50,-20)(387,50)
\thicklines
\multirput(0,0)(52,0){8}{\psframe[linewidth=0pt,fillstyle=solid,
  fillcolor=yellow!40!white](-18,-7.5)(18,37.5)}
\multirput(0,0)(52,0){8}{\psline[linewidth=1pt](-20,37.5)(20,37.5)}
\multiput(-7,0)(52,0){8}{\multiput(0,0)(0,15){3}{\psarc[linecolor=black,
  linewidth=.5pt,fillstyle=solid,fillcolor=white](0,0){3}{0}{360}}}
\multiput(7,0)(52,0){8}{\multiput(0,0)(0,15){3}{\psarc[linecolor=black,
  linewidth=.5pt,fillstyle=solid,fillcolor=white](0,0){3}{0}{360}}}
\multiput(-7,0)(0,15){3}{\psarc[linecolor=red,linewidth=.5pt,fillstyle=solid,
  fillcolor=red](0,0){3}{0}{360}}
\multiput(45,0)(0,15){2}{\psarc[linecolor=red,linewidth=.5pt,fillstyle=solid,
  fillcolor=red](0,0){3}{0}{360}}
\multiput(97,0)(0,15){1}{\psarc[linecolor=red,linewidth=.5pt,fillstyle=solid,
  fillcolor=red](0,0){3}{0}{360}}
\multiput(267,0)(0,15){1}{\psarc[linecolor=blue,linewidth=.5pt,fillstyle=solid,
  fillcolor=blue](0,0){3}{0}{360}}
\multiput(319,0)(0,15){2}{\psarc[linecolor=blue,linewidth=.5pt,fillstyle=solid,
  fillcolor=blue](0,0){3}{0}{360}}
\multiput(371,0)(0,15){3}{\psarc[linecolor=blue,linewidth=.5pt,fillstyle=solid,
  fillcolor=blue](0,0){3}{0}{360}}
%
\rput[B](2,50){$\qbin77q$}
\rput[B](54,50){$\qbin76q$}
\rput[B](106,50){$\qbin75q$}
\rput[B](158,50){$\qbin74q$}
\rput[B](210,50){$\qbin73q$}
\rput[B](262,50){$\qbin72q$}
\rput[B](314,50){$\qbin71q$}
\rput[B](366,50){$\qbin70q$}
\rput[B](-40,-25){$\sigma$}
\rput[B](0,-25){$-4$}
\rput[B](52,-25){$-3$}
\rput[B](104,-25){$-2$}
\rput[B](156,-25){$-1$}
\rput[B](208,-25){$0$}
\rput[B](260,-25){$1$}
\rput[B](312,-25){$2$}
\rput[B](364,-25){$3$}
\rput[B](-40,-42){$\sigma_{\text{min}}$}
\rput[B](0,-42){$-3$}
\rput[B](52,-42){$-2$}
\rput[B](104,-42){$-1$}
\rput[B](156,-42){$0$}
\rput[B](208,-42){$0$}
\rput[B](260,-42){$1$}
\rput[B](312,-42){$2$}
\rput[B](364,-42){$3$}
\rput[B](-40,-3){\scriptsize $j=1$}
\rput[B](-40,12){\scriptsize $j=2$}
\rput[B](-40,27){\scriptsize $j=3$}
\end{pspicture}
\end{center}
\smallskip
\caption{Neveu-Schwarz sectors ($N$ even, $\ell/2$ odd): Minimal configurations of 
double-columns within the binomials $\qbin nmq=\qbin7mq$. The energy is $E(\sigma)=\frac{1}{2}\sigma(\sigma+1)$ where
the quantum number is $\sigma=\floor{n/2}-m$. The excess of blue (right) over red (left) 
1-strings in these minimal configurations is $\sigma_{\text{min}}$ as given in (\ref{sigmamin}).  At each position $j$, 
the number of 1-strings $m_j$ plus the number of 2-strings $n_j$ is 2. This analyticity strip is in 
the upper-half complex $u$-plane rotated by 180 degrees so that position $j=1$ 
(furthest from the real axis) is at the bottom.
\label{EsigmaR2}}
\end{figure}

\begin{figure}[p]
\setlength{\unitlength}{.7pt}
\psset{unit=.7pt}
\begin{center}
\begin{pspicture}(-60,-30)(545,170)
\thicklines
\multirput(0,0)(52,0){11}{\psframe[linewidth=0pt,fillstyle=solid,
  fillcolor=yellow!40!white](-18,-7.5)(18,37.5)}
\multirput(0,0)(52,0){11}{\psline[linewidth=1pt](-20,37.5)(20,37.5)}
\multirput(104,60)(52,0){7}{\psframe[linewidth=0pt,fillstyle=solid,
  fillcolor=yellow!40!white](-18,-7.5)(18,37.5)}
\multirput(104,60)(52,0){7}{\psline[linewidth=1pt](-20,37.5)(20,37.5)}
\multirput(208,120)(52,0){3}{\psframe[linewidth=0pt,fillstyle=solid,
  fillcolor=yellow!40!white](-18,-7.5)(18,37.5)}
\multirput(208,120)(52,0){3}{\psline[linewidth=1pt](-20,37.5)(20,37.5)}
%
\multiput(-7,0)(52,0){11}{\multiput(0,0)(0,15){3}{\psarc[linecolor=black,
  linewidth=.5pt,fillstyle=solid,fillcolor=white](0,0){3}{0}{360}}}
\multiput(7,0)(52,0){11}{\multiput(0,0)(0,15){3}{\psarc[linecolor=black,
  linewidth=.5pt,fillstyle=solid,fillcolor=white](0,0){3}{0}{360}}}
\multiput(97,60)(52,0){7}{\multiput(0,0)(0,15){3}{\psarc[linecolor=black,
  linewidth=.5pt,fillstyle=solid,fillcolor=white](0,0){3}{0}{360}}}
\multiput(111,60)(52,0){7}{\multiput(0,0)(0,15){3}{\psarc[linecolor=black,
  linewidth=.5pt,fillstyle=solid,fillcolor=white](0,0){3}{0}{360}}}
\multiput(201,120)(52,0){3}{\multiput(0,0)(0,15){3}{\psarc[linecolor=black,
  linewidth=.5pt,fillstyle=solid,fillcolor=white](0,0){3}{0}{360}}}
\multiput(215,120)(52,0){3}{\multiput(0,0)(0,15){3}{\psarc[linecolor=black,
  linewidth=.5pt,fillstyle=solid,fillcolor=white](0,0){3}{0}{360}}}
\psarc[linecolor=blue,linewidth=.5pt,fillstyle=solid,fillcolor=blue](7,0){3}{0}{360}
\psarc[linecolor=blue,linewidth=.5pt,fillstyle=solid,fillcolor=blue](59,15){3}{0}{360}
\psarc[linecolor=blue,linewidth=.5pt,fillstyle=solid,fillcolor=blue](111,0){3}{0}{360}
\psarc[linecolor=blue,linewidth=.5pt,fillstyle=solid,fillcolor=blue](111,15){3}{0}{360}
\psarc[linecolor=blue,linewidth=.5pt,fillstyle=solid,fillcolor=blue](111,90){3}{0}{360}
\psarc[linecolor=red,linewidth=.5pt,fillstyle=solid,fillcolor=red](149,60){3}{0}{360}
\psarc[linecolor=blue,linewidth=.5pt,fillstyle=solid,fillcolor=blue](163,0){3}{0}{360}
\psarc[linecolor=blue,linewidth=.5pt,fillstyle=solid,fillcolor=blue](163,30){3}{0}{360}
\psarc[linecolor=blue,linewidth=.5pt,fillstyle=solid,fillcolor=blue](163,60){3}{0}{360}
\psarc[linecolor=blue,linewidth=.5pt,fillstyle=solid,fillcolor=blue](163,75){3}{0}{360}
\psarc[linecolor=red,linewidth=.5pt,fillstyle=solid,fillcolor=red](201,0){3}{0}{360}
\psarc[linecolor=red,linewidth=.5pt,fillstyle=solid,fillcolor=red](201,135){3}{0}{360}
\psarc[linecolor=blue,linewidth=.5pt,fillstyle=solid,fillcolor=blue](215,0){3}{0}{360}
\psarc[linecolor=blue,linewidth=.5pt,fillstyle=solid,fillcolor=blue](215,30){3}{0}{360}
\psarc[linecolor=blue,linewidth=.5pt,fillstyle=solid,fillcolor=blue](215,75){3}{0}{360}
\psarc[linecolor=blue,linewidth=.5pt,fillstyle=solid,fillcolor=blue](215,90){3}{0}{360}
\psarc[linecolor=blue,linewidth=.5pt,fillstyle=solid,fillcolor=blue](215,120){3}{0}{360}
\psarc[linecolor=blue,linewidth=.5pt,fillstyle=solid,fillcolor=blue](215,135){3}{0}{360}
\psarc[linecolor=red,linewidth=.5pt,fillstyle=solid,fillcolor=red](253,0){3}{0}{360}
\psarc[linecolor=red,linewidth=.5pt,fillstyle=solid,fillcolor=red](253,75){3}{0}{360}
\psarc[linecolor=red,linewidth=.5pt,fillstyle=solid,fillcolor=red](253,150){3}{0}{360}
\psarc[linecolor=blue,linewidth=.5pt,fillstyle=solid,fillcolor=blue](267,15){3}{0}{360}
\psarc[linecolor=blue,linewidth=.5pt,fillstyle=solid,fillcolor=blue](267,30){3}{0}{360}
\psarc[linecolor=blue,linewidth=.5pt,fillstyle=solid,fillcolor=blue](267,60){3}{0}{360}
\psarc[linecolor=blue,linewidth=.5pt,fillstyle=solid,fillcolor=blue](267,90){3}{0}{360}
\psarc[linecolor=blue,linewidth=.5pt,fillstyle=solid,fillcolor=blue](267,120){3}{0}{360}
\psarc[linecolor=blue,linewidth=.5pt,fillstyle=solid,fillcolor=blue](267,135){3}{0}{360}
\psarc[linecolor=red,linewidth=.5pt,fillstyle=solid,fillcolor=red](305,15){3}{0}{360}
\psarc[linecolor=red,linewidth=.5pt,fillstyle=solid,fillcolor=red](305,90){3}{0}{360}
\psarc[linecolor=red,linewidth=.5pt,fillstyle=solid,fillcolor=red](305,120){3}{0}{360}
\psarc[linecolor=blue,linewidth=.5pt,fillstyle=solid,fillcolor=blue](319,15){3}{0}{360}
\psarc[linecolor=blue,linewidth=.5pt,fillstyle=solid,fillcolor=blue](319,30){3}{0}{360}
\psarc[linecolor=blue,linewidth=.5pt,fillstyle=solid,fillcolor=blue](319,60){3}{0}{360}
\psarc[linecolor=blue,linewidth=.5pt,fillstyle=solid,fillcolor=blue](319,90){3}{0}{360}
\psarc[linecolor=blue,linewidth=.5pt,fillstyle=solid,fillcolor=blue](319,120){3}{0}{360}
\psarc[linecolor=blue,linewidth=.5pt,fillstyle=solid,fillcolor=blue](319,135){3}{0}{360}
\psarc[linecolor=blue,linewidth=.5pt,fillstyle=solid,fillcolor=blue](319,150){3}{0}{360}
\psarc[linecolor=red,linewidth=.5pt,fillstyle=solid,fillcolor=red](357,30){3}{0}{360}
\psarc[linecolor=red,linewidth=.5pt,fillstyle=solid,fillcolor=red](357,75){3}{0}{360}
\psarc[linecolor=blue,linewidth=.5pt,fillstyle=solid,fillcolor=blue](371,15){3}{0}{360}
\psarc[linecolor=blue,linewidth=.5pt,fillstyle=solid,fillcolor=blue](371,30){3}{0}{360}
\psarc[linecolor=blue,linewidth=.5pt,fillstyle=solid,fillcolor=blue](371,60){3}{0}{360}
\psarc[linecolor=blue,linewidth=.5pt,fillstyle=solid,fillcolor=blue](371,75){3}{0}{360}
\psarc[linecolor=blue,linewidth=.5pt,fillstyle=solid,fillcolor=blue](371,90){3}{0}{360}
\psarc[linecolor=red,linewidth=.5pt,fillstyle=solid,fillcolor=red](409,30){3}{0}{360}
\psarc[linecolor=red,linewidth=.5pt,fillstyle=solid,fillcolor=red](409,60){3}{0}{360}
\psarc[linecolor=red,linewidth=.5pt,fillstyle=solid,fillcolor=red](409,75){3}{0}{360}
\psarc[linecolor=blue,linewidth=.5pt,fillstyle=solid,fillcolor=blue](423,0){3}{0}{360}
\psarc[linecolor=blue,linewidth=.5pt,fillstyle=solid,fillcolor=blue](423,15){3}{0}{360}
\psarc[linecolor=blue,linewidth=.5pt,fillstyle=solid,fillcolor=blue](423,30){3}{0}{360}
\psarc[linecolor=blue,linewidth=.5pt,fillstyle=solid,fillcolor=blue](423,60){3}{0}{360}
\psarc[linecolor=blue,linewidth=.5pt,fillstyle=solid,fillcolor=blue](423,75){3}{0}{360}
\psarc[linecolor=blue,linewidth=.5pt,fillstyle=solid,fillcolor=blue](423,90){3}{0}{360}
\psarc[linecolor=red,linewidth=.5pt,fillstyle=solid,fillcolor=red](461,0){3}{0}{360}
\psarc[linecolor=red,linewidth=.5pt,fillstyle=solid,fillcolor=red](461,30){3}{0}{360}
\psarc[linecolor=blue,linewidth=.5pt,fillstyle=solid,fillcolor=blue](475,0){3}{0}{360}
\psarc[linecolor=blue,linewidth=.5pt,fillstyle=solid,fillcolor=blue](475,15){3}{0}{360}
\psarc[linecolor=blue,linewidth=.5pt,fillstyle=solid,fillcolor=blue](475,30){3}{0}{360}
\psarc[linecolor=red,linewidth=.5pt,fillstyle=solid,fillcolor=red](513,15){3}{0}{360}
\psarc[linecolor=red,linewidth=.5pt,fillstyle=solid,fillcolor=red](513,30){3}{0}{360}
\psarc[linecolor=blue,linewidth=.5pt,fillstyle=solid,fillcolor=blue](527,0){3}{0}{360}
\psarc[linecolor=blue,linewidth=.5pt,fillstyle=solid,fillcolor=blue](527,15){3}{0}{360}
\psarc[linecolor=blue,linewidth=.5pt,fillstyle=solid,fillcolor=blue](527,30){3}{0}{360}
\rput[B](-43,-32){$\qbin72q\;=$}
\multiput(21,-32)(52,0){10}{$+$}
\rput[B](0,-32){$1$}
\rput[B](52,-32){$q$}
\rput[B](104,-32){$2q^2$}
\rput[B](156,-32){$2q^3$}
\rput[B](208,-32){$3q^4$}
\rput[B](260,-32){$3q^5$}
\rput[B](312,-32){$3q^6$}
\rput[B](364,-32){$2q^7$}
\rput[B](416,-32){$2q^8$}
\rput[B](468,-32){$q^9$}
\rput[B](520,-32){$q^{10}$}
\rput[B](-45,-3){\scriptsize $j=1$}
\rput[B](-45,12){\scriptsize $j=2$}
\rput[B](-45,27){\scriptsize $j=3$}
\end{pspicture}
\end{center}
\smallskip
\caption{Neveu-Schwarz sectors ($N$ even, $\ell/2$ odd): Combinatorial enumeration by 
double-columns of the $q$-binomial $\sqbin nmq=\sqbin 72q
=q^{-1} \sum q^{\sum_j m_j E_j}$.  The number of positions is \mbox{$(n-1)/2=3$}. The quantum number is $\sigma=\floor{n/2}-m=1$.  
The excess of blue (right) over red (left) 1-strings is $\sigma=1$ or $\sigma+1=2$. The 
elementary excitation energy of a 1-string at position $j$ is $E_j=j$. The lowest energy 
configuration has energy $E(\sigma)=\frac{1}{2}\sigma(\sigma+1)=1$. 
At each position $j$, there are $m_j$ 1-strings and $n_j=2-m_j$ 2-strings. This analyticity strip 
is in the upper-half complex $u$-plane rotated by 180 degrees so that position $j=1$ 
(furthest from the real axis) is at the bottom. The elementary excitations (of energy 1) are 
generated by either inserting a left or right 1-string at position $j=1$ or promoting a 1-string 
at position $j$ to position $j+1$.  Notice that $\sqbin nmq=\sqbin n{n-m}q$ as $q$-polynomials 
but they admit different combinatorial interpretations because they have different quantum 
numbers $\sigma$ and $\sigma'=-\sigma-1$. In calculating $\sigma'$, we have used the fact that $n$ in (\ref{RZ2}) is odd. 
In the lower half-plane, $q$ is replaced with $\qbar$ and no rotation is 
required. The value $\bar\sigma$ of the quantum number in the lower half-plane is related to 
$\sigma$ in the upper half-plane by the selection rules $\sigma+\bar\sigma=(\ell-2)/2$ and 
$\half(\sigma-\bar\sigma)\in\mathbb{Z}$.
\label{binomR2}}
\end{figure}

The combinatorial description of the spectra follows precisely as in \cite{PRV}. 
The building blocks of the spectra in the upper half-plane consist of the ``symplectic" $q$-binomials
\bea
\qbin nmq=\qbin n{\floor{n/2}-\sigma}q=
\begin{cases}
\disp  q^{-\frac{1}{2}\sigma(\sigma+\frac{1}{2})} 
  \sum_{\genfrac{}{}{0pt}{}{\text{$\sigma$-single}}{\text{columns}}} q^{\sum_j m_j E_j},&\mbox{$\mathbb{Z}_4$:\ \ $N,\ell$ odd}\\[22pt]
\disp q^{-\frac{1}{2}\sigma^2} 
\sum_{\genfrac{}{}{0pt}{}{\text{$\sigma$-double}}{\text{columns}}} q^{\sum_j m_j E_j},&\mbox{R:\ \ $N,\ell/2$ even}\\[22pt]
\disp q^{-\frac{1}{2}\sigma(\sigma+1)} 
  \sum_{\genfrac{}{}{0pt}{}{\text{$\sigma$-double}}{\text{columns}}} q^{\sum_j m_j E_j},&\mbox{NS:\ \ $N$ even, $\ell/2$ odd}
\end{cases}
\eea
as shown in Figures~\ref{binomZ4}, \ref{binomR1} and \ref{binomR2}. 
For the one-column diagrams in the $\mathbb{Z}_4$ sectors, the excess $\sigma$ of the number $m_{\text{even}}$ of even 1-strings over the number $m_{\text{odd}}$ of odd 1-strings is
\begin{align}
\sigma&=m_{\text{even}}-m_{\text{odd}}
  =\sum_{k=1}^{\floor{n/2}}m_{2k}-\sum_{k=1}^{\floor{(n+1)/2}}m_{2k-1},\qquad \mbox{$\mathbb{Z}_4$:\ \ $N,\ell$ odd}
\end{align}
For R, NS sectors, the excess $\sigma=\floor{n/2}-m$ of the number of 1-strings in the right column minus the number of 1-strings in the left column of the double-column diagram is given by 
\begin{align}
&m_{\text{right}}-m_{\text{left}}=\begin{cases}
\sigma,&\mbox{R:\ \ $\ell/2$ even}\\
\mbox{$\sigma$ or $\sigma+1$},&\mbox{NS:\ \ $\ell/2$ odd}\end{cases}
\end{align}
The minimal configurations in the upper half-plane with energy $E(\sigma)$ are as shown in Figures~\ref{EsigmaZ4}, \ref{EsigmaR1} and \ref{EsigmaR2}. 

There are similar building blocks in the lower half-plane with $\sigma$ replaced by $\bar\sigma$. 
In each half-plane, the {\em minimum} energy configurations satisfy
\bea
 m_{\text{right}}-m_{\text{left}}=\sigma_{\text{min}}=\begin{cases}
\sigma,&\mbox{$\sigma\ge 0$}\\
\sigma+1,&\mbox{$\sigma<0$}
\end{cases}\label{sigmamin}
\eea
By convention, any zeros on the real $u$ axis are pushed into the upper half plane. 
Empirically determined selection rules dictate that, in a sector with $\ell$ defects, the quantum numbers of the groundstate satisfy
\begin{align}
 \sigma&=\bar\sigma\;=\;\begin{cases}
  (\ell-1)/4,&\mbox{$\ell=1$ mod 4},\quad \mbox{$\mathbb{Z}_4$:\ \ $N$ odd}\\
  -(\ell+1)/4,&\mbox{$\ell=3$ mod 4},\quad \mbox{$\mathbb{Z}_4$:\ \ $N$ odd}\\
\ell/4,\qquad &\mbox{R:\ \ $\ell/2$ even}\\
(\ell-2)/4,\qquad &\mbox{NS:\ \ $\ell/2$ odd}
\end{cases}
\end{align}
with $ \ell=|4\sigma+1|=1,3,5,7,\ldots$ in the $\mathbb{Z}_4$ sectors. 
Similarly, it is found that all excitations satisfy the selection rules
\begin{align}
&\sigma+\bar\sigma=
\begin{cases}
\half(\ell-1),&\mbox{$\ell=1$ mod 4}\\
-\half(\ell+1),&\mbox{$\ell=3$ mod 4}
\end{cases}
\qquad\quad \mbox{$\mathbb{Z}_4$:\ \ $N$ odd}\qquad \half(\sigma-\bar\sigma)\in
\mathbb{Z}&\\
& \sigma+\bar\sigma=\begin{cases}
 \ell/2,&\mbox{\ R:\ \ $\ell/2$ even}\\
 (\ell-2)/2,&\mbox{\ NS:\ \ $\ell/2$ odd}
 \end{cases}\qquad\ \  \half(\sigma-\bar\sigma)\in\mathbb{Z}&
\end{align}
These selection rules hold~\cite{MDPR2013} equally for critical dense polymers and the free-fermion six-vertex model. 

\subsection{Modular invariant partition function}
Using the $q$-binomial building blocks and empirical selection rules for $N$ odd or even, gives  
the finitized partition functions as in \cite{PRV}. In the $\mathbb{Z}_4$ sectors with $N,\ell$ odd
\bea
Z_\ell^{(N)}(q)&\!\!=\!\!&
\begin{cases}
\displaystyle(q\qbar)^{-c/24}\sum_{k\in\mathbb{Z}} 
  q^{\Delta_{2k+\ell/2}}\qbin{\sc{\frac{N+1}{2}}}{\sc{\frac{N-\ell}{4}}-k}q  
\qbar^{\Delta_{2k-\ell/2}}\qbin{\sc{\frac{N-1}{2}}}{\sc{\frac{N-\ell}{4}}+k}\qbar\\[12pt]
\displaystyle(q\qbar)^{-c/24}\sum_{k\in\mathbb{Z}} 
q^{\Delta_{2k+\ell/2}}\qbin{\sc{\frac{N+1}{2}}}{\sc{\frac{N+\ell+2}{4}}\!+\!k}q  
\qbar^{\Delta_{2k-\ell/2}}\qbin{\sc{\frac{N-1}{2}}}{\sc{\frac{N+\ell-2}{4}}\!-\!k}\qbar
\end{cases}\mbox{$N\!-\!\ell=0$ mod 4}\qquad\label{Z4Z1}\\[2pt]
Z_\ell^{(N)}(q)&\!\!=\!\!&
\begin{cases}
\displaystyle(q\qbar)^{-c/24}\sum_{k\in\mathbb{Z}} 
  q^{\Delta_{2k+\ell/2}}\qbin{\sc{\frac{N+1}{2}}}{\sc{\frac{N-\ell+2}{4}}\!-\!k}q  
\qbar^{\Delta_{2k-\ell/2}}\qbin{\sc{\frac{N-1}{2}}}{\sc{\frac{N-\ell-2}{4}}\!+\!k}\qbar\\[12pt]
\displaystyle(q\qbar)^{-c/24}\sum_{k\in\mathbb{Z}} 
q^{\Delta_{2k+\ell/2}}\qbin{\sc{\frac{N+1}{2}}}{\sc{\frac{N+\ell}{4}}+k}q  
\qbar^{\Delta_{2k-\ell/2}}\qbin{\sc{\frac{N-1}{2}}}{\sc{\frac{N+\ell}{4}}-k}\qbar
\end{cases}\mbox{$N\!-\!\ell=2$ mod 4}\qquad\label{Z4Z3}
\eea
For given mod 4 parities of $N-\ell$, these expressions are equivalent as partition functions but, 
in each case, the first form is used for the combinatorial interpretation when $\ell=1$ mod 4 
and the second form when $\ell=3$ mod 4.
As in \cite{PRV}, this leads to
\bea
\sum_{\ell\in2\mathbb{N}-1}^{\ell\le N}\!\!\!Z_\ell^{(N)}(q)=\half(q\qbar)^{-\frac{c}{24}-\frac{3}{32}}\!\bigg[\!
\prod_{n=1}^{\frac{N+1}{2}} (1\!+\!q^{\frac{2n-1}{4}})\!\prod_{n=1}^{\frac{N-1}{2}} (1\!+\!\qbar^{\frac{2n-1}{4}})
+\!\prod_{n=1}^{\frac{N+1}{2}} (1\!-\!q^{\frac{2n-1}{4}})\!\prod_{n=1}^{\frac{N-1}{2}} (1\!-\!\qbar^{\frac{2n-1}{4}})\!\bigg]
\eea
In the R and NS sectors with $N$ even
\bea
Z_\ell^{(N)}(q)=\begin{cases}
\disp (q\qbar)^{-c/24}\sum_{k\in\mathbb{Z}} 
q^{\Delta_{2k+\ell/2}}\qbin{\sc{2\floor{\frac{N+2}{4}}}}{\sc{\floor{\frac{N+2-\ell}{4}}}\!-\!k}q  
\qbar^{\Delta_{2k-\ell/2}}\qbin{\sc{2\floor{\frac{N}{4}}}}{\sc{\floor{\frac{N-\ell}{4}}}\!+\!k}\qbar,
\qquad &\mbox{R:\ \ $\ell/2$ even}\qquad\label{RZ0}\\[16pt]
\disp (q\qbar)^{-c/24}\sum_{k\in\mathbb{Z}} 
q^{\Delta_{2k+\ell/2}}\qbin{\sc{2\floor{\frac{N}{4}}}+1}{\sc{\floor{\frac{N+2-\ell}{4}}}\!-\!k}q  
\qbar^{\Delta_{2k-\ell/2}}\qbin{\sc{2\floor{\frac{N+2}{4}}}-1}{\sc{\floor{\frac{N-\ell}{4}}}\!+\!k}\qbar,
\qquad  &\mbox{NS:\ \ $\ell/2$ odd}
\label{RZ2}
\end{cases}
\eea
In these formulas the central charge and conformal weights, given by the Euler-Maclaurin formula, are
\bea
c=-2,\qquad \Delta=\bar\Delta=\Delta_j=-\frac{1}{8}, 0, \frac{3}{8},\qquad j=0,1,2,\qquad \Delta_j=\frac{j^2-1}{8}
\eea

The modular invariant partition function $Z(q)$ of the free-fermion six-vertex model is given by taking the trace over all $S_z$ sectors with $N$ even. 
Using the explicit expressions in terms of products, as in \cite{PRV}, and summing over the $\ell$ sectors with multiplicities 2 arising from $S_z=\pm\ell$ yields the finitized partition function of the free-fermion six-vertex model
\begin{subequations}
\begin{align}
Z^{N}(q)&=Z_0^{(N)}+2\sum_{\ell\in4\mathbb{N}}^{\ell\le N}Z_\ell^{(N)}(q)+2\sum_{\ell\in 4\mathbb{N}-2}^{\ell\le N}Z_\ell^{(N)}(q)\\
&=\half(q\qbar)^{-\frac{c}{24}-\frac{1}{8}}\bigg[
\prod_{n=1}^{\floor{\frac{N+2}{4}}} (1+q^{n-\frac{1}{2}})^2\prod_{n=1}^{\floor{\frac{N}{4}}} (1+\qbar^{n-\frac{1}{2}})^2\nonumber
+\prod_{n=1}^{\floor{\frac{N+2}{4}}} (1-q^{n-\frac{1}{2}})^2\prod_{n=1}^{\floor{\frac{N}{4}}} (1-\qbar^{n-\frac{1}{2}})^2\bigg]\\
&\qquad\qquad\mbox{}+2(q\qbar)^{-\frac{c}{24}}\prod_{n=1}^{\floor{\frac{N}{4}}} (1+q^n)^2\,\prod_{n=1}^{\floor{\frac{N-2}{4}}} (1+\qbar^n)^2\label{finitized}
\end{align}
\end{subequations}
It is easily checked that the counting of states $Z^{N}(1)=2^N$ is correct at $q=\bar q=1$.

Taking the thermodynamic limit $N\to\infty$ gives the conformal modular invariant partition function
\bea
\begin{array}{rcl}
\disp Z_0(q)\!+\!2\sum_{\ell\in 4\mathbb{N}} Z_\ell(q)\!\!\!&=&\!\!\!
  \disp{\frac{|\vartheta_{0,2}(q)|^2+|\vartheta_{2,2}(q)|^2}{|\eta(q)|^2}}
=|\hat\chi_{-1/8}(q)|^2+|\hat\chi_{3/8}(q)|^2\\
\disp2\!\!\sum_{\ell\in 4\mathbb{N}-2} Z_\ell(q)\!\!\!&=&\!\!\!
  \disp{\frac{|\vartheta_{1,2}(q)|^2+|\vartheta_{3,2}(q)|^2}{|\eta(q)|^2}}
  =\frac{2|\vartheta_{1,2}(q)|^2}{|\eta(q)|^2}
=2|\hat\chi_0(q)+\hat\chi_1(q)|^2\qquad\\
\disp Z(q)=Z_0(q)\!+\!2\sum_{\ell\in 2\mathbb{N}} Z_\ell(q)\!\!\!&=&\!\!\!
  \disp{\frac{1}{|\eta(q)|^2}} \sum_{j=0}^3|\vartheta_{j,2}(q)|^2
=|\hat\chi_{-1/8}(q)|^2+2|\hat\chi_0(q)\!+\!\hat\chi_1(q)|^2+|\hat\chi_{3/8}(q)|^2\\[18pt]
&=&|\varkappa_0^2(q)|^2+2|\varkappa_1^2(q)|^2+|\varkappa_2^2(q)|^2
\end{array}
\eea
where the $u(1)$ and ${\cal W}$-irreducible characters are
\bea
\varkappa_j^n(q)=\frac{1}{\eta(q)}\,\vartheta_{j,n}(q),\qquad 
\begin{array}{rclrcl}
\hat\chi_{-1/8}(q)&\!\!=\!\!&\disp{\frac{1}{\eta(q)}}\,\vartheta_{0,2}(q),\qquad&
\hat\chi_0(q)&\!\!=\!\!& \disp{\frac{1}{2\eta(q)}}[\vartheta_{1,2}(q)+\eta(q)^3]\\[10pt]
\hat\chi_{3/8}(q)&\!\!=\!\!& \disp{\frac{1}{\eta(q)}}\,\vartheta_{2,2}(q),\qquad&
\hat\chi_1(q)&\!\!=\!\!& \disp{\frac{1}{2\eta(q)}}[\vartheta_{1,2}(q)-\eta(q)^3]\qquad
\end{array}
\eea
and the Dedekind eta and theta functions are
\be
\eta(q)=q^{1/24}\prod_{n=1}^\infty (1-q^n),\qquad \vartheta_{j,n}(q)=\sum_{k\in\mathbb{Z}} q^{\frac{(j+2kn)^2}{4n}}
\ee

The MIPF $Z(q)$ of the free-fermion six-vertex model thus precisely agrees with the MIPF of dimers in the usual orientation~\cite{IzPRHu2005}
and critical dense polymers~\cite{MDPR2013}. The latter coincidence is nontrivial as a modified trace is needed to close the cylinder to a torus for this lattice loop model. Although the MIPF agrees with symplectic fermions~\cite{Symplectic}, which is a logarithmic theory, there are no Jordan cells and no indication of logarithmic behaviour for dimers on the cylinder. Indeed, viewing the free-fermion model as the critical eight-vertex model at the decoupling point~\cite{CKP1989}, the MIPF reduces to the square of the Ising model MIPF with central charge $c=\half$
\bea
Z(q)=Z_\text{Ising}(q)^2
\eea
Comparing (\ref{finitized}) with \cite{OPW1996} shows that this relation also holds at the level of the finitized MIPFs. 
To see Jordan cells for dimers, we consider the vacuum boundary condition on the strip in Section~\ref{secStrip}.


\def\leftzig{
\psline[linewidth=1.5pt,linestyle=solid,linecolor=red](0,0)(.5,.5)
\psline[linewidth=1.5pt,linestyle=solid,linecolor=red](0,1)(.5,.5)}
\def\rightzig{
\psline[linewidth=1.5pt,linestyle=solid,linecolor=red](.5,.5)(1,1)
\psline[linewidth=1.5pt,linestyle=solid,linecolor=red](.5,.5)(1,0)}
\def\leftzag{
\psline[linewidth=1.5pt,linestyle=solid,linecolor=red](0,0)(-.5,.5)(0,1)}
\def\rightzag{
\psline[linewidth=1.5pt,linestyle=solid,linecolor=red](1,0)(1.5,.5)(1,1)}


\section{Periodic Dimers on a Finite $M\times N$ Rectangular Lattice}
\label{FiniteRect}

The problem of counting of periodic dimers on a finite $M\times N$ rectangular lattice, in the usual orientation, has been solved exactly~\cite{Kasteleyn1961,IzOganHu2003}. 
The number of periodic dimer configurations is given by 
\bea
\tilde Z_{M\times N}=\half\big(\tilde Z^{1/2,1/2}_{M\times N}+\tilde Z^{0,1/2}_{M\times N}+\tilde Z^{1/2,0}_{M\times N}\big)
\eea
where 
\bea
\tilde Z^{\alpha,\beta}_{M\times N}=\prod_{n=0}^{N/2-1}\prod_{m=0}^{M/2-1} 4\Big(\sin^2\frac{2\pi(n+\alpha)}{N}+\sin^2\frac{2\pi(m+\beta)}{M}\Big),\qquad M,N=2,4,6,\ldots
\eea
Explicitly, arranging the entries in a symmetric matrix gives
\bea
(\tilde Z_{M\times N})=\begin{pmatrix}
8&36&200&1156&\cdots\\
36&272&3,\!108&39,\!952&\cdots\\
200&3,\!108&90,\!176&3,\!113,\!860&\cdots\\
1,\!156&39,\!952&3,\!113,\!860&311,\!853,\!312&\cdots\\
\vdots&\vdots&\vdots&\vdots&\ddots
\end{pmatrix},\qquad M,N=2,4,6,\ldots
\eea

The exact counting of periodic dimer configurations on a finite $M\times N$ rectangular lattice, in the 45 degree rotated orientation, is given by taking the trace of the $M$th power of the transfer matrix (\ref{Tmatrix}) with eigenvalues (\ref{expSoln}). The expressions, however, are more involved than for the usual orientation. 
Explicitly, setting $\rho=\sqrt{2}$ at the isotropic point $u=\pi/4$ with $\epsilon_j=\pm 1$, the number of periodic dimer configurations with the rotated orientation is 
\bea
\hspace{-10pt}Z_{M\times N}\!=\!\begin{cases}\disp 2^{MN+1}\sum_{s=-N+2;4}^N\, \sum_{\sum_{j=1}^N \epsilon_j=s} (-1)^{\frac{M(N-s)}{4}}
\prod_{j=1}^N \cos^M\!\big(\epsilon_j t_j-\tfrac{\pi}{4}\big),\qquad \mbox{$N$ odd}\\[20pt]
\disp 2^{MN}\!\!\!\!\!\!\mathop{\disp\sum_{s=-N}^{N}}_\text{$s=0$ mod 4}\, 
\sum_{\sum_{j=1}^{N} \epsilon_j=-|s|} (-1)^{\frac{M(2N+s)}{4}}\,
\prod_{j=1}^{N} \cos^M\!\big(\epsilon_j t_j^\text{R}-\tfrac{\pi}{4}\big)\\[26pt]
\qquad\mbox{}+\,\disp 2^{MN}\!\!\!\!\!\!\disp\mathop{\disp\sum_{s=-N}^{N}}_\text{$s=2$ mod 4}\,
\sum_{\sum_{j=1}^{N} \epsilon_j=-|s|}\!\!
(-1)^{\frac{M(2N+|s|+2)}{4}}
\disp\prod_{j=1}^{N} \cos^M\!\big(\epsilon_j t_j^\text{NS}-\tfrac{\pi}{4}\big),\qquad\mbox{$N$ even}\end{cases}\hspace{-40pt}\label{RotCounting}
\eea
where $s=S_z$ in the sums increments in steps of 4 as indicated and
\bea
t_j=\frac{(2j-1)\pi}{4N},\qquad t_j^\text{R}=\frac{(2j-1)\pi}{2N},\qquad t_j^{\text{NS}}=
\begin{cases}\frac{j\pi}{N},&j\ne N/2\\ 0,&j=N/2\end{cases}
\label{teta}
\eea
The restrictions on $s$ are compatible with the selection rules and the signs $\epsilon$ in (\ref{epsfix})
ensure that the eigenvalues contribute with the correct overall sign. 
The trigonometric identities~\cite{Jolley}
\bea
\prod_{j=1}^N \cos t_j=2^{1/2-N},\qquad \prod_{j=1}^{N} \cos t_j^\text{R}=(-1)^{N/2}\,2^{1-N},\qquad 
\prod_{j=1,j\ne N/2}^{N} \cos t_j^\text{NS}=(-1)^{N/2} N\,2^{1-N}
\eea
are used to evaluate the products in the denominators arising from the simplification
\bea
1+\epsilon_j \tan t_j=\frac{\cos t_j+\epsilon_j \sin t_j}{\cos t_j}=\sqrt{2}\,\frac{\cos(\epsilon_jt_j-\tfrac{\pi}{4})}{\cos t_j},\qquad \epsilon_j=\pm 1
\eea
For $N$ even, precisely half the eigenvalues in (\ref{RotCounting}) come from the Ramond sectors and half from the Neveu-Schwarz sectors in accord with the binomial identity
\bea
\sum_{s=-N;4}^N 
\genfrac{(}{)}{0pt}{}{N}{\frac{N-s}{2}}=\sum_{s=-N+2;4}^{N-2} 
\genfrac{(}{)}{0pt}{}{N}{\frac{N-s}{2}}=2^{N-1}, 
\qquad s=S_z
\eea

Arranging the entries in a symmetric matrix, the number of periodic dimer configurations for the rotated orientation is
\bea
(Z_{M\times N})=\begin{pmatrix}
4&8&16&32&64&\ldots\\
8&24&80&288&1,\!088&\ldots\\
16&80&448&2,\!624&15,\!616&\ldots\\
32&288&2,\!624&26,\!752&280,\!832&\ldots\\
64&1,\!088&15,\!616&280,\!832&5,\!080,\!064&\ldots\\
\vdots&\vdots&\vdots&\vdots&\vdots&\ddots
\end{pmatrix},\qquad M,N=1,2,3,\ldots
\eea
It is easy to recognize the integer sequences~\cite{oeis} in the first 3 rows. 
The formulas (\ref{RotCounting}) look unwieldy but are straightforward to code in Mathematica~\cite{Wolfram}. In particular, for comparison, the number of periodic dimer configurations $\tilde Z_{8\times 8}$ on an $8\times 8$ square lattice in the usual orientation and the number $Z_{8\times 8}$ in the rotated orientation are
\bea
\tilde Z_{8\times 8}=311,\!853,\!312,\qquad Z_{8\times 8}=38,\!735,\!278,\!017,\!380,\!352
\eea
The difference in magnitude observed here is due to the difference in the unit cells by a linear factor $\sqrt{2}$ on each edge of the rectangle. 
This is in accord with the fact that an $M\times N$ rectangle in the original orientation has $MN/2$ dimers compared to $MN$ dimers in the rotated orientation. 
While the precise counting of dimer configurations differs in the two orientations, the asymptotic growth (\ref{CatalanG}) per dimer coincide
\bea
(\tilde{Z}_{2M,N})^{\frac{1}{MN}}\sim (\tilde{Z}_{M,2N})^{\frac{1}{MN}}\sim (Z_{M,N})^{\frac{1}{MN}}\sim \exp(\tfrac{2G}{\pi})
\eea

\psset{unit=1.3cm}
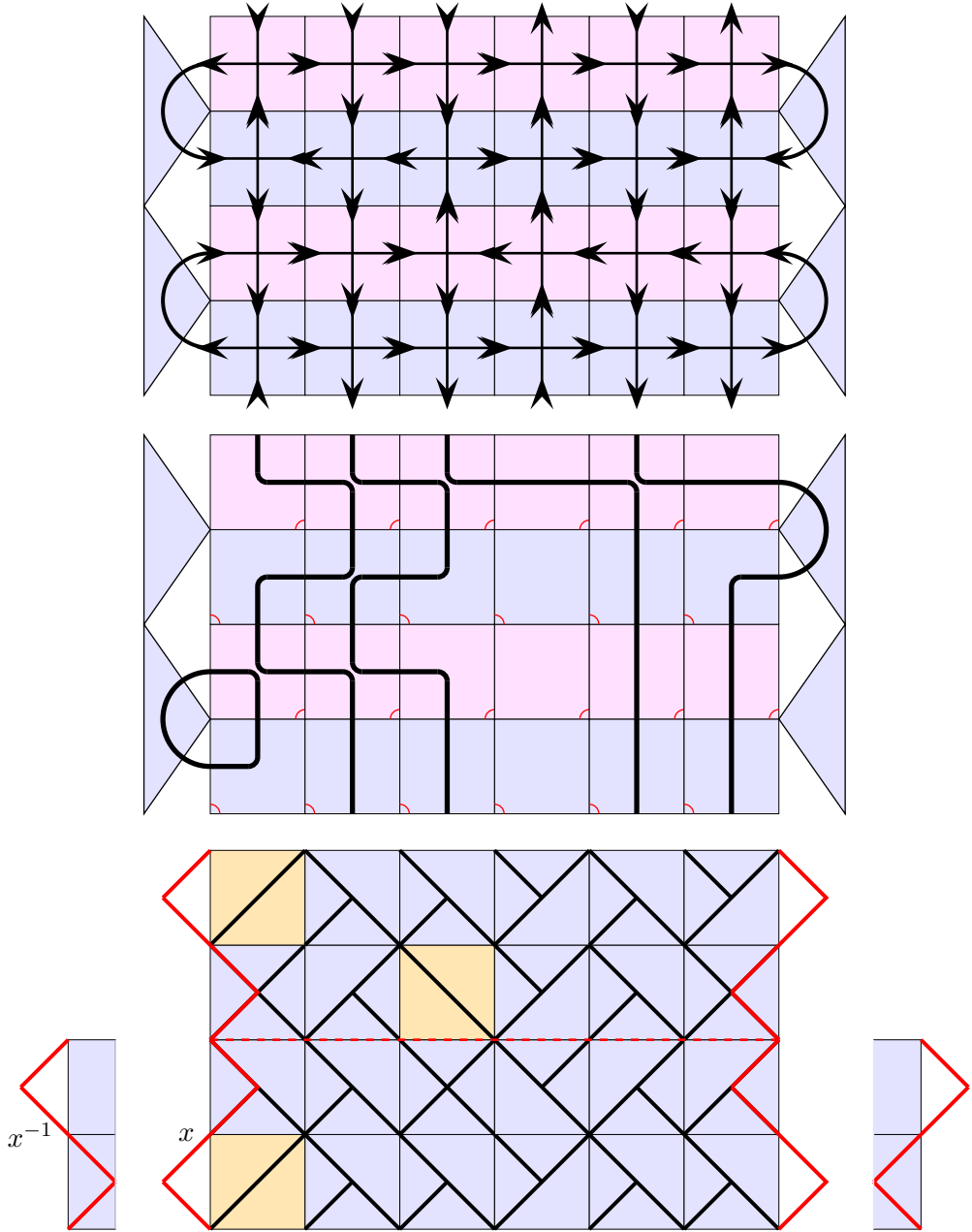
\begin{figure}
\begin{center}
\begin{pspicture}(0,0)(6,4)
\pspolygon[fillstyle=solid,fillcolor=lightlightblue,linewidth=.5pt](-.7,4)(-.7,2)(0,3)
\pspolygon[fillstyle=solid,fillcolor=lightlightblue,linewidth=.5pt](-.7,2)(-.7,0)(0,1)
\pspolygon[fillstyle=solid,fillcolor=lightlightblue,linewidth=.5pt](6.7,4)(6.7,2)(6,3)
\pspolygon[fillstyle=solid,fillcolor=lightlightblue,linewidth=.5pt](6.7,2)(6.7,0)(6,1)
\rput(0,0){\ve}
\rput(1,0){\vd}
\rput(2,0){\vd}
\rput(3,0){\va}
\rput(4,0){\vd}
\rput(5,0){\vd}
\rput(0,1){\vvd}
\rput(1,1){\vvd}
\rput(2,1){\vvf}
\rput(3,1){\vvc}
\rput(4,1){\vvb}
\rput(5,1){\vvb}
\rput(0,2){\vf}
\rput(1,2){\vb}
\rput(2,2){\ve}
\rput(3,2){\va}
\rput(4,2){\vd}
\rput(5,2){\vf}
\rput(0,3){\vve}
\rput(1,3){\vvd}
\rput(2,3){\vvd}
\rput(3,3){\vva}
\rput(4,3){\vvd}
\rput(5,3){\vva}
\psarc[linewidth=1.5pt](0,1){.5}{90}{270}
\psarc[linewidth=1.5pt](0,3){.5}{90}{270}
\psarc[linewidth=1.5pt](6,1){.5}{-90}{90}
\psarc[linewidth=1.5pt](6,3){.5}{-90}{90}
\end{pspicture}

\vspace{.2in}
\mbox{}\vspace{.8cm}\mbox{}\!\!
\begin{pspicture}(0,0)(6,4)
\pspolygon[fillstyle=solid,fillcolor=lightlightblue,linewidth=.5pt](-.7,4)(-.7,2)(0,3)
\pspolygon[fillstyle=solid,fillcolor=lightlightblue,linewidth=.5pt](-.7,2)(-.7,0)(0,1)
\pspolygon[fillstyle=solid,fillcolor=lightlightblue,linewidth=.5pt](6.7,4)(6.7,2)(6,3)
\pspolygon[fillstyle=solid,fillcolor=lightlightblue,linewidth=.5pt](6.7,2)(6.7,0)(6,1)
\rput(0,0){\pe}
\rput(1,0){\pd}
\rput(2,0){\pd}
\rput(3,0){\pa}
\rput(4,0){\pd}
\rput(5,0){\pd}
\rput(0,1){\qb}
\rput(1,1){\qb}
\rput(2,1){\qh}
\rput(3,1){\qa}
\rput(4,1){\qd}
\rput(5,1){\qd}
\rput(0,2){\pf}
\rput(1,2){\pb}
\rput(2,2){\pe}
\rput(3,2){\pa}
\rput(4,2){\pd}
\rput(5,2){\pf}
\rput(0,3){\qg}
\rput(1,3){\qb}
\rput(2,3){\qb}
\rput(3,3){\qc}
\rput(4,3){\qb}
\rput(5,3){\qc}
\psarc[linewidth=2pt](0,1){.5}{90}{270}
\psarc[linewidth=2pt](6,3){.5}{-90}{90}
\end{pspicture}

\vspace{-.13in}
\begin{pspicture}(0,0)(6,4)
\rput(0,0){\de}
\rput(1,0){\dd}
\rput(2,0){\dd}
\rput(3,0){\da}
\rput(4,0){\dd}
\rput(5,0){\dd}
\rput(0,1){\dd}
\rput(1,1){\dd}
\rput(2,1){\df}
\rput(3,1){\dc}
\rput(4,1){\db}
\rput(5,1){\db}
\rput(0,2){\df}
\rput(1,2){\db}
\rput(2,2){\dee}
\rput(3,2){\da}
\rput(4,2){\dd}
\rput(5,2){\df}
\rput(0,3){\de}
\rput(1,3){\dd}
\rput(2,3){\dd}
\rput(3,3){\da}
\rput(4,3){\dd}
\rput(5,3){\da}
\psframe[fillstyle=solid,fillcolor=lightlightblue,linewidth=0pt](-1.5,0)(-1.,1)
\psframe[fillstyle=solid,fillcolor=lightlightblue,linewidth=0pt](-1.5,1)(-1.,2)
\psframe[fillstyle=solid,fillcolor=lightlightblue,linewidth=0pt](7,0)(7.5,1)
\psframe[fillstyle=solid,fillcolor=lightlightblue,linewidth=0pt](7,1)(7.5,2)
\rput(-1,0){\rightzig}
\rput(-1,1){\rightzag}
\rput(-1,2){\rightzag}
\rput(-1,3){\rightzig}
\rput(-1.5,0){\leftzig}
\rput(-2.5,1){\rightzig}
\rput(5,0){\rightzag}
\rput(5,1){\rightzig}
\rput(5,2){\rightzig}
\rput(5,3){\rightzag}
\rput(6.5,0){\rightzig}
\rput(7.5,1){\leftzig}
\psline[linewidth=.5pt,linecolor=lightlightblue](-1.,0)(-1.,2)
\psline[linewidth=.5pt,linecolor=lightlightblue](7.,0)(7.,2)
\psline[linewidth=1pt,linecolor=red,linestyle=dashed,dash=3pt 3pt](0,2)(6,2)
\rput(-.25,1){$x$}
\rput(-1.9,1){$x^{-1}$}
\end{pspicture}
\end{center}
\caption{\label{vacBdy}Typical dimer configuration on a $6\times 4$ strip with vacuum boundary conditions in the vertex, particle and dimer representations. For the vertex representation, the boundary arrows can be in either one of the two possible directions (corresponding to a particle or vacancy in the particle representation). Particles move up and right on odd rows and up and left on even rows. The number of particles/down arrows inside the strip is conserved from double row to double row but not necessarily in intermediate rows. For dimers, there are two different zigzag edges allowed independently on the left and right edges of each double row. The left boundary zigzags have weights $x,x^{-1}$ as shown. The right boundary zigzags have weight 1. The mapping between arrow and dimer configurations at the upper and lower edges is one-to-many since an apricot face can allow two dimer configurations locally both of which should be summed over to get the correct face weights. This is automatically accounted for in taking the trace in the vertex representation.}
\end{figure}

\section{Vacuum Boundary Conditions on the Strip and Jordan Cells}
\label{secStrip}


The vacuum boundary condition for dimers is shown in Figure~\ref{vacBdy}. This is the ``vacuum" in the sense that it is the Kac $(r,s)=(1,1)$ boundary condition with no boundary seams. The normalized double row transfer matrix is
\psset{unit=.9cm}
\setlength{\unitlength}{.9cm}
\bea
\vec D(u)\,=\,\frac{1}{\sin2u} \ \ 
\begin{pspicture}[shift=-1.1](-.7,.75)(8.7,3)
\facegrid{(0,1)}{(8,3)}
\pspolygon[fillstyle=solid,fillcolor=lightlightblue](0,2)(-.7,1)(-.7,3)
\pspolygon[fillstyle=solid,fillcolor=lightlightblue](8,2)(8.7,1)(8.7,3)
\leftarc{(0,2)}
\rightarc{(8,2)}
\rput(0.5,1.5){\small $u$}
\rput(7.5,1.5){\small $u$}
\rput(0.5,2.5){\small $u$}
\rput(7.5,2.5){\small $u$}
\rput(2.5,1.5){\small $\dots$}
\rput(5.5,1.5){\small $\dots$}
\rput(2.5,2.5){\small $\dots$}
\rput(5.5,2.5){\small $\dots$}
\psarc[linecolor=red,linewidth=.5pt](0,1){.15}{0}{90}
\psarc[linecolor=red,linewidth=.5pt](7,1){.15}{0}{90}
\psarc[linecolor=red,linewidth=.5pt](1,2){.15}{90}{180}
\psarc[linecolor=red,linewidth=.5pt](8,2){.15}{90}{180}
\end{pspicture}\quad
\label{D}
\eea
where the left and right triangular boundary weights are
\bea
\psset{unit=.8cm}
\quad
\begin{pspicture}[shift=-.9](-.7,0)(.2,2)
\pspolygon[fillstyle=solid,fillcolor=lightlightblue,linewidth=.75pt](-.7,2)(-.7,0)(0,1)
\psarc[linewidth=1.5pt](0,1){.5}{90}{270}
\psline[linewidth=1.5pt,arrowsize=6pt]{->}(0,1.5)(.3,1.5)
\psline[linewidth=1.5pt,arrowsize=6pt]{-<}(0,.5)(.3,.5)
\end{pspicture}\;=\;x,\qquad
\begin{pspicture}[shift=-.9](-.7,0)(.2,2)
\pspolygon[fillstyle=solid,fillcolor=lightlightblue,linewidth=.75pt](-.7,2)(-.7,0)(0,1)
\psarc[linewidth=1.5pt](0,1){.5}{90}{270}
\psline[linewidth=1.5pt,arrowsize=6pt]{-<}(0,1.5)(.3,1.5)
\psline[linewidth=1.5pt,arrowsize=6pt]{->}(0,.5)(.3,.5)
\end{pspicture}\;=\;x^{-1},\qquad
\begin{pspicture}[shift=-.9](-.7,0)(.2,2)
\pspolygon[fillstyle=solid,fillcolor=lightlightblue,linewidth=.75pt](0,2)(0,0)(-.7,1)
\psarc[linewidth=1.5pt](-.7,1){.5}{-90}{90}
\psline[linewidth=1.5pt,arrowsize=6pt]{->}(-.7,1.5)(-1,1.5)
\psline[linewidth=1.5pt,arrowsize=6pt]{-<}(-.7,.5)(-1,.5)
\end{pspicture}\;=\;1,\qquad
\begin{pspicture}[shift=-.9](-.7,0)(.2,2)
\pspolygon[fillstyle=solid,fillcolor=lightlightblue,linewidth=.75pt](0,2)(0,0)(-.7,1)
\psarc[linewidth=1.5pt](-.7,1){.5}{-90}{90}
\psline[linewidth=1.5pt,arrowsize=6pt]{-<}(-.7,1.5)(-1,1.5)
\psline[linewidth=1.5pt,arrowsize=6pt]{->}(-.7,.5)(-1,.5)
\end{pspicture}\;=\;1
\eea

The commuting double row transfer matrices~\cite{Sklyanin} satisfy the same inversion identity~\cite{PRpolymers,PRVKac} as critical dense polymers with vacuum boundary conditions
\bea
 \vec D(u)\vec D(u+\lambda)\,=\,\left(\frac{\cos^{2N}\!u-\sin^{2N}\!u}{\cos^2\!u-\sin^2\!u}
  \right)^{\!2}\Ib
\label{DDI}
\eea
Similarly, the ordinates $y_j$ of the 1-strings and their conformal excitation energies $E_j$ are given by the same expressions
\bea
y_j\,=\,-\frac{i}{2}\ln\tan\frac{E_j\pi}{2N},\qquad 
E_j=\begin{cases}j,&\mbox{$N$ even}\\ j-\half,&\mbox{$N$ odd}\end{cases}
\label{yj}
\eea
Moreover, the ground state patterns of zeros coincide for $N$ even ($\ell=|S_z|=0$, $(r,s)=(1,1)$) and $N$ odd 
($\ell=|S_z|=1$, $(r,s)=(1,2)$). It follows that the same calculation, based on Euler-Maclaurin, applies with the expected results $c=-2$, $\Delta_{1,1}=0$ and $\Delta_{1,2}=-\frac{1}{8}$. The difference between dimers and critical dense polymers on the strip with vacuum boundary conditions resides in the counting and classification of states.

The Hamiltonian for dimers with the vacuum boundary condition on the strip is given by
\bea
{\cal H}=-\half\,\frac{d}{du} \log \vec D(u)\Big|_{u=0}
\eea
It coincides with the $U_q(sl(2))$-invariant XX Hamiltonian~\cite{GST2014} of the free-fermion six-vertex model
\begin{align}
{\cal H}&=-\sum_{j=1}^{N-1}e_j=-\half\sum_{j=1}^{N-1} (\sigma_j^x\sigma_{j+1}^x+\sigma_j^y\sigma_{j+1}^y)-\half i(\sigma_1^z-\sigma_N^z)\\
&=-\sum_{j=1}^{N-1}(f_j^\dagger f_{j+1}+f_{j+1}^\dagger f_j)-i( f_1^\dagger f_1- f_N^\dagger f_N)
\end{align}
where $\sigma_j^{x,y,z}$ are Pauli matrices and $f_j=\half(\sigma_j^x-i\sigma_j^y)$, $f_j^\dagger=\half(\sigma_j^x+i\sigma_j^y)$. This Hamiltonian is manifestly not Hermitian. Nevertheless, the spectra of this Hamiltonian is real~\cite{MDRRSA2015}. 
The Jordan canonical forms for $N=2$ and $N=4$ respectively are
\begin{align}
&0\oplus\mbox{\scriptsize $\begin{pmatrix}0&1\\ 0&0\end{pmatrix}$}\oplus 0\\
&0\oplus\mbox{\scriptsize $\begin{pmatrix}0&1\\ 0&0\end{pmatrix}$}\oplus 0\oplus 0\oplus
\mbox{\scriptsize $\begin{pmatrix}0&1\\ 0&0\end{pmatrix}$}\oplus 0
\oplus( -\sqrt{2})\oplus\mbox{\scriptsize $\begin{pmatrix}-\sqrt{2}&1\\ 0&-\sqrt{2}\end{pmatrix}$}\oplus (-\sqrt{2})
\oplus \sqrt{2}\oplus\mbox{\scriptsize $\begin{pmatrix}\sqrt{2}&1\\ 0&\sqrt{2}\end{pmatrix}$}\oplus \sqrt{2}
\end{align}
In the continuum scaling limit, the Hamiltonian gives the Virasoro dilatation operator $L_0$. Assuming that the Jordan cells persist in this scaling limit, the representation is reducible yet indecomposable and so, as a CFT, dimers is logarithmic.

\section{Conclusion}

It is often stated that two-dimensional lattice models are exactly solvable if their Boltzmann weights satisfy the local Yang-Baxter equation so that they admit a family of commuting transfer matrices with an infinite number of conserved quantities. In a sense, Yang-Baxter integrability is the {\em gold standard\/} for solvability on the lattice. Until now, dimers has been solved exactly by Pfaffian and other techniques but not by Yang-Baxter methods. Now, dimers is brought firmly into the framework of Yang-Baxter integrability. For periodic transfer matrices, through the special inversion identity, this has enabled the detailed calculation of the dimer model spectra on the cylinder in the $\mathbb{Z}_4$, Ramond and Neveu-Schwarz sectors for arbitrary finite sizes. Taking a trace to form a torus and combining these sectors at the isotropic point $u=\frac{\lambda}{2}=\frac{\pi}{4}$ yields explicit formulas for the counting of dimer configurations on arbitrary periodic $M\times N$ rectangular lattices. Because the orientation of the dimers is rotated by $45\degree$, the precise counting of these states differs from the counting of configurations for the usual orientation on the square lattice even though the residual entropies coincide. 

The inclusion of spatial anisotropy and the spectral parameter $u$ enables the analytic calculation of the complete finite-size spectra of dimers yielding the central charge $c=-2$ and conformal weights $\Delta=-\frac{1}{8},0,\frac{3}{8}$. Remarkably, the modular invariant partition function precisely coincides with that of critical dense polymers sector-by-sector even though critical dense polymers requires the implementation of a modified trace. Because dimers (six-vertex model at $\lambda=\frac{\pi}{2}$) exhibits Jordan cells on the strip with the vacuum boundary conditions, we argue that dimers is best described as a logarithmic CFT with central charge $c=-2$ and effective central charge $c_\text{eff}=1$. 

Since the bulk CFTs appear to be the same, at least in terms of spectra, it is tempting to argue that dimers and critical dense polymers lie in the same {\em universality class}. But, in considering universality, it should be borne in mind that, since these two theories exhibit different Jordan cell structures, they should be regarded as different logarithmic CFTs.
Of course, critical dense polymers ${\cal LM}(1,2)$ is just the first member of the family of logarithmic minimal models ${\cal LM}(p,p')$~\cite{PRZ2006}. It is therefore natural to ask whether the coincidence, at $\lambda=\frac{(p'-p)\pi}{p'}=\frac{\pi}{2}$, between the ${\cal LM}(p,p')$ and six-vertex model MIPFs extends to other values of $(p,p')$. 

Yang-Baxter integrability opens up further avenues for future research. Commuting double row transfer matrices and inversion identities can now be used to elucidate the role of different dimer boundary conditions on the strip. These are expected to include Kac and Robin $(r,s)$ type boundary conditions. Insight may also be gained into boundary conditions analogous to the ``current" boundary conditions (left and right boundary arrows both point to the right) and to domain wall boundary conditions and Aztec diamonds (left and right boundary arrows both point out). We plan to study these questions in a future paper~\cite{PRVinprogress}. Finally, it is known that the inversion identity methods extend off-criticality to the elliptic eight-vertex free-fermion model~\cite{Felderhof}. It would be of interest to study the off-critical dimer model given by the free-fermion eight-vertex model at $\lambda=\frac{\pi}{2}$. An extension of the mapping of \cite{KorepinZJ} and Figure~\ref{vpd} suggests that this should involve horizontal and vertical dimers in addition to their $45\degree$  rotated counterparts. Interestingly, through the mappings in Figure~\ref{vpd}, the critical six-vertex model (\ref{6Vwts}) can also be viewed as a critical model of interacting dimers with anisotropic 3-dimer interactions associated with the face weights $a(u)$ and $b(u)$.

\subsection*{Acknowledgments}

AVO is supported by a Melbourne International Research Scholarship and a Melbourne International Fee Remission Scholarship. We gratefully acknowledge the hospitality of Holger Frahm at Hannover University where part of this work was carried out. PAP also thanks the Asia Pacific Center for Theoretical Physics, POSTECH, Pohang, South Korea and the International Center for Theoretical Physics, Trieste, Italy for support during a visit to the APCTP as an ICTP Visiting Scholar. We thank Alexi Morin-Duchesne and Jorgen Rasmussen for useful discussions.

\appendix
\section{Proof of Inversion Identities on the Cylinder}

\label{CylInvProof}
For completeness, in this appendix, we present the derivation following Felderhof~\cite{Felderhof} of the inversion identities (\ref{InvIdCyl}) for periodic boundary conditions on the cylinder 
\begin{subequations}
\begin{align}
\vec T_d(u)\vec T_d(u+\lambda)&=\big(\cos^{2N} u- \sin^{2N} u\big) I,\qquad &\mbox{$N$ odd}\label{AInvIdCylOdd}\\[4pt]
\vec T_d(u)\vec T_d(u+\lambda)&=\big(\cos^{2N} u+ \sin^{2N} u+2(-1)^d \sin^N u\cos^N u\big) I,\quad &\mbox{$N$ even}\label{AInvIdCylEven}
\end{align}
\label{AInvIdCyl}
\end{subequations}
For simplicity, since the transfer matrix is independent of the gauge, we work in the gauge $g=\rho=1$.

For a 2-column at position $j$ with fixed $a_j,b_j$, let us define the following four $4\times 4$ matrices
\bea
\psset{unit=1cm}
R\!\begin{pmatrix}b_j\\ a_j\end{pmatrix}=\begin{pspicture}[shift=-1](-.4,0)(1.3,2)
\rput(0,0){\oface u}
\rput(0,1){\oface {u\!+\!\lambda}}
\rput[t](.5,-.1){\small $a_j$}
\rput[b](.5,2.1){\small $b_j$}
\rput[r](-.1,.5){\small $c$}
\rput[r](-.1,1.5){\small $c'$}
\rput[l](1.1,.5){\small $d$}
\rput[l](1.1,1.5){\small $d'$}
\end{pspicture}
\eea 
Ordering the four intermediate basis states as
\bea
\begin{pmatrix}c'\\ c\end{pmatrix}=\begin{pmatrix}0\\ 0\end{pmatrix}, \begin{pmatrix}1\\ 0\end{pmatrix}, \begin{pmatrix}0\\ 1\end{pmatrix}, \begin{pmatrix}1\\ 1\end{pmatrix}
\eea
the explicit form of these $R$ matrices is
\begin{align}
R\!\begin{pmatrix}0\\ 0\end{pmatrix}&=\mbox{\scriptsize $\begin{pmatrix}-\sin u\cos u&0&0&0\\ 0&\cos^2 u&0&0\\0&1&-\sin^2 u&0\\ 0&0&0&\sin u\cos u\end{pmatrix}$},&
R\!\begin{pmatrix}1\\ 1\end{pmatrix}&=\mbox{\scriptsize $\begin{pmatrix}\sin u\cos u&0&0&0\\ 0&-\sin^2 u&1&0\\0&0&\cos^2 u&0\\ 0&0&0&-\sin u\cos u\end{pmatrix}$}&\nonumber\\
R\!\begin{pmatrix}1\\ 0\end{pmatrix}&=\mbox{\scriptsize $\begin{pmatrix}0&0&0&0\\ \cos u&0&0&0\\ \cos u&0&0&0\\ 0&-\sin u&\sin u&0\end{pmatrix}$},&
R\!\begin{pmatrix}0\\ 1\end{pmatrix}&=\mbox{\scriptsize $\begin{pmatrix}0&\sin u&-\sin u&0\\ 0&0&0&\cos u\\0&0&0&\cos u\\ 0&0&0&0\end{pmatrix}$}&
\end{align}
It follows that the matrix entries of the left-side of the inversion identity are given by the trace of an ordered matrix product
\bea
[\vec T_d(u)\vec T_d(u+\lambda)]_{\svec a,\svec b}=\Tr \prod_{j=1}^N R\!\begin{pmatrix}b_j\\ a_j\end{pmatrix},\qquad a_j,b_j=0,1\label{MatrixEntry}
\eea
where the lower and upper row configurations are $\vec a=\{a_1,a_2,\ldots,a_N\}$, $\vec b=\{b_1,b_2,\ldots,b_N\}$. 

Carrying out a similarity transformation with the matrices
\begin{align}
S&=\mbox{\scriptsize $\begin{pmatrix}0&x_1&x_2&0\\ x_3&0&0&x_4\\ x_5&0&0&x_6\\ 0&x_7&x_8&0\end{pmatrix}$}
=\mbox{\scriptsize $\begin{pmatrix}0&0&1&0\\ 0&0&0&-1\\ 1&0&0&0\\ 0&1&-1&0\end{pmatrix}$},\qquad
x_1=x_3=x_6=0,\quad x_2=x_5=x_7=1,\quad x_4=x_8=-1\\
S^{-1}&=\mbox{\scriptsize $\begin{pmatrix}0&y_1&y_2&0\\ y_3&0&0&y_4\\ y_5&0&0&y_6\\ 0&y_7&y_8&0\end{pmatrix}$}
=\mbox{\scriptsize $\begin{pmatrix}0&0&1&0\\ 1&0&0&1\\ 1&0&0&0\\ 0&-1&0&0\end{pmatrix}$},\qquad
y_1=y_6=y_8=0,\quad y_2=y_3=y_4=y_5=1,\quad y_7=-1
\end{align}
brings the four ``diagonal" $R$ matrices simultaneously to upper triangular form
\begin{subequations}
\begin{align}
SR\!\begin{pmatrix}0\\ 0\end{pmatrix}\!S^{-1}&=\mbox{\scriptsize $\begin{pmatrix}\cos^2 u&0&0&1\\ 0&\sin u\cos u&0&0\\ 0&0&-\sin u\cos u&0\\ 0&0&0&-\sin^2 u\end{pmatrix}$}\\ 
SR\!\begin{pmatrix}1\\ 1\end{pmatrix}\!S^{-1}&=\mbox{\scriptsize $\begin{pmatrix}\cos^2 u&0&0&0\\ 0&-\sin u\cos u&0&0\\ 0&0&\sin u\cos u&0\\ 0&0&0&-\sin^2 u\end{pmatrix}$} \\
SR\!\begin{pmatrix}1\\ 0\end{pmatrix}\!S^{-1}&=\mbox{\scriptsize $\begin{pmatrix}0&0&\cos u&0\\ 0&0&0&\sin u\\ 0&0&0&0\\ 0&0&0&0\end{pmatrix}$} \\
SR\!\begin{pmatrix}0\\ 1\end{pmatrix}\!S^{-1}&=\mbox{\scriptsize $\begin{pmatrix}0&-\cos u&0&0\\ 0&0&0&0\\ 0&0&0&\sin u\\ 0&0&0&0\end{pmatrix}$} 
\end{align}
\end{subequations}
The required inversion identities (\ref{AInvIdCyl}) then follow immediately 
\bea
\vec T_d(u)\vec T_d(u+\lambda)=\big[(\cos^2 u)^N+(-\sin^2 u)^N+(-1)^{N-d}(\sin u\cos u)^N+(-1)^d(\sin u\cos u)^N\big]I
\eea

\goodbreak 


\end{document}